\documentclass[a4paper,11pt]{article}
\pdfoutput=1

\usepackage{jheppub}
\usepackage[T1]{fontenc}
\usepackage{bm}
\usepackage{booktabs}
\usepackage{graphicx}
\usepackage{tikzexternal}

\DeclareMathOperator{\diag}{diag}

\newcommand{\vket}[3]{
\biggl\lvert #1
\begin{matrix} #2 \\ #3 \end{matrix}
\biggr\rangle}

\newcommand{\vbra}[3]{
\biggl\langle #1
\begin{matrix} #2 \\ #3 \end{matrix}
\biggr\rvert}

\title{Hidden-heavy Pentaquarks\texorpdfstring{\\}{ }in the Born-Oppenheimer Approximation for QCD}

\author{Fareed Alasiri,}
\author{Eric Braaten}
\author{and Roberto Bruschini}
\affiliation{Department of Physics,
         The Ohio State University, Columbus, OH\ 43210, USA}

\emailAdd{alasiri.6@osu.edu}
\emailAdd{braaten.1@osu.edu}
\emailAdd{roberto.bruschini@tum.de}
\preprint{October 5, 2026}

\abstract{
Hidden-heavy pentaquarks can be identified as bound states in diabatic Born-Oppenheimer potentials that are repulsive color-Coulomb potentials at short distance, cross below a heavy-hadron-pair threshold, and then approach that threshold at long distance.
We identify the pentaquarks $P_{c\bar{c}}(4312)^+$, $P_{c\bar{c}}(4337)^+$, $P_{c\bar{c}}(4440)^+$, and $P_{c\bar{c}}(4457)^+$ as the ground-state quartet for $c \bar c$ in the two coupled potentials associated with a spin-$3/2$ adjoint baryon. 
We predict that their $J^P$ quantum numbers are $\frac{1}{2}^-$, $\frac{3}{2}^-$, $\frac{3}{2}^-$, and $\frac{5}{2}^-$, respectively. 
Our prediction for $P_{c\bar{c}}(4457)^+$ differs from almost all previous predictions.
Our prediction that there are only four hidden-charm pentaquarks differs qualitatively from that of simple molecular models that there should be seven.
We identify the strange hidden-charm pentaquarks $P_{c\bar{c}s}(4338)^0$ and $P_{c\bar{c}s}(4459)^0$ as two members of the ground-state triplet for $c \bar c$ in the potential associated with a spin-$1/2$ strange adjoint baryon. 
We also present model-independent predictions for pentaquark decays into charmonium plus a light baryon and into a charm baryon and a charm meson.
}


\begin{document} 
\maketitle
\flushbottom

\section{Introduction}

The discovery of exotic heavy hadrons has ushered in a new era in hadron spectroscopy. 
Most of the observed states are \emph{exotic hidden-heavy hadrons} containing a charm quark-antiquark pair ($c\bar{c}$) or a bottom quark-antiquark pair ($b\bar{b}$). 
The list includes about four dozen $c\bar{c}$ tetraquark mesons, about a half dozen $b\bar{b}$ tetraquark mesons, and about a half dozen $c\bar{c}$ pentaquark baryons.  

There have been numerous attempts using constituent models to explain the exotic hidden-heavy hadrons, but none have yet succeeded; see, for instance, Refs.~\cite{Ali17,Esp17,Leb17,Guo18,Kar18,Ols18,Bra20,Che23} and references therein.
In quark models, the constituents are the heavy quark, the heavy antiquark, and light quarks and antiquarks, which all interact through strong color forces.
In diquark models, the constituents are colored quark clusters that interact through strong color forces.
In molecular models, the constituents are color-singlet hadrons that interact through light-hadron exchange.
Constituent models have many parameters whose relation to the fundamental parameters of QCD is unknown.

The Born-Oppenheimer (B\nobreakdash-O) approximation provides an approach to hidden-heavy hadrons that is firmly based on QCD. In the B\nobreakdash-O approach, the problem is separated into two steps. 
One step is calculating the B\nobreakdash-O potentials using lattice QCD. 
The other step is solving the Schr\"odinger equation for the heavy quark and antiquark in those potentials. 
The B\nobreakdash-O approximation for QCD is a compelling theoretical framework for the study of any hadron containing multiple heavy quarks and/or antiquarks \cite{Bra18a,Sot20a,Ber24}.
The B\nobreakdash-O approximation has been used in conjunction with lattice QCD  to study double-bottom tetraquarks \cite{Bic17b,Hof24}.
The diabatic representation of the B\nobreakdash-O approximation has been used to derive model-independent selection rules and ratios of decay rates for hidden-heavy hadrons and double-heavy hadrons into pairs of heavy hadrons \cite{Braa24a,Bru24}.
The techniques of the B\nobreakdash-O approximation have also been used in constituent models, including models with diquark and meson constituents \cite{Gir19,Leb25} and models with meson and quark constituents \cite{Zha25}.
There are also diquark models that are motivated by the general philosophy of the B\nobreakdash-O approximation but do not use its techniques \cite{Liu14,Mai22,Ger25,Kan25}.

The B\nobreakdash-O potentials for exotic hidden-heavy hadrons are discrete energy levels of QCD with static color-triplet and color-antitriplet sources separated by a variable distance.
At short  distances, the potential approaches a repulsive color-Coulomb potential offset by the energy of an adjoint hadron, which is a discrete energy level of QCD with a static color-octet source. 
At long distance, the potential approaches the threshold for a pair of heavy hadrons.
In Ref.~\cite{Braa24b}, the exotic hidden-heavy hadrons were identified as bound states or resonances in B\nobreakdash-O potentials that cross below a heavy-hadron-pair threshold before approaching it. 
In Ref.~\cite{Alas25}, we applied this identification to hidden-heavy pentaquarks.
A mismatch between the numbers of ground-state adjoint-baryon potentials at short distance and heavy-hadron-pair  potentials at long distance leads to remarkable predictions for hidden-heavy pentaquarks.
In this paper, we discuss the application of the B\nobreakdash-O approximation to the hidden-charm pentaquarks in greater detail.
In Ref.~\cite{Bram25}, the B\nobreakdash-O approximation was applied to the hidden-heavy pentaquarks using a more flexible approach that ignored the mismatch in the numbers of potentials.

The contents of this paper are organized as follows. In Sec.~\ref{pentaquarkbaryons}, we review the existing data on hidden-heavy pentaquarks. 
In Sec.~\ref{statichadrons}, we determine the quantum numbers of the lowest-energy adjoint baryons.  
In Sec.~\ref{pentaquarkbopotentials}, we determine the qualitative behaviors of the B\nobreakdash-O potentials for hidden-heavy pentaquarks. 
In Sec.~\ref{sec:pentaquarkEnergies}, we predict the quantum numbers of the hidden-charm pentaquarks and we estimate their masses by including the effects of charm-hadron spin splittings as a first-order perturbation. 
In Sec.~\ref{sec:Decays}, we present model-independent predictions for ratios of decay rates of the hidden-charm pentaquarks. 
We show that our identifications of the pentaquarks are compatible with their narrow widths.
In Sec.~\ref{sec:Summary}, we discuss these results and summarize our conclusions.
Details of the calculations of pentaquark spin splittings, amplitudes, and decays that involve angular-momentum algebra with Wigner $6j$ and $9j$ symbols are relegated to Appendices~\ref{app:PentaquarkSplittingsAmplitudes} and \ref{app:PentaquarkDecays}.

\section{Observed Pentaquarks}
\label{pentaquarkbaryons}

The first pentaquark was discovered by the LHCb Collaboration in 2015 as a peak near 4450~MeV in the $J\!/\!\psi \, p$ invariant-mass distribution from the decay $\Lambda_b^0\to J\!/\!\psi \, p \, K^-$ \cite{LHCb15}.
A broad peak near 4380~MeV was also reported in Ref.~\cite{LHCb15} with mass $4380\pm 8 \pm 29$~MeV and large width $205 \pm 18 \pm 86$~MeV. 
In 2019, LHCb resolved the peak near 4450~MeV into two separate peaks near 4440 and 4457~MeV and discovered a new peak near 4312~MeV \cite{LHCb19}.
In 2021, LHCb presented evidence for an additional pentaquark as a peak near 4337~MeV in the $J\!/\!\psi \, p$ invariant-mass distribution from the decay $B^0\to J\!/\!\psi \, p \, \bar{p}$ with a significance of at most $3.7 \sigma$ \cite{LHCb21}.
The properties of the four nonstrange hidden-charm ($c\bar{c}$) pentaquarks $P_{c\bar{c}}(4312)^+$, $P_{c\bar{c}}(4337)^+$, $P_{c\bar{c}}(4440)^+$, and $P_{c\bar{c}}(4457)^+$ are listed in Table~\ref{tab:nonstrangepentaquarks}.
We do not include the broad peak near 4380~MeV reported in Ref.~\cite{LHCb15} for the reasons indicated in the PDG Review on Pentaquarks \cite{PDG24}.
Instead, we include the peak near 4337~MeV that was reported in Ref.~\cite{LHCb21}.
In Ref.~\cite{Alas25}, we did not consider $P_{c\bar{c}}(4337)^+$ or $P_{c\bar{c}}(4380)^+$.
In Ref.~\cite{Bram25}, the authors considered $P_{c\bar{c}}(4380)^+$ but not $P_{c\bar{c}}(4337)^+$.

\begin{table}
\centering
\begin{tabular}{ccccc}
\toprule
Pentaquark & $J^P$ &  $M$ [MeV]  & $\Gamma$ [MeV]  &Ref.\\
\midrule
$P_{c\bar{c}}(4312)^+$   &      $?^?$       & $4311.9 \pm 0.7 ^{+6.8}_{-0.6}$ & $9.8  \pm 2.7^{+3.7}_{-4.5}$   & \cite{LHCb19}   \\
$P_{c\bar{c}}(4337)^+$	&	$?^?$	& $4337^{+7}_{-4}\pm 2$	& $29^{+26}_{-12}\pm14$ 	& \cite{LHCb21} \\	
$P_{c\bar{c}}(4440)^+$   &      $?^?$       & $4440.3 \pm 1.3^{+4.1}_{-4.7}$  & $20.6 \pm 4.9^{+8.7}_{-10.1}$ & \cite{LHCb19} \\
$P_{c\bar{c}}(4457)^+$   &      $?^?$       & $4457.3 \pm 0.6^{+4.1}_{-1.7}$  & $6.4  \pm 2.0^{+5.7}_{-1.9}$    & \cite{LHCb19}\\
\bottomrule
\end{tabular}
\caption{The observed $c\bar{c}$ pentaquarks, their (unknown) $J^P$ quantum numbers, their masses $M$, and their widths $\Gamma$. }
\label{tab:nonstrangepentaquarks}
\end{table}

In 2020, the LHCb Collaboration presented evidence with a significance of $3.1\sigma$ for a strange hidden-charm ($c\bar{c}s$) pentaquark as a peak near 4450~MeV in the $J\!/\!\psi \, \Lambda$ invariant-mass distribution from the decay $\Xi_b^-\to J\!/\!\psi \, \Lambda \, K^-$ \cite{LHCb20}. 
In 2022, LHCb discovered a $c\bar{c}s$ pentaquark as a peak near 4338~MeV in the $J\!/\!\psi \, \Lambda$ invariant-mass distribution from the decay $B^-\to J\!/\!\psi \, \Lambda \, \bar{p}$ \cite{LHCb23}. 
Its $J^P$ quantum numbers were determined to be $\frac{1}{2}^-$. 
The properties of the two $c\bar{c}s$ pentaquarks $P_{c\bar{c}s}(4338)^0$ and $P_{c\bar{c}s}(4459)^0$ are listed in Table~\ref{tab:strangepentaquarks}.

\begin{table}
\centering
\begin{tabular}{ccccc}
\toprule
Pentaquark & $J^P$ &  $M$ [MeV]  & $\Gamma$ [MeV]  & Ref. \\
\midrule
$P_{c\bar{c}s}(4338)^0$  & $\frac{1}{2}^-$  & $4338.2 \pm 0.7 \pm 0.4$           & ~$7.0 \pm 1.2 \pm 1.3$          & \cite{LHCb23} \\
$P_{c\bar{c}s}(4459)^0$  &       $?^?$         & $4458.8 \pm 2.9 ^{+4.7}_{-1.1}$ & $17.3 \pm 6.5^{+8.0}_{-5.7}$ &  \cite{LHCb20} \\
\bottomrule
\end{tabular}
\caption{The observed $c\bar{c}s$ pentaquarks, their $J^P$ quantum numbers, their masses $M$, and their widths $\Gamma$. }
\label{tab:strangepentaquarks}
\end{table}

\begin{figure}
\tikzsetnextfilename{experimental-pentaquark-energy-levels}
\centering
\begin{tikzpicture}
\pgfmathsetmacro{\MSigma}{2.453}
\pgfmathsetmacro{\MSigmastar}{2.518}
\pgfmathsetmacro{\MXi}{2.469}
\pgfmathsetmacro{\MD}{1.867}
\pgfmathsetmacro{\MDstar}{2.009}
\pgfplotsset{
	/pgfplots/error bars/draw error bar/.code 2 args={%
		\node[inner sep=0pt] (a) at #1 {};
		\node[inner sep=0pt] (b) at #2 {};
		\path[draw=none, fill, opacity=.4] ($(a) - (22pt,0)$) rectangle ($(b) + (22pt,0)$);
	}
}
\pgfplotstableread{
x 	y 	ey+ 	ey-
0	4.312	0.007	0.001
0	4.337	0.007	0.004
0	4.440	0.004	0.005
0	4.457	0.004	0.002
}{\nonstrangeexp}
\begin{axis}[
	at = {(0cm,0cm)},
	anchor=west,
	width=5cm,
	height=9cm,
	only marks,
	black,
	clip mode=individual,
	ylabel={$E$},
	y unit=eV,
	y SI prefix=giga,
	axis y line=left,
	axis x line=bottom,
	ymin=4.27,
	ymax=4.57,
	xmin=-.5,
	xmax=.5,
	hide x axis,
	y tick style={black},
	ytick distance=0.05,
	yticklabel style={/pgf/number format/.cd,fixed,zerofill,precision=2},
	cycle list/Dark2
	]
	\begin{scope}[dashed]
		\draw (-.5,{\MSigma+\MD}) -- (.5,{\MSigma+\MD})
		node[anchor=west] {$\Sigma_c\bar{D}$};
		\draw (-.5,{\MSigmastar+\MD}) -- (.5,{\MSigmastar+\MD})
		node[anchor=west] {$\Sigma_c^\ast\bar{D}$};
		\draw (-.5,{\MSigma+\MDstar}) -- (.5,{\MSigma+\MDstar})
		node[anchor=west] {$\Sigma_c\bar{D}^\ast$};
            		\draw (-.5,{\MSigmastar+\MDstar}) -- (.5,{\MSigmastar+\MDstar})
		node[anchor=west] {$\Sigma_c^\ast\bar{D}^\ast$};
	\end{scope}
	\addplot+[mark=-,thick,mark size=22pt,
		error bars/.cd, y dir=both, y explicit]
		table[y error plus=ey+, y error minus=ey-] {\nonstrangeexp};
	\node[anchor=north] at (0, 4.311) {\footnotesize $P_{c\bar{c}}(4312)^+$};
	\node[anchor=south] at (0, 4.340) {\footnotesize $P_{c\bar{c}}(4337)^+$};
	\node[anchor=north] at (0, 4.440) {\footnotesize $P_{c\bar{c}}(4440)^+$};
	\node[anchor=south] at (0, 4.457) {\footnotesize $P_{c\bar{c}}(4457)^+$};
\end{axis}
\pgfplotstableread{
x 	y 	ey+ 	ey-
0	4.338	0.001	0.001
0	4.459	0.006	0.003
}{\strangeexp}
\begin{axis}[
	at={(8cm,0cm)},
	anchor=west,
	width=5cm,
	height=9cm,
	only marks,
	black,
	clip mode=individual,
	ylabel={$E$},
	y unit=eV,
	y SI prefix=giga,
	axis y line=left,
	axis x line=bottom,
	ymin=4.27,
	ymax=4.57,
	xmin=-.5,
	xmax=.5,
	hide x axis,
	y tick style={black},
	ytick distance=0.05,
	yticklabel style={/pgf/number format/.cd,fixed,zerofill,precision=2},
	cycle list/Dark2
	]
	\begin{scope}[dashed]
		\draw (-.5,{\MXi+\MD}) -- (.5,{\MXi+\MD})
		node[anchor=west] {$\Xi_c\bar{D}$};
		\draw (-.5,{\MXi+\MDstar}) -- (.5,{\MXi+\MDstar})
		node[anchor=west] {$\Xi_c\bar{D}^\ast$};
	\end{scope}
	\addplot+[mark=-,thick,mark size=22pt,
		error bars/.cd, y dir=both, y explicit]
		table[y error plus=ey+, y error minus=ey-] {\strangeexp};
	\node[anchor=north] at (0, 4.337) {\footnotesize $P_{c\bar{c}s}(4338)^0$};
	\node[anchor=north] at (0, 4.456) {\footnotesize $P_{c\bar{c}s}(4459)^0$};
\end{axis}
\end{tikzpicture}
\caption{\label{fig:ExperimentalPentaquarks} The observed $c\bar{c}$ pentaquarks (left) and $c\bar{c}s$ pentaquarks (right) considered in this work and the relevant baryon-meson thresholds.
}
\end{figure}
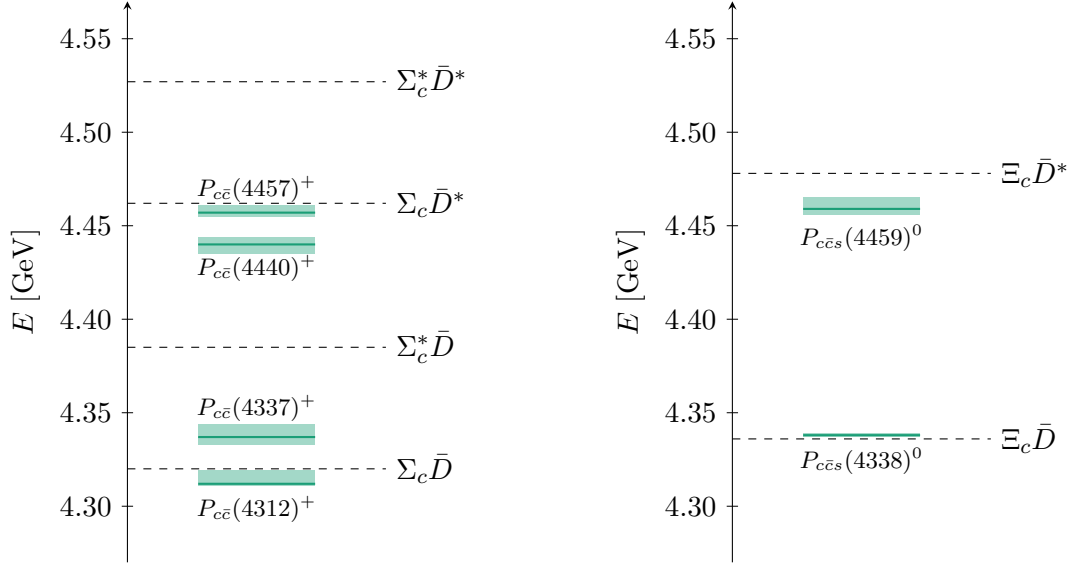

Each of the observed pentaquarks is near the threshold for a charm baryon and a charm meson, as illustrated in Fig.~\ref{fig:ExperimentalPentaquarks}.
The thresholds for some of the lowest-energy pairs of nonstrange charm baryons and charm mesons are listed in Table~\ref{tab:nonstrangepairs}. 
The central values of the masses of $P_{c\bar{c}}(4312)^+$ and $P_{c\bar{c}}(4337)^+$ are about 9~MeV below and about 16~MeV above the $\Sigma_c\bar{D}$ threshold. 
The central values of the masses of $P_{c\bar{c}}(4440)^+$ and $P_{c\bar{c}}(4457)^+$ are below the $\Sigma_c\bar{D}^\ast$ threshold by about 22~MeV and about 5~MeV. 
The widths of each of the four pentaquarks is comparable to the difference between the mass and the threshold, so their line shapes may extend beyond the threshold.
The thresholds for some of the lowest-energy charm-hadron pairs with strangeness $-1$ are listed in Table~\ref{tab:strangepairs}. 
The central value of the mass of $P_{c\bar{c}s}(4338)^0$ is above the $\Xi_c \bar{D}$ threshold by about 2~MeV. 
The central value of the mass of $P_{c\bar{c}s}(4459)^0$ is below the $\Xi_c \bar{D}^\ast$ threshold by about 19~MeV.
The widths of the two pentaquarks are comparable to the difference between the mass and the threshold, so their line shapes may extend beyond the threshold.
The proximity of all these pentaquarks to charm-hadron-pair thresholds has motivated their interpretation as hadronic molecules consisting of a charm baryon and a charm meson.

\begin{table}
\centering
\begin{tabular}{ccc}
\toprule
~Baryon-Meson~ & $J^P$	& $M_\mathrm{B}+M_\mathrm{M}$ [MeV]   \\ 
\midrule
$\Lambda_c\bar{D}$          	  & $\frac{1}{2}^-$	                             & 4153.7 \\ 
$\Lambda_c\bar{D}^\ast$            & $\frac{1}{2}^-,\frac{3}{2}^-$               & 4295.0 \\
$\Sigma_c\bar{D}$                      & $\frac{1}{2}^-$	                             & 4320.7 \\ 
$\Sigma_c^\ast\bar{D}$            & $\frac{3}{2}^-$	                             & 4385.3 \\ 
$\Sigma_c\bar{D}^\ast$            & $\frac{1}{2}^-,\frac{3}{2}^-$                  & 4462.0   \\ 
$\Sigma_c^\ast\bar{D}^\ast$  & $\frac{1}{2}^-,\frac{3}{2}^-,\frac{5}{2}^-$ & 4526.7  \\ 
\bottomrule
\end{tabular}
\caption{Lowest-energy nonstrange pairs of a charm baryon and a charm meson, the $J^P$ quantum numbers of $S$-wave pairs, and the sum $M_\mathrm{B}+M_\mathrm{M}$ of their masses.  }
\label{tab:nonstrangepairs}
\end{table}

\begin{table}
\centering
\begin{tabular}{ccc}
\toprule
Baryon-Meson & $J^P$ 	& $M_\mathrm{B}+M_\mathrm{M}$ [MeV]  \\ 
\midrule
$\Lambda_c\bar{D}_s$           & $\frac{1}{2}^-$	                 & 4254.8 \\ 
$\Lambda_c\bar{D}_s^\ast$ & $\frac{1}{2}^-,\frac{3}{2}^-$ & 4398.7  \\ 
$\Xi_c\bar{D}$                         & $\frac{1}{2}^-$	                 & 4336.3  \\ 
$\Xi_c\bar{D}^\ast$               & $\frac{1}{2}^-,\frac{3}{2}^-$ & 4477.6   \\
$\Sigma_c\bar{D}_s$                      & $\frac{1}{2}^-$	                                  & 4421.8  \\ 
$\Sigma_c^\ast\bar{D}_s$            & $\frac{3}{2}^-$	                                  & 4486.4 \\ 
$\Sigma_c\bar{D}^\ast_s$            & $\frac{1}{2}^-,\frac{3}{2}^-$                   & 4565.7 \\ 
$\Sigma_c^\ast\bar{D}^\ast_s$  & $\frac{1}{2}^-,\frac{3}{2}^-,\frac{5}{2}^-$ & 4630.3 \\ 
$\Xi_c^\prime\bar{D}$                      & $\frac{1}{2}^-$	                                  & 4445.7 \\ 
$\Xi_c^{\prime\ast}\bar{D}$            & $\frac{3}{2}^-$	                                  & 4512.9 \\ 
$\Xi_c^\prime\bar{D}^\ast$            & $\frac{1}{2}^-,\frac{3}{2}^-$                   & 4587.0 \\ 
$\Xi_c^{\prime\ast}\bar{D}^\ast$  & $\frac{1}{2}^-,\frac{3}{2}^-,\frac{5}{2}^-$ & 4654.2  \\ 
\bottomrule
\end{tabular}
\caption{
Lowest-energy pairs of a charm baryon and a charm meson with total strangeness $-1$, the $J^P$ quantum numbers of $S$-wave pairs, and the sum of their masses.  }
\label{tab:strangepairs}
\end{table}

\section{Static Hadrons}
\label{statichadrons}

\noindent
A \emph{static hadron} is a discrete energy level of QCD bound to a single local color source. 
In this section, we discuss the static hadrons relevant to hidden-heavy pentaquarks and we determine the lowest-energy adjoint baryons.

\subsection{QCD and Color Sources}
\label{sec:staticquantumnumbers}

The parameter space of QCD includes the masses $m_u$, $m_d$, and $m_s$ of the  up, down, and strange quarks.
We refer to the point in QCD parameter space in which $m_s$ has its physical value and $m_u =m_d = m_s$  as the \emph{SU(3)-flavor point}.
The SU(3) flavor symmetry is exact at the SU(3)-flavor point.
We refer to the line in QCD parameter space in which $m_s$ has its physical value and $m_u$ and $m_d$ are equal and smaller than $m_s$  as the \emph{(2+1)-flavor line}. 
The SU(2) isospin symmetry is exact along the (2+1)-flavor line.
Physical QCD is a point close to  the (2+1)-flavor line.

QCD with a single local color source has rotational and parity symmetries. 
The static hadrons are therefore labeled by angular-momentum and parity quantum numbers $j^\pi$. 
Static hadrons form SU(3)-flavor multiplets labeled by their multiplicity. 
They can be decomposed into isospin multiplets labeled by flavor quantum numbers $(I,S)$, with $I$ for isospin and $S$ for strangeness, and with multiplicity $2I+1$.

We proceed to describe our terminology for the static hadrons relevant to hidden-heavy pentaquarks. 
A static hadron can be labeled by the color representation of the source to which it is bound. 
It is labeled a triplet, antitriplet, or adjoint hadron if it is bound to a triplet ($\bm{3}$), antitriplet ($\bm{\bar{3}}$), or adjoint ($\bm{8}$) color source.
Alternatively, it can be labeled a $\bm{3}$-hadron, $\bm{\bar{3}}$-hadron, or $\bm{8}$-hadron, respectively. 
We can specify a static hadron further by its light quark content.
A $\bm{3}$-hadron is  a $\bm{3}$-meson or a $\bm{3}$-baryon if the light quark content is $\bar q$ or $qq$.
An $\bm{8}$-hadron is an $\bm{8}$-meson or an $\bm{8}$-baryon if the light quark content is $q\bar q$ or $qqq$. 
An $\bm{8}$-hadron with only gluon fields bound to the static source is a gluelump. 
There is mixing between gluelumps and $\bm{8}$-mesons with flavor $u \bar u + d \bar d + s \bar s$.
Triplet baryons and antitriplet mesons are relevant to the Born-Oppenheimer potentials for hidden-heavy pentaquark baryons at long distance.
Adjoint baryons are relevant to the Born-Oppenheimer potentials for hidden-heavy pentaquarks at short distance.

A heavy hadron is one whose constituents include a single heavy (charm or bottom) quark or antiquark. 
We refer to QCD with the heavy-quark fields omitted as \emph{light QCD}.
The light-QCD fields in the heavy hadron can be approximated by those in a $\bm{3}$-hadron or $\bm{\bar{3}}$-hadron.
Heavy hadrons form heavy-quark spin symmetry (HQSS) multiplets labeled by the quantum numbers $j^\pi$ and $(I,S)$ of the static hadron.
The members of the HQSS multiplets are labeled by spin and parity quantum numbers $J^P$. 
The multiplet is a singlet with spin $J=\frac{1}{2}$ if $j=0$ and a doublet with spins $J=j-\frac{1}{2},j+\frac{1}{2}$ if $j\geq\frac{1}{2}$. 
The parity $P$ is $+\pi$ or $-\pi$ for a ground-state hadron with a heavy quark or antiquark.

The energies of static hadrons can be calculated using lattice QCD with a $\bm{3}$ source or using lattice NRQCD with a heavy particle with a $\bm{3}$ color charge.
The relative energies of $\bm{3}$-hadrons can also be determined phenomenologically by extrapolating mass differences of charm and bottom hadrons to the limit of large $m_Q$.

We refer to the energy increase of a hadron from replacing an up or down quark by a strange quark as a \emph{strangeness splitting}.
Strangeness splittings can be calculated using lattice QCD or lattice NRQCD.
We refer to the energy difference between two members of an HQSS multiplet as a \emph{spin splitting}.
The spin splittings of heavy hadrons scale as $1/m_Q$, where $m_Q$ is the heavy quark mass.
Spin splittings of heavy hadrons cannot be calculated using lattice simulations of light QCD only.

\subsection{Antitriplet Mesons}
\label{sec:AntitripletMesons}

\begin{figure}
\tikzsetnextfilename{antitriplet-meson-multiplet}
\centering
\begin{tikzpicture}[bigdot/.style={circle,fill,
inner sep=0pt,minimum size=3mm},
scale=.75]
\node (dbar) at (-1,0) [bigdot,label=above left:{$d$}] {};
\node (ubar) at (1,0) [bigdot,label=above right:{$u$}] {};
\node (sbar) at (0,{-sqrt(3)}) [bigdot,label=below:{$s$}] {};
\draw [ultra thick] (dbar.center) -- (ubar.center) -- (sbar.center) -- cycle;
\end{tikzpicture}
\caption{\label{fig:3astmesmult}
The light-quark flavors in the $j^\pi=\frac{1}{2}^+$ $\bm{\bar{3}}$-mesons form a triplet in the $(I_3,S)$ plane.}
\end{figure}
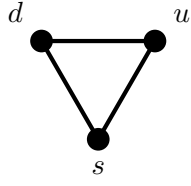

The ground-state $\bm{3}$-hadrons are $\bm{3}$-mesons with $j^\pi=\frac{1}{2}^-$.
A $\bm{\bar{3}}$-hadron is the charge conjugate of the corresponding $\bm{3}$-hadron. 
The SU(3)-flavor multiplet of ground-state $\bm{\bar{3}}$-mesons is displayed in Fig.~\ref{fig:3astmesmult}. 
They form an SU(3) triplet consisting of an isospin doublet $\bigl(\frac{1}{2},0\bigr)$ and a singlet $(0,-1)$.
 
In the ground-state heavy mesons with a heavy antiquark, the light-QCD fields can be approximated by a $j^\pi = \frac{1}{2}^+$ $\bm{\bar{3}}$-meson.
These mesons form an SU(3) triplet of HQSS doublets with quantum numbers $J^P=0^-,1^-$.
For mesons containing a charm antiquark, the  HQSS doublets with flavor $(I,S) = \bigl(\frac{1}{2}, 0\bigr)$ and $(0, -1)$ are $(\bar{D},\bar{D}^\ast)$ and $(\bar{D}_s,\bar{D}_s^\ast)$. 
We label these doublets by $\bar{D}^{(\ast)}$ and $\bar{D}_s^{(\ast)}$.
Their isospin multiplets form the two rows of the SU(3) triplet in Fig.~\ref{fig:3astmesmult}.  
The spin splittings are $\Delta_\mathrm{M}=142.1$~MeV  for charm mesons and $\Delta_\mathrm{M}=46.3$~MeV  for bottom mesons.
The estimates of the spin-averaged strangeness splittings for $\bm{3}$-mesons from charm and bottom spectroscopy are 103.0~MeV and 89.8~MeV.  
The strangeness splitting in the static limit  determined by lattice HQET is $85.0\pm 5.6$~MeV \cite{Bern15}. 

\subsection{Triplet Baryons}

\begin{figure}
\tikzsetnextfilename{triplet-baryon-multiplets}
\centering
\begin{tikzpicture}[%
	bigdot/.style={circle,fill,
	inner sep=0pt,minimum size=3mm},
	scale=.75
	]
	\begin{scope}[shift={(-6,0)}]
		\node (ud) at (0,0) [bigdot,label=above:{$[ud]$}] {};
		\node (ds) at (-1,{-sqrt(3)}) [bigdot,label=below left:{$[ds]$}] {};
		\node (us) at (1,{-sqrt(3)}) [bigdot,label=below right:{$[su]$}] {};
		\draw [ultra thick] (ds.center) -- (us.center) -- (ud.center) -- cycle;
	\end{scope}
	\node (dd) at (-2,0) [bigdot,label=above left:{$dd$}] {};
	\node at (0,0) [bigdot,label=above:{$\{ud\}$}] {};
	\node (uu) at (2,0) [bigdot,label=above right:{$uu$}] {};
	\node at (-1,{-sqrt(3)}) [bigdot,label=below left:{$\{ds\}$}] {};
	\node at (1,{-sqrt(3)}) [bigdot,label=below right:{$\{su\}$}] {};
	\node (ss) at (0, {-2 * sqrt(3)}) [bigdot,label=below:{$ss$}] {};
	\draw [ultra thick] (dd.center) -- (uu.center) -- (ss.center) -- cycle;
\end{tikzpicture}
\caption{\label{fig:3barmult}
The light-quark flavors in the $j^\pi=0^+$ and $1^+$ $\bm{3}$-baryons form an antitriplet (left) and a sextet (right) in the $(I_3,S)$ plane. 
Brackets $\{\,\cdots\}$ or $[\,\cdots]$ indicate a symmetric or antisymmetric combination of the flavors enclosed.}
\end{figure}
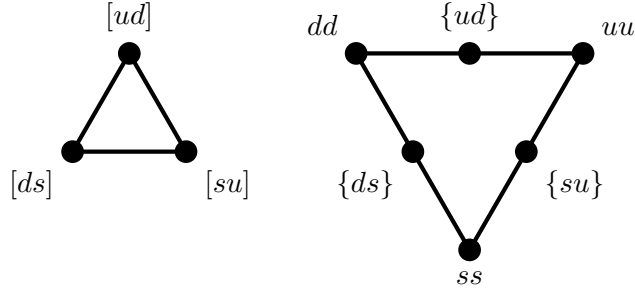

The SU(3)-flavor multiplets of the ground-state $\bm{3}$-baryons are displayed in Fig.~\ref{fig:3barmult}. 
The lower-energy multiplet has $j^\pi=0^+$ and is an SU(3) antitriplet consisting of isospin multiplets with $(I,S)=(0,0)$ and $\bigl(\frac{1}{2},-1\bigr)$.
The  average strangeness splittings in the antitriplet  from charm and bottom spectroscopy are 182.6~MeV and 174.9~MeV. 
The higher-energy multiplet of ground-state $\bm{3}$-baryons has  $j^\pi=1^+$ and is an SU(3) sextet consisting of isospin multiplets with $(I,S)=(1,0)$, $\bigl(\frac{1}{2},-1\bigr)$, and $(0,-2)$. 
The average strangeness splittings in the sextet from charm and bottom spectroscopy are 122.9~MeV and 119.8~MeV. 

In the ground-state heavy baryons, the light-QCD fields can be approximated by a $0^+$ $\bm{3}$-baryon or a  $1^+$  $\bm{3}$-baryon bound to a heavy quark.
The heavy baryons that can be approximated by a $0^+$ $\bm{3}$-baryon form an SU(3) antitriplet of HQSS singlets with $J^P=\frac{1}{2}^+$.
For charm baryons, the HQSS singlets with $(I,S)= (0,0)$ and $(\frac{1}{2}, -1)$ are $\Lambda_c$ and $\Xi_c$. 
They form the two rows of the antitriplet in Fig.~\ref{fig:3barmult}. 
The heavy baryons  that can be approximated by a $1^+$ $\bm{3}$-baryon form an SU(3) sextet of HQSS doublets with $J^P=\frac{1}{2}^+,\frac{3}{2}^+$.
For charm baryons, the HQSS doublets with $(I,S)=(1,0)$, $(\frac{1}{2}, -1)$, and $(0, -2)$  are $(\Sigma_c,\Sigma_c^\ast)=(\Sigma_c(2455),\Sigma_c(2520)$), $(\Xi_c^\prime,\Xi_c^{\prime\ast})=(\Xi_c^\prime,\Xi_c(2645)$), and $(\Omega_c,\Omega_c^\ast) =(\Omega_c^0,\Omega_c(2770)$).
We label these three HQSS doublets by $\Sigma_c^{(\ast)}$, $\Xi_c^{\prime(\ast)}$, and $\Omega_c^{(\ast)}$, respectively.
The isospin multiplets for these HQSS doublets form the three rows of the sextet in Fig.~\ref{fig:3barmult}.
The spin splittings in the HQSS doublets are $\Delta_\mathrm{B}=66.5$~MeV  for charm baryons and  $\Delta_\mathrm{B}=19.2$~MeV  for bottom baryons.

\subsection{Adjoint Baryons}
\label{sec:AdjointBaryons}

The spectrum of $\bm{8}$-hadrons in QCD has not yet been established. 
The only lattice gauge theory calculations for $\bm{8}$-hadrons are for gluelumps in SU(3) gauge theory \cite{Fos99,Her24} and in QCD with two dynamical light quarks \cite{Mar14}. 
The possibility of a long-lived heavy gluino (the supersymmetric partner of the gluon) has motivated the study of the spectrum of $\bm{8}$-hadrons using various constituent models. 
A simple summary of the results from these models is given in a conference proceeding by the ATLAS collaboration \cite{ATLAS19review}. 
Depending on the model, the lowest-energy $\bm{8}$-hadron is either a gluelump, a nonstrange $\bm{8}$-meson with isospin zero or one, or a strange $\bm{8}$-baryon. 

The expected spin-flavor multiplets of the lowest-energy $\bm{8}$-baryons at the SU(3)-flavor point  can be deduced from the requirement that the state be antisymmetric under exchange of identical fermions.
Each of the three light quarks  in an $\bm{8}$-baryon is in the fundamental representation of SU(3) color, SU(3) flavor, and SU(2) spin. 
The overall state of three light quarks must be antisymmetric under the exchange of their position, color, flavor, and spin degrees of freedom. 
The lowest-energy $\bm{8}$-baryons are expected to be symmetric in the positions of the three light quarks and therefore have even parity $\pi=+$. 
They must therefore be antisymmetric under exchange of their color, flavor, and spin degrees of freedom.

To saturate the color of the static $\bm{8}$ source, the three light quarks must be in an overall color-octet state with mixed symmetry that we denote by $\bm{8}_\mathrm{M}$.  
The direct product of the three SU(3)-flavor triplets can be decomposed into one antisymmetric singlet, two mixed-symmetry octets, and one  symmetric decuplet: $\bm{3}\otimes\bm{3}\otimes\bm{3}=\bm{1}_\mathrm{A}\oplus\bm{8}_\mathrm{M}\oplus\bm{8}_\mathrm{M}\oplus\bm{10}_\mathrm{S}$. 
The direct product of three SU(2)-spin doublets can be decomposed into two mixed-symmetry doublets ($j=\frac{1}{2}$) and one symmetric quartet ($j=\frac{3}{2}$): $\bm{2}\otimes\bm{2}\otimes\bm{2}=\bm{2}_\mathrm{M}\oplus\bm{2}_\mathrm{M}\oplus\bm{4}_\mathrm{S}$. 
Given the mixed-symmetry color state $\bm{8}_\mathrm{M}$, the $\bm{8}$-baryons must have mixed symmetry under the exchanges of both flavors and spins. 
If the spin state is a mixed-symmetry $\bm{2}_\mathrm{M}$, then the possible flavor states are the antisymmetric $\bm{1}_\mathrm{A}$, a mixed-symmetry $\bm{8}_\mathrm{M}$, and the symmetric $\bm{10}_\mathrm{S}$. 
If the spin state is the symmetric $\bm{4}_\mathrm{S}$, then the only possible flavor state is a mixed-symmetry $\bm{8}_\mathrm{M}$.

\begin{figure*}
\tikzsetnextfilename{octet-baryon-multiplets}
\begin{tikzpicture}[bigdot/.style={circle,fill,
inner sep=0pt,minimum size=3mm},
scale=.75]
\node at (-15,{-sqrt(3)}) [bigdot,label=above:{$[uds]$}] {};
\begin{scope}[shift={(-8,0)}]
\node (udd) at (-1,0) [bigdot,label=above left:{$(udd)$}] {};
\node (uud) at (1,0) [bigdot,label=above right:{$(uud)$}] {};
\node (dds) at (-2,{-sqrt(3)}) [bigdot,label=left:{$(dds)$}] {};
\node at (0,{-sqrt(3)}) [bigdot,label={[label distance=2mm]above:{$(uds)$}},label={[label distance=2mm]below:{$(uds)^\prime$}}] {};
\node (uus) at (2,{-sqrt(3)}) [bigdot,label=right:{$(uus)$}] {};
\node (dss) at (-1,{-2 * sqrt(3)}) [bigdot,label=below left:{$(dss)$}] {};
\node (uss) at (1,{-2 * sqrt(3)}) [bigdot,label=below right:{$(uss)$}] {};
\draw [ultra thick] (udd.center) -- (uud.center) -- (uus.center) -- (uss.center) -- (dss.center) -- (dds.center) -- cycle;
\draw[very thick] (0,{-sqrt(3)}) circle [radius=3mm];
\end{scope}
\node (ddd) at (-3,0) [bigdot,label=above left:{$ddd$}] {};
\node at (-1,0) [bigdot,label=above:{$\{udd\}$}] {};
\node at (1,0) [bigdot,label=above:{$\{uud\}$}] {};
\node (uuu) at (3,0) [bigdot,label=above right:{$uuu$}] {};
\node at (-2,{-sqrt(3)}) [bigdot,label=below left:{$\{dds\}$}] {};
\node at (0,{-sqrt(3)}) [bigdot,label=above:{$\{uds\}$}] {};
\node at (2,{-sqrt(3)}) [bigdot,label=below right:{$\{uus\}$}] {};
\node at (-1,{-2 * sqrt(3)}) [bigdot,label=below left:{$\{dss\}$}] {};
\node at (1,{-2 * sqrt(3)}) [bigdot,label=below right:{$\{uss\}$}] {};
\node (sss) at (0, {-3 * sqrt(3)}) [bigdot,label=below:{$sss$}] {};
\draw [ultra thick] (ddd.center) -- (uuu.center) -- (sss.center) -- cycle;
\end{tikzpicture}
\caption{\label{fig:8barmult}
The light-quark flavors in the $j^\pi=\frac{1}{2}^+$ $\bm{8}$-baryons form a singlet $\bm{1}_\mathrm{A}$ (left), an octet $\bm{8}_\mathrm{M}$ (middle), and a decuplet $\bm{10}_\mathrm{S}$ (right) in the $(I_3,S)$ plane. 
The light-quark flavors in the $j^\pi=\frac{3}{2}^+$ $\bm{8}$-baryons form an additional octet $\bm{8}_\mathrm{M}$. 
Brackets $\{\,\cdots\}$, $[\,\cdots]$, or $(\,\cdots)$ indicate a symmetric, antisymmetric, or mixed-symmetry combination of the flavors enclosed, respectively.}
\end{figure*}
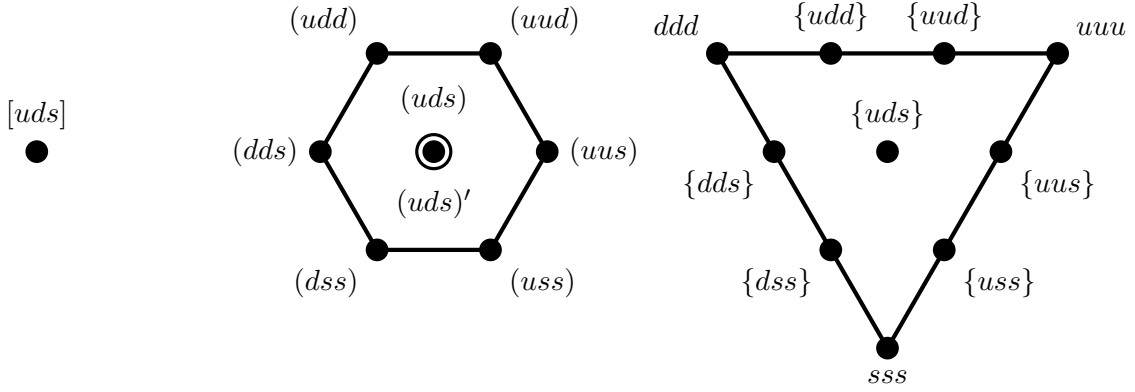

The expected SU(3)-flavor multiplets of the lowest-energy $\bm{8}$-baryons are displayed in Fig.~\ref{fig:8barmult}. 
The $j^\pi = \frac{1}{2}^+$ $\bm{8}$-baryons form an SU(3) singlet with $(I,S)=(0, -1)$, an SU(3) octet consisting of $\bigl(\frac{1}{2}, 0\bigr),(0,-1),(1,-1)$, and $\bigl(\frac{1}{2},-2\bigr)$, and an SU(3) decuplet consisting of $\bigl(\frac{3}{2},0\bigr),(1,-1),(\frac{1}{2},-2)$, and $(0,-3)$. 
The $\frac{3}{2}^+$ $\bm{8}$-baryons form an additional SU(3) octet consisting of $\bigl(\frac{1}{2}, 0\bigr),(0,-1),(1,-1)$, and $\bigl(\frac{1}{2},-2\bigr)$.

At the SU(3)-flavor point, each of the four SU(3) multiplets of low-energy $\bm{8}$-baryons consists of states that are degenerate in energy. 
On the (2+1)-flavor line, there is a strangeness splitting between the rows of each multiplet.  
There can also be additional energy shifts from mixing of $j^\pi=\frac{1}{2}^+$ states in different SU(3) multiplets with the same strangeness and isospin. 
On the (2+1)-flavor line, the two strange isospin singlets are orthogonal linear combinations of those in the SU(3) singlet and octet. 
The two strange isospin triplets are orthogonal linear combinations of those in the SU(3) octet and decuplet. 
Finally, the two double-strange isospin doublets are orthogonal linear combinations of those in the SU(3) octet and decuplet.

\section{Pentaquark Born-Oppenheimer Potentials}
\label{pentaquarkbopotentials}

In this section, we point out that the Born-Oppenheimer (B\nobreakdash-O) potentials for hidden-heavy pentaquarks are adjoint-baryon potentials at short distance and static-hadron pair potentials at long distance that connect smoothly at intermediate distances.  
We discuss in detail the B\nobreakdash-O potentials for isospin-$\frac{1}{2}$ $c\bar{c}$ pentaquarks and for isospin-0 $c\bar{c}s$ pentaquarks.

\subsection{Adjoint-Baryon and Static-Hadron-Pair Potentials}

An adiabatic B\nobreakdash-O potential for hidden-heavy pentaquarks is a discrete energy level of QCD with the flavor of three light quarks bound to static $\bm{3}$ and $\bm{\bar{3}}$ color sources separated by a vector $\bm{r}$. 
The adiabatic potentials are functions of the distance $r$. 
The continuous symmetries are rotations around the axis $\bm{\hat{r}}$ generated by $\bm{j}\cdot\bm{\hat{r}}$, where $\bm{j}$ is the angular-momentum vector for the light-QCD fields.
The discrete symmetries are  the combined transformation of charge conjugation and  parity and reflections through any plane containing $\bm{\hat{r}}$.
The B\nobreakdash-O potentials are labeled by B\nobreakdash-O quantum numbers traditionally denoted by $\Lambda_\eta^\epsilon$, where $\Lambda$ is the quantum number of $\lvert\bm{j}\cdot\bm{\hat{r}}\rvert$, the subscript $\eta$ is $g$ or $u$ if the $CP$ quantum number is $+1$ or $-1$, and the superscript $\epsilon$ is $+$ or $-$ if the reflection quantum number is $+1$ or $-1$. 
If the baryon number is nonzero, the B\nobreakdash-O potentials with $\eta=g,u$ are linear combinations with opposite baryon number and equal energies. 
Since a state with three light quarks has baryon number 1, the subscript $\eta$  can  be omitted. If $\Lambda$ is nonzero, the B\nobreakdash-O potentials  with $\epsilon=+,-$ are linear combinations with opposite $\bm{j}\cdot\bm{\hat{r}}$ and equal energies.
Since a state with three light quarks has  half-integer $\Lambda$, the superscript $\epsilon$ can  be omitted. The B\nobreakdash-O quantum numbers for hidden-heavy pentaquarks therefore reduce simply to $\Lambda$. 
The B\nobreakdash-O potentials are also labeled by flavor quantum numbers $(I,S)$.

At small $r$, an adiabatic B\nobreakdash-O potential for hidden-heavy pentaquarks is an \emph{adjoint-baryon potential} that approaches a repulsive color-Coulomb potential offset by the energy of an $\bm{8}$-baryon.
If the $\bm{8}$-baryon has angular-momentum and parity $j^\pi$, there are $j+\frac{1}{2}$ coupled adjoint-baryon potentials with $\Lambda$ ranging from $\frac{1}{2}$ to $j$ that become degenerate as $r\to0$, where they are related by rotations. 
We denote these coupled B\nobreakdash-O potentials by $(j^\pi) j \big/ \ldots \big/ \frac{1}{2}$.
The flavor $(I,S)$ of these potentials is the flavor of the $\bf{8}$-baryon.

At large $r$, an adiabatic B\nobreakdash-O potential for hidden-heavy pentaquarks is a \emph{static-hadron-pair potential} that approaches the sum of the energies of a $\bm{3}$-baryon and a $\bm{\bar{3}}$-meson. 
The static-hadron-pair potentials can be labeled by the quantum numbers $j_\mathrm{B}^{\pi_\mathrm{B}}$ and $j_\mathrm{M}^{\pi_\mathrm{M}}$ and the flavors $(I_\mathrm{B},S_\mathrm{B})$ and $(I_\mathrm{M},S_\mathrm{M})$ of the $\bm{3}$-baryon and $\bm{\bar{3}}$-meson.
Their quantum number and flavors can alternatively be specified by the HQSS multiplets of the corresponding charm baryons and charm mesons.
Thus we denote the nonstrange static-hadron-pair potentials with the lowest energies at large $r$ as $\Lambda_c\bar{D}^{(\ast)}$ and $\Sigma_c^{(\ast)} \bar{D}^{(\ast)}$.
We denote the strange static-hadron-pair potentials with the lowest energies at large $r$ as $\Lambda_c \bar{D}_s^{(\ast)}$, $\Xi_c \bar{D}^{(\ast)}$, $\Sigma_c^{(\ast)} \bar{D}_s^{(\ast)}$, and $\Xi_c^{\prime(\ast)} \bar{D}^{(\ast)}$.
Given the quantum numbers $j_\mathrm{B}^{\pi_\mathrm{B}}$ and $j_\mathrm{M}^{\pi_\mathrm{M}}$ of a $\bm{3}$-baryon and a $\bm{\bar{3}}$-meson, there is a multiplet of $(j_\mathrm{B} + 1) (j_\mathrm{M} + \frac{1}{2})$  static-hadron-pair potentials that become degenerate as $r\to\infty$.
That multiplet can be decomposed into sub-multiplets labeled by $j^\prime{}^{\pi^\prime}$ with $j^\prime$ ranging from $\lvert j_\mathrm{B} - j_\mathrm{M}\rvert$ to $j_\mathrm{B} + j_\mathrm{M}$ and with $\pi^\prime = \pi_\mathrm{B} \pi_\mathrm{M}$.
Each sub-multiplet consists of coupled potentials with $\Lambda$ ranging from $\frac{1}{2}$ to $j^\prime$ that are related by rotations. 
We denote the coupled static-hadron-pair potentials in the sub-multiplet by $(j^\prime{}^{\pi^\prime}) j^\prime \big/ \ldots \big/ \frac{1}{2}$.
The flavor $(I,S)$ of these potentials is the total flavor of the static-hadron pair.

The adiabatic B\nobreakdash-O potentials for hidden-heavy pentaquarks interpolate smoothly between adjoint-baryon potentials $(j^\pi)\Lambda$ at small $r$ and static-hadron-pair potentials $(j^\prime{}^{\pi^\prime}) \Lambda$ at large $r$ with the same B\nobreakdash-O quantum number $\Lambda$ and flavor $(I,S)$.
Adiabatic potentials with identical $\Lambda$ and $(I,S)$ cannot cross each other.
For each $\Lambda$ and flavor, the spectrum of QCD in the presence of static sources must be a smooth  function of  $r$.
The lowest adjoint-baryon potential must therefore connect to the lowest static-hadron-pair potential, the next lowest adjoint-baryon potential must connect to the next lowest static-hadron-pair potential, and so on.
An adiabatic adjoint-baryon potential $(j^\pi)\Lambda$ at small $r$ could connect to an adiabatic static-hadron-pair potential $(j^\prime{}^{\pi^\prime}) \Lambda$ at large $r$ with the same $\Lambda$ but $j^\prime{}^{\pi^\prime}$ different from $j^\pi$.
However, using the diabatic representation of the B\nobreakdash-O approximation, it is possible to define \emph{diabatic} potentials $(j^\pi)\Lambda$ that are related by rotations at all $r$  \cite{Bru23a}.
The diabatic potentials $(j^\pi)\Lambda$ with $\Lambda=\frac{1}{2},\dotsc,j$ approach adiabatic adjoint-baryon potentials $(j^\pi)\Lambda$  at small $r$ and they approach adiabatic static-hadron-pair potentials $(j^\pi)\Lambda$ at large $r$.
Unlike the adiabatic potentials, diabatic potentials with identical $\Lambda$ and $(I,S)$ may cross each other.
For the sake of simplicity, we shall however assume there are no such crossings.\footnote{The validity of this assumption can be tested by calculating the adiabatic B\nobreakdash-O potentials using lattice QCD.  $\mathrm{A}$ narrow avoided crossing between two adiabatic potentials with the same $\Lambda$ and $(I,S)$ indicates that the corresponding diabatic B\nobreakdash-O potentials cross.}

\subsection{Nonstrange Isospin-\texorpdfstring{$\frac{1}{2}$}{1/2} Potentials}

The $c\bar{c}$ pentaquarks were discovered through their decays into $J/\psi\, p$, which implies that they have isospin $\frac{1}{2}$. 
We therefore consider in detail the B\nobreakdash-O potentials with isospin $\frac{1}{2}$ and strangeness zero.
From the SU(3)-flavor multiplets in Fig.~\ref{fig:8barmult}, we expect the lowest-energy $(I,S)= \bigl(\frac{1}{2}, 0\bigr)$ $\bm{8}$-baryons to be the top rows of the $j^\pi=\frac{1}{2}^+$ octet and $\frac{3}{2}^+$ octet.
The single adjoint-baryon potential associated with the $\frac{1}{2}^+$ $\bm{8}$-baryon is $\bigl(\frac{1}{2}^+\bigr)\frac{1}{2}$. 
The doublet of adjoint-baryon potentials associated with the $\frac{3}{2}^+$  $\bm{8}$-baryon is $\bigl(\frac{3}{2}^+\bigr)\frac{3}{2} \big/ \frac{1}{2}$.
These adjoint-baryon potentials are listed in Table~\ref{tab:8barpot}.

\begin{table}
\centering
\begin{tabular}{cccc}
\toprule
$j^\pi$	& $(I,S)$	& $\mathrm{SU(3)}$	& $(j^\pi)\Lambda$	\\
\midrule
$\frac{1}{2}^+$	& $\bigl(\frac{1}{2},0\bigr)$ & Octet      & $\bigl(\frac{1}{2}^+\bigr)\frac{1}{2}$                                                            \\
$\frac{3}{2}^+$	& $\bigl(\frac{1}{2},0\bigr)$ & Octet      & $\bigl(\frac{3}{2}^+\bigr)\frac{1}{2}$, $\bigl(\frac{3}{2}^+\bigr)\frac{3}{2}$ \\
$\frac{1}{2}^+$	& $\bigl(\frac{3}{2},0\bigr)$ & Decuplet & $\bigl(\frac{1}{2}^+\bigr)\frac{1}{2}$                                                          \\
$\frac{1}{2}^+$	& $(0,-1)$	& ``Octet''	& $\bigl(\frac{1}{2}^+\bigr)\frac{1}{2}$                                                           \\
$\frac{1}{2}^+$	& $(0,-1)$	& ``Singlet''	& $\bigl(\frac{1}{2}^{+\prime}\bigr)\frac{1}{2}$                                       \\
$\frac{3}{2}^+$	& $(0,-1)$	& Octet 	& $\bigl(\frac{3}{2}^+\bigr)\frac{3}{2}$, $\bigl(\frac{3}{2}^+\bigr)\frac{1}{2}$  \\
$\frac{1}{2}^+$	& $(1,-1)$	& ``Octet''	& $\bigl(\frac{1}{2}^+\bigr)\frac{1}{2}$                                                           \\
$\frac{1}{2}^+$	& $(1,-1)$	& ``Decuplet''	& $\bigl(\frac{1}{2}^{+\prime}\bigr)\frac{1}{2}$                                      \\
$\frac{3}{2}^+$	& $(1,-1)$	& Octet 	& $\bigl(\frac{3}{2}^+\bigr)\frac{1}{2}$, $\bigl(\frac{3}{2}^+\bigr)\frac{3}{2}$ \\
$\frac{1}{2}^+$	&$\bigl(\frac{1}{2},-2\bigr)$	& ``Octet''	& $\bigl(\frac{1}{2}^+\bigr)\frac{1}{2}$                         \\
$\frac{1}{2}^+$	& $\bigl(\frac{1}{2},-2\bigr)$	& ``Decuplet''	& $\bigl(\frac{1}{2}^{+\prime}\bigr)\frac{1}{2}$    \\
$\frac{3}{2}^+$	& $\bigl(\frac{1}{2},-2\bigr)$	& Octet 	& $\bigl(\frac{3}{2}^+\bigr)\frac{1}{2}$, $\bigl(\frac{3}{2}^+\bigr)\frac{3}{2}$  \\
$\frac{1}{2}^+$	& $\bigl(0,-3\bigr)$	& Decuplet 	& $\bigl(\frac{1}{2}^+\bigr)\frac{1}{2}$  \\
\bottomrule
\end{tabular}
\caption{
The 13 low-energy multiplets of adjoint-baryon potentials at small $r$ consisting of a total of 17  distinct B-O potentials $(j^\pi)\Lambda$.
The associated $\bm{8}$-baryons have quantum numbers $j^\pi$ and $(I,S)$ and they belong to SU(3) multiplets.
The SU(3) multiplets labeled ``Octet'' and ``Singlet'' are pure Octet and pure Singlet at the SU(3)-flavor point but they are linear combinations on the (2+1)-flavor line.
The SU(3) multiplets labeled ``Decuplet'' and ``Octet'' are pure Decuplet and pure Octet at the SU(3)-flavor point but they are linear combinations on the (2+1)-flavor line. }
\label{tab:8barpot}
\end{table}

The lowest-energy nonstrange charm-hadron-pair multiplets are $\Lambda_c \bar{D}^{(\ast)}$ and $\Sigma_c^{(\ast)} \bar{D}^{(\ast)}$. 
Given that $j^\pi = 0^+$ for $\Lambda_c$ and $\frac{1}{2}^+$ for $\bar{D}^{(\ast)}$, the only $\Lambda_c \bar{D}^{(\ast)}$ potential is $\bigl(\frac{1}{2}^+\bigr)\frac{1}{2}$, which must connect to the potential associated with the ground-state $\frac{1}{2}^+$ $\bm{8}$-baryon.
Given that $j^\pi = 1^+$ for $\Sigma_c^{(\ast)}$,  the three $\Sigma_c^{(\ast)} \bar{D}^{(\ast)}$ potentials are $\bigl(\frac{3}{2}^+\bigr)\frac{3}{2} \big/ \frac{1}{2}$ and $\bigl(\frac{1}{2}^{+\prime}\bigr)\frac{1}{2}$. 
The prime on the $\bigl(\frac{1}{2}^{+\prime}\bigr)\frac{1}{2}$ potential indicates that it is higher in energy than the $\bigl(\frac{1}{2}^+\bigr)\frac{1}{2}$ potential.
The $\bigl(\frac{3}{2}^+\bigr)\frac{3}{2} \big/ \frac{1}{2}$ potentials must connect to the potentials associated with the ground-state $\frac{3}{2}^+$ $\bm{8}$-baryon.
The remaining $\bigl(\frac{1}{2}^{+\prime}\bigr)\frac{1}{2}$  potential must connect to the potential associated with a same-parity orbitally excited $\frac{1}{2}^+$ $\bm{8}$-baryon.
These static-hadron-pair potentials are listed in Table~\ref{tab:hadronpairpot}.

\begin{table}
\centering
\begin{tabular}{cccc}
\toprule
Baryon-Meson & $\overline{M}_\mathrm{B} + \overline{M}_\mathrm{M}$ [MeV] & $(I,S)$ & $(j^\pi)\Lambda$	\\
\midrule
$\Lambda_c \, \bar{D}^{(\ast)}$	         & 4259.7 	& $\bigl(\frac{1}{2},0\bigr)$ & $\bigl(\frac{1}{2}^+\bigr)\frac{1}{2}$ \\
$\Sigma_c^{(\ast)} \, \bar{D}^{(\ast)}$	 & 4469.8 	& $\bigl(\frac{1}{2},0\bigr)$ & $\bigl(\frac{1}{2}^{+\prime}\bigr)\frac{1}{2}^\star$, $\bigl(\frac{3}{2}^+\bigr)\frac{1}{2}$, $\bigl(\frac{3}{2}^+\bigr)\frac{3}{2}$  \\
$\Sigma_c^{(\ast)} \, \bar{D}^{(\ast)}$ & 4469.8 	& $\bigl(\frac{3}{2},0\bigr)$ & $\bigl(\frac{1}{2}^+\bigr)\frac{1}{2}$, $\bigl(\frac{3}{2}^+\bigr)\frac{1}{2}^\star$, $\bigl(\frac{3}{2}^+\bigr)\frac{3}{2}^\star$ \\
$\Lambda_c \, \bar{D}_s^{(\ast)}$	    & 4362.7  & $(0,-1)$	& $\bigl(\frac{1}{2}^+\bigr)\frac{1}{2}$ \\
$\Xi_c \, \bar{D}^{(\ast)}$	                     & 4442.3  & $(0,-1)$	& $\bigl(\frac{1}{2}^{+\prime}\bigr)\frac{1}{2}$  \\
$\Xi_c^{\prime(\ast)} \, \bar{D}^{(\ast)}$ & 4596.5  & $(0,-1)$	&  $\bigl(\frac{1}{2}^{+\prime\prime}\bigr)\frac{1}{2}^\star$, $\bigl(\frac{3}{2}^+\bigr)\frac{1}{2}$, $\bigl(\frac{3}{2}^+\bigr)\frac{3}{2}$  \\
$\Xi_c \, \bar{D}^{(\ast)}$	                     & 4442.3  & $(1,-1)$	& $\bigl(\frac{1}{2}^+\bigr)\frac{1}{2}$  \\
$\Sigma_c^{(\ast)} \, \bar{D}_s^{(\ast)}$ & 4572.8 & $(1,-1)$	& $\bigl(\frac{1}{2}^{+\prime}\bigr)\frac{1}{2}$, $\bigl(\frac{3}{2}^+\bigr)\frac{1}{2}$, $\bigl(\frac{3}{2}^+\bigr)\frac{3}{2}$ \\
$\Xi_c^{\prime(\ast)} \, \bar{D}^{(\ast)}$ & 4596.5 & $(1,-1)$ & $\bigl(\frac{1}{2}^{+\prime\prime}\bigr)\frac{1}{2}^\star$, $\bigl(\frac{3}{2}^{+\prime}\bigr)\frac{1}{2}^\star$, $\bigl(\frac{3}{2}^{+\prime}\bigr)\frac{3}{2}^\star$ \\
$\Xi_c \, \bar{D}^{(\ast)}_s$                         & 4545.3 & $\bigl(\frac{1}{2},-2\bigr)$ & $\bigl(\frac{1}{2}^+\bigr)\frac{1}{2}$ \\
$\Xi_c^{\prime(\ast)}  \, \bar{D}^{(\ast)}_s$	 & 4699.5 & $\bigl(\frac{1}{2},-2\bigr)$ & $\bigl(\frac{1}{2}^{+\prime}\bigr)\frac{1}{2}$, $\bigl(\frac{3}{2}^+\bigr)\frac{1}{2}$, $\bigl(\frac{3}{2}^+\bigr)\frac{3}{2}$ \\
$\Omega_c^{(\ast)} \, \bar{D}^{(\ast)}$        & 4715.6 & $\bigl(\frac{1}{2},-2\bigr)$ & $\bigl(\frac{1}{2}^{+\prime\prime}\bigr)\frac{1}{2}^\star$, $\bigl(\frac{3}{2}^{+\prime}\bigr)\frac{1}{2}^\star$, $\bigl(\frac{3}{2}^{+\prime}\bigr)\frac{3}{2}^\star$ \\
$\Omega_c^{(\ast)} \, \bar{D}^{(\ast)}_s$ &4818.6 	& $(0,-3)$ & $\bigl(\frac{1}{2}^+\bigr)\frac{1}{2}$, $\bigl(\frac{3}{2}^+\bigr)\frac{1}{2}^\star$, $\bigl(\frac{3}{2}^+\bigr)\frac{3}{2}^\star$  \\
\bottomrule
\end{tabular}
\caption{The 13 low-energy multiplets of static-hadron-pair potentials at large $r$ consisting of a total of 29 distinct B-O potentials $(j^\pi)\Lambda$. 
Stars mark the 12 potentials that do not correspond to any potential in Table~\ref{tab:8barpot}.
The HQSS multiplets of charm baryons and charm mesons have total flavor $(I,S)$.
The threshold $\overline{M}_\mathrm{B} + \overline{M}_\mathrm{M}$ is the sum of the spin-weighted masses for the charm-baryon and charm-meson multiplets.}
\label{tab:hadronpairpot}
\end{table}

The lack of evidence for $c \bar{c}$ pentaquarks with flavor $(I,S)=\bigl(\frac{1}{2},0\bigr)$ and masses near the $\Lambda_c \bar{D}$ and $\Lambda_c \bar{D}^\ast$ thresholds suggests that the $\bigl(\frac{1}{2}^+\bigr)\frac{1}{2}$ potential does not support bound states.
Each of the four observed $c \bar{c}$ pentaquarks with flavor $\bigl(\frac{1}{2},0\bigr)$ listed in Table~\ref{tab:nonstrangepentaquarks} has mass near one of the $\Sigma_c \bar{D}$, $\Sigma_c^\ast \bar{D}$, and $\Sigma_c \bar{D}^\ast$ thresholds.
This suggests that either the $\bigl(\frac{3}{2}^+\bigr)\frac{3}{2} \big/ \frac{1}{2}$ potentials or the $\bigl(\frac{1}{2}^{+\prime}\bigr)\frac{1}{2}$ potential or both support bound states.

At small $r$, an adjoint-baryon potential is dominated by the repulsive color-Coulomb potential between the $\bm{3}$ and $\bm{\bar{3}}$ sources.
Since the color-Coulomb potential is the same for all adjoint-baryon potentials, it cancels out from the energy differences between potentials at $r\to0$.
Therefore, the difference between two adjoint-baryon potentials changes smoothly from the adjoint-baryon energy difference at small $r$ to the difference between static-meson-pair thresholds at large $r$.
At large $r$, the difference between the $\bigl(\frac{1}{2}^{+\prime}\bigr)\frac{1}{2}$ potential that approaches the $\Sigma_c^{(\ast)} \bar{D}^{(\ast)}$ threshold and the $\bigl(\frac{1}{2}^+\bigr)\frac{1}{2}$ potential that approaches the $\Lambda_c \bar{D}^{(\ast)}$ threshold is the difference between the spin-weighted mass of $\Sigma_c^{(\ast)}$ and the mass of $\Lambda_c$, which is 210.1~MeV. 
At small $r$, the difference between the $\bigl(\frac{1}{2}^{+\prime}\bigr)\frac{1}{2}$ and $\bigl(\frac{1}{2}^+\bigr)\frac{1}{2}$ potentials is that between the energies of the same-parity orbital excitation and the ground state of the $\bm{8}$-baryons with quantum numbers $j^\pi=\frac{1}{2}^+$ and flavor $(I,S)=(\frac{1}{2},0)$, which is probably much larger than 210~MeV. 
The order of magnitude of this energy difference can be estimated using the Particle Listings for heavy hadrons in Ref.~\cite{PDG24}.
The same-parity orbital excitations of the pseudoscalar charm mesons and the vector charm mesons have energies larger by about 600~MeV.

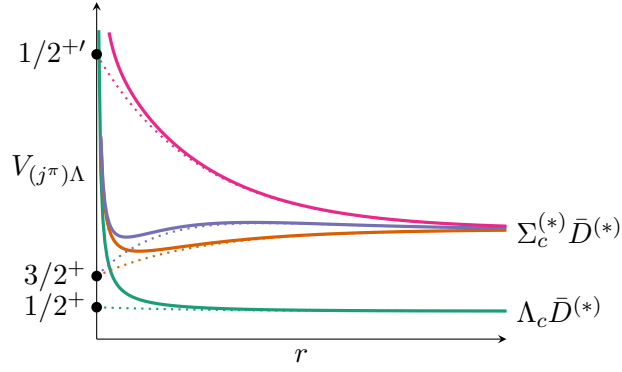
\begin{figure}
\tikzsetnextfilename{qualitative-pentaquark-potentials-nonstrange}
\centering
\begin{tikzpicture}
\pgfmathsetmacro{\hbar}{0.1973}
\pgfmathsetmacro{\keight}{0.037}
\pgfmathsetmacro{\sommer}{0.5}
\begin{axis}[%
	width=7cm,
	xlabel={$r$},
	ylabel={$V_{(j^\pi)\Lambda}$},
	ylabel style={rotate=-90},
	xmajorticks=false,
	ymajorticks=false,
	axis y line=left,
	axis x line=bottom,
	restrict y to domain=0:5,
	enlarge y limits=true,
	cycle list/Dark2,
	clip mode=individual
	]
	\begin{scope}[domain=0:2, samples=200]
		\addplot+[dotted, thick, forget plot]
			{0.01 * exp(-x / \sommer) + 4.26};
		\addplot+[very thick]
			{(\keight * \hbar / x + 0.01) * exp(-x / \sommer) + 4.26};
		\addplot+[dotted, thick, forget plot]
			{(- 0.12) * exp(-x / \sommer) + 4.47};
		\addplot+[very thick]
			{(\keight * \hbar / x - 0.12) * exp(-x / \sommer) + 4.47};
		\addplot+[dotted, thick, forget plot]
			{(- 0.12 + .3 * tanh(x)) * exp(-x / \sommer) + 4.47};
		\addplot+[very thick]
			{(\keight * \hbar / x - 0.12 + .3 * tanh(x)) * exp(-x / \sommer) + 4.47};
		\addplot+[dotted, thick, forget plot]
			{(0.45) * exp(-x / \sommer) + 4.47};
		\addplot+[very thick]
			{(\keight * \hbar / x + 0.45) * exp(-x / \sommer) + 4.47};
	\end{scope}
	\coordinate[label=left:$1/2^+$] (A) at (0,4.27);
	\coordinate[label=left:$3/2^+$] (B) at (0,4.35);
	\coordinate[label=left:$1/2^{+\prime}$] (C) at (0,4.92);
	\fill (A) circle (2pt);
	\fill (B) circle (2pt);
	\fill (C) circle (2pt);
	\node[anchor=west] at (2,4.26) {$\Lambda_c \bar{D}^{(\ast)}$};
	\node[anchor=west] at (2,4.47) {$\Sigma_c^{(\ast)} \bar{D}^{(\ast)}$};
\end{axis}
\end{tikzpicture}
\caption{\label{fig:plotS0}
Plausible qualitative behaviors of the lowest-energy B-O potentials with $(I,S)=\bigl(\frac{1}{2},0\bigr)$.
The $\bigl(\frac{1}{2}^+\bigr)\frac{1}{2}$ and $\bigl(\frac{1}{2}^{+\prime}\bigr)\frac{1}{2}$ potentials approach the $\Lambda_c \bar{D}^{(\ast)}$ and $\Sigma_c^{(\ast)} \bar{D}^{(\ast)}$  thresholds from above at large $r$.
The $\bigl(\frac{3}{2}^+\bigr)\frac{1}{2}$ and $\bigl(\frac{3}{2}^+\bigr)\frac{3}{2}$ potentials cross below the $\Sigma_c^{(\ast)} \bar{D}^{(\ast)}$ threshold before approaching it at large $r$.
The dots on the vertical axis are adjoint-baryon energies.
The dotted curves are the potentials with the color-Coulomb potential set to zero, which change smoothly from the adjoint-baryon energy at small $r$ to a static-hadron-pair threshold at large $r$.}
\end{figure}

The simplest possibility for the $\bigl(\frac{1}{2}^+\bigr)\frac{1}{2}$ potential is that it approaches the $\Lambda_c \bar{D}^{(\ast)}$ threshold from above.
A particularly simple possibility is that the difference between the $\bigl(\frac{1}{2}^{+\prime}\bigr)\frac{1}{2}$ and $\bigl(\frac{1}{2}^+\bigr)\frac{1}{2}$ potentials is a monotonically decreasing function of $r$, in which case the $\bigl(\frac{1}{2}^{+\prime}\bigr)\frac{1}{2}$ potential also approaches the threshold from above.
This would leave the $\bigl(\frac{3}{2}^+\bigr)\frac{3}{2} \big/ \frac{1}{2}$ potentials as the only ones supporting bound states.
A simple possibility is that these two potentials both cross below the $\Sigma_c^{(\ast)} \bar{D}^{(\ast)}$ threshold and then approach it from below.
These simple qualitative behaviors of the B\nobreakdash-O potentials with flavor $\bigl(\frac{1}{2},0\bigr)$ are illustrated in Fig.~\ref{fig:plotS0}.
We will assume they have these qualitative behaviors.

\subsection{Strange Isospin-0 Potentials}

The $c\bar{c}s$ pentaquarks were discovered through their decays into $J/\psi\, \Lambda$, which implies that they have isospin zero. 
We therefore consider in detail the B\nobreakdash-O potentials with isospin zero and strangeness $-1$.
From the SU(3)-flavor multiplets in Fig.~\ref{fig:8barmult}, we expect the lowest-energy $(I,S)= (0, -1)$ $\bm{8}$-baryons to be the $j^\pi=\frac{1}{2}^+$ singlet, the $I=0$ state at the center of the $\frac{1}{2}^+$ octet, and the $I=0$ state at the center of the $\frac{3}{2}^+$ octet.
The two adjoint-baryon potentials associated with the $\frac{1}{2}^+$ $\bm{8}$-baryons are $\bigl(\frac{1}{2}^+\bigr)\frac{1}{2}$ and $\bigl(\frac{1}{2}^{+\prime}\bigr)\frac{1}{2}$, with flavors that are linear combinations of SU(3) octet and singlet.
The prime on the $\bigl(\frac{1}{2}^{+\prime}\bigr)\frac{1}{2}$ potential indicates that it is higher in energy than the $\bigl(\frac{1}{2}^+\bigr)\frac{1}{2}$ potential.
The adjoint-baryon potentials associated with the  $\frac{3}{2}^+$ $\bm{8}$-baryon are coupled $\bigl(\frac{3}{2}^+\bigr)\frac{3}{2} \big/ \frac{1}{2}$ potentials.
All of these adjoint-baryon potentials are listed in Table~\ref{tab:8barpot}.

The lowest-energy charm-hadron-pair multiplets with $(I,S)=(0,-1)$ are $\Lambda_c \bar{D}_s^{(\ast)}$, $\Xi_c \bar{D}^{(\ast)}$, and $\Xi_c^{\prime(\ast)} \bar{D}^{(\ast)}$.
Given that $j^\pi = 0^+$ for $\Lambda_c$ and $\frac{1}{2}^+$  for $\bar{D}_s^{(\ast)}$, the only $\Lambda_c \bar{D}_s^{(\ast)}$ potential is $\bigl(\frac{1}{2}^+\bigr)\frac{1}{2}$, which must connect to the potential associated with the ground-state strange $\frac{1}{2}^+$ $\bm{8}$-baryon.
Given that $j^\pi=0^+$ for $\Xi_c$ and $\frac{1}{2}^+$  for $\bar{D}^{(\ast)}$, the only $\Xi_c \bar{D}^{(\ast)}$ potential is $\bigl(\frac{1}{2}^{+\prime}\bigr)\frac{1}{2}$, which must connect to the potential associated with the first excited strange $\frac{1}{2}^+$ $\bm{8}$-baryon.
Note that this excited $\bm{8}$-baryon differs from the ground state by its flavor, so it need not be much higher in energy.
Given that $j^\pi = 1^+$ for $\Xi_c^{\prime(\ast)}$, the three $\Xi_c^{\prime(\ast)} \bar{D}^{(\ast)}$ potentials are $\bigl(\frac{3}{2}^+\bigr)\frac{3}{2} \big/ \frac{1}{2}$ and $\bigl(\frac{1}{2}^{+\prime\prime}\bigr)\frac{1}{2}$. 
The $\bigl(\frac{3}{2}^+\bigr)\frac{3}{2} \big/ \frac{1}{2}$ potentials must connect to the potentials associated with the ground-state strange $\frac{3}{2}^+$ $\bm{8}$-baryon.
The remaining $\bigl(\frac{1}{2}^{+\prime\prime}\bigr)\frac{1}{2}$ potential must connect to the  potential associated with a same-parity orbitally-excited strange $\frac{1}{2}^+$ $\bm{8}$-baryon.
All of these static-hadron-pair potentials are listed in Table~\ref{tab:hadronpairpot}.

\begin{figure}
\tikzsetnextfilename{qualitative-pentaquark-potentials-strange}
\centering
\begin{tikzpicture}
\pgfmathsetmacro{\hbar}{0.1973}
\pgfmathsetmacro{\keight}{0.037}
\pgfmathsetmacro{\sommer}{0.5}
\begin{axis}[%
	width=7cm,
	xlabel={$r$},
	ylabel={$V_{(j^\pi)\Lambda}$},
	ylabel style={rotate=-90},
	xmajorticks=false,
	ymajorticks=false,
	axis y line=left,
	axis x line=bottom,
	restrict y to domain=0:4.5,
	enlarge y limits=true,
	cycle list/Dark2,
	clip mode=individual
	]
	\begin{scope}[domain=0:2, samples=200]
		\addplot+[dotted, thick, forget plot]
			{ 0.01 * exp(-x / \sommer) + 4.363};
		\addplot+[very thick]
			{(\keight * \hbar / x + 0.01) * exp(-x / \sommer) + 4.363};
		\addplot+[dotted, thick, forget plot]
			{- 0.05 * exp(-x / \sommer) + 4.442};
		\addplot+[very thick]
			{(\keight * \hbar / x - 0.05) * exp(-x / \sommer) + 4.442};
	\end{scope}
	\coordinate[label=left:$1/2^+$] (A) at (0,4.373);
	\coordinate[label=left:$1/2^{+\prime}$] (B) at (0,4.392);
	\fill (A) circle (2pt);
	\fill (B) circle (2pt);
	\node[anchor=west] at (2,4.363) {$\Lambda_c \bar{D}_s^{(\ast)}$};
	\node[anchor=west] at (2,4.442) {$\Xi_c \bar{D}^{(\ast)}$};
\end{axis}
\end{tikzpicture}
\caption{\label{fig:plotS1}
Plausible qualitative behavior of the lowest-energy B-O potentials with $(I,S)=(0,-1)$.
The $\bigl(\frac{1}{2}^+\bigr)\frac{1}{2}$ potential approaches the $\Lambda_c \bar{D}_s^{(\ast)}$ threshold from above at large $r$.
The $\bigl(\frac{1}{2}^{+\prime}\bigr)\frac{1}{2}$ potential crosses below the $\Xi_c \bar{D}^{(\ast)}$ threshold before approaching it at large $r$.
The dots on the vertical axis are adjoint-baryon energies.
The dotted curves are the potentials with the color-Coulomb potential set to zero, which change smoothly from the adjoint-baryon energy at small $r$ to a  static-hadron-pair threshold at large $r$.}
\end{figure}
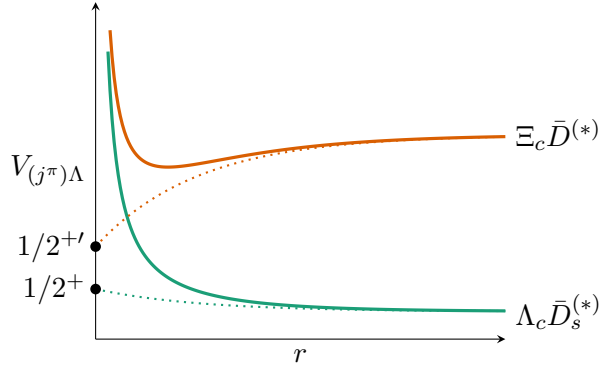

The lack of evidence for $c \bar{c}$ pentaquarks with flavor $(I,S)=(0,-1)$ and masses near the $\Lambda_c \bar{D}_s$ and $\Lambda_c \bar{D}_s^\ast$ thresholds suggests that the $\bigl(\frac{1}{2}^+\bigr)\frac{1}{2}$ potential does not support bound states.  
The simplest possibility is that it approaches the $\Lambda_c \bar{D}_s^{(\ast)}$ threshold from above.
The two observed $c \bar{c}s$ pentaquarks with flavor $(0,-1)$ listed in Table~\ref{tab:strangepentaquarks} have mass near the $\Xi_c \bar{D}$ and $\Xi_c \bar{D}^\ast$ thresholds.
This suggests that the $\bigl(\frac{1}{2}^{+\prime}\bigr)\frac{1}{2}$ potential supports bound states.
A simple possibility is that it crosses below the $\Xi_c \bar{D}^{(\ast)}$ threshold and then approaches it from below.
These simple qualitative behaviors of the B\nobreakdash-O  potentials with $(I,S)= \bigl(\frac{1}{2}, 0\bigr)$ are illustrated in Fig.~\ref{fig:plotS1}.
We will assume they have these qualitative behaviors.

The $\bigl(\frac{1}{2}^+\bigr)\frac{1}{2}$ and $\bigl(\frac{1}{2}^{+\prime}\bigr)\frac{1}{2}$ potentials have the flavors $\Lambda_c \bar{D}_s^{(\ast)}$ and $\Xi_c \bar{D}^{(\ast)}$ at large $r$.
They may have significant flavor dependence as functions of $r$.
We denote their flavor states as $r \to \infty$ by $\lvert\Lambda_c \bar{D}_s^{(\ast)}\rangle $  and  $\lvert\Xi_c \bar{D}^{(\ast)}\rangle$, where $\lvert\Xi_c \bar{D}^{(\ast)}\rangle$ is the odd linear combination of  $\lvert\Xi_c^{+} \bar{D}^{(\ast)-}\rangle$ and $\lvert\Xi_c^{0} \bar{D}^{(\ast)0}\rangle$.
At the SU(3)-flavor point, the $\Lambda_c \bar{D}_s^{(\ast)}$ and $\Xi_c \bar{D}^{(\ast)}$ thresholds are degenerate and the flavors of the B\nobreakdash-O potentials do not change with $r$.
In physical QCD, the $\Xi_c \bar{D}^{(\ast)}$ threshold is higher in energy than the $\Lambda_c \bar{D}_s^{(\ast)}$ threshold by 79.6~MeV.  
At large $r$, the $\bigl(\frac{1}{2}^{+\prime}\bigr)\frac{1}{2}$ potential is therefore larger than the $\bigl(\frac{1}{2}^+\bigr)\frac{1}{2}$ potential by that energy.
At small $r$, the flavors of the $\bigl(\frac{1}{2}^+\bigr)\frac{1}{2}$ and $\bigl(\frac{1}{2}^{+\prime}\bigr)\frac{1}{2}$ potentials approach those of the $\frac{1}{2}^+$ and $\frac{1}{2}^{+\prime}$ $\bm{8}$-baryons with $(I,S)=(0,-1)$.
At the SU(3)-flavor point, the two lowest-energy $\bm{8}$-baryons are an SU(3) singlet and the center of an SU(3) octet. We denote the corresponding flavor states by $\lvert\mathrm{Singlet}\rangle$ and $\lvert\mathrm{Octet}\rangle$. 
They can be expressed as orthogonal linear combinations of flavor states of a charm baryon and a charm meson:
\begin{subequations}
\begin{align}
\bigl\lvert \mathrm{Singlet} \bigr\rangle &=  \sqrt{\tfrac{2}{3}}\,  \bigl\lvert \Xi_c \bar{D}^{(\ast)} \bigr\rangle
+\sqrt{\tfrac{1}{3}}\, \bigl\lvert \Lambda_c \bar{D}_s^{(\ast)} \bigr\rangle,
\label{Singlet}
\\
\bigl\lvert \mathrm{Octet} \bigr\rangle &= - \sqrt{\tfrac{1}{3}}\, \bigl\lvert \Xi_c \bar{D}^{(\ast)} \bigr\rangle 
+\sqrt{\tfrac{2}{3}}\, \bigl\lvert \Lambda_c \bar{D}_s^{(\ast)} \bigr\rangle.
\label{Octet}
\end{align}
\label{SingletOctet}%
\end{subequations}
The octet $\bm{8}$-baryon is expected to be lower in energy, because it has the higher probability for $\Lambda_c \bar{D}_s^{(\ast)}$, which has a lower energy than $\Xi_c \bar{D}^{(\ast)}$ in physical QCD. 

In $(2+1)$-flavor QCD, the flavors of the $\bigl(\frac{1}{2}^+\bigr)\frac{1}{2}$ and $\bigl(\frac{1}{2}^{+ \prime}\bigr)\frac{1}{2}$ potentials change from those of the $\bm{8}$-baryons at small $r$ to those of the hadron pairs $\Lambda_c \bar{D}_s^{(\ast)}$ and $\Xi_c \bar{D}^{(\ast)}$ at large $r$.
There are two extreme possibilities for the flavors of the $\bm{8}$-baryons as $m_u/m_s$ decreases along the (2+1)-flavor line towards the physical QCD point.
The first extreme is that the $\bm{8}$-baryon flavors remain pure SU(3) octet and pure SU(3) singlet.  
In this case, the flavors of the $\bigl(\frac{1}{2}^+\bigr)\frac{1}{2}$ and $\bigl(\frac{1}{2}^{+ \prime}\bigr)\frac{1}{2}$ potentials have significant dependence on $r$, approaching those in Eqs.~\eqref{SingletOctet} at small $r$ and approaching $\Lambda_c \bar{D}_s^{(\ast)}$ and $\Xi_c \bar{D}^{(\ast)}$  at large $r$.
The second extreme is that the $\bm{8}$-baryon flavors approach those of $\Lambda_c \bar{D}_s^{(\ast)}$ and $\Xi_c \bar{D}^{(\ast)}$.
In this case, the flavors of the $\bigl(\frac{1}{2}^+\bigr)\frac{1}{2}$ and $\bigl(\frac{1}{2}^{+\prime}\bigr)\frac{1}{2}$ potentials do not change as functions of $r$. 
Thus the probability for the $\bigl(\frac{1}{2}^{+ \prime}\bigr)\frac{1}{2}$ potential to have flavor $\Xi_c \bar{D}^{(\ast)}$ is 1 at large $r$ but it could decrease to as small as $\frac{2}{3}$ at small $r$. 
Similarly, the probability for the $\bigl(\frac{1}{2}^{+ \prime}\bigr)\frac{1}{2}$ potential to have flavor $\Lambda_c \bar{D}_s^{(\ast)}$ is zero at large $r$ but it could increase to as large as $\frac{1}{3}$ at small $r$.

\subsection{Other Pentaquark Potentials}
\label{sec:OtherPotentials}

We have discussed pentaquark potentials with flavor $(I,S) = \bigl( \frac{1}{2},0 \bigr)$ and $( 0, -1 )$.
We will briefly discuss potentials with a larger value of $I$ or a larger value of $\lvert S\rvert$.
Note that there are 17 adjoint-baryon potentials listed in Table~\ref{tab:8barpot} while there are 29 static-hadron-pair potentials listed in Table~\ref{tab:hadronpairpot}. 
Thus almost half the static-hadron-pair potentials must connect to adjoint-baryon potentials associated with same-parity orbitally-excited $\bm{8}$-baryons.
These potentials are likely to approach the static-hadron-pair potential from above and therefore are unlikely to support bound states.

Nonstrange  hidden-charm pentaquarks with isospin $\frac{3}{2}$ could be discovered through their decays into $J/\psi\, \Delta^{++}$, with $\Delta^{++}$ decaying into $p\, \pi^+$ resulting in two charged hadrons.
The lowest $(I,S) = \bigl( \frac{3}{2},0 \bigr)$ potential has short-distance behavior determined by the $\frac{1}{2}^+$ $\bm{8}$-baryons that form the  isospin quartet in the top row of the $\frac{1}{2}^+$ decuplet in Fig.~\ref{fig:8barmult}.
At long distance, it must approach the $\Sigma_c^{(\ast)} \bar{D}^{(\ast)}$ threshold.
The potentials that approach that threshold are those in the row of Table~\ref{tab:hadronpairpot} for $\Sigma_c^{(\ast)} \bar{D}^{(\ast)}$ with $(I,S) = \bigl( \frac{3}{2},0 \bigr)$, which are $\bigl( \frac{1}{2}^+ \bigr)\frac{1}{2}$ and $\bigl(\frac{3}{2}^+\bigr)\frac{3}{2} \big/ \frac{1}{2}$.
The only potential that can connect at short distance to the adjoint-baryon potential associated with the  $\frac{1}{2}^+$ $\bm{8}$-baryon with $(I,S) = \bigl( \frac{3}{2},0 \bigr)$ is $\bigl( \frac{1}{2}^+ \bigr)\frac{1}{2}$.
The $\bigl(\frac{3}{2}^+\bigr)\frac{3}{2} \big/ \frac{1}{2}$ potentials must connect at short distance to potentials associated with a higher-energy $\bm{8}$-baryon.
They are likely to approach the $\Sigma^{(\ast)} \bar{D}^{(\ast)}$ threshold from above,
in which case they cannot support bound states or resonances.
If the energy of the $\bm{8}$-baryon with $j^\pi =\frac{1}{2}^+$ and $(I,S) = \bigl(\frac{3}{2},0\bigr)$ is low enough, the $\bigl( \frac{1}{2}^+ \bigr)\frac{1}{2}$ potential could cross below the $\Sigma_c^{(\ast)} \bar{D}^{(\ast)}$ threshold before approaching it,
in which case it could support bound states or resonances.

Strange hidden-charm pentaquarks with isospin 1 could decay into $J/\psi\, \Sigma$,  but they would be difficult to observe because none of the strange baryons $\Sigma$ have any decay modes with all charged hadrons.
Double-strange $c \bar{c}$ pentaquarks with  isospin $\frac{1}{2}$ could be discovered through their decays into $J/\psi\, \Xi^-$, with $\Xi^-$ decaying into $\Lambda\, \pi^-$ and $\Lambda$  decaying into $p\, \pi^-$ resulting in three charged hadrons.
The lowest $(I,S) = \bigl( \frac{1}{2},-2 \bigr)$ potentials have short-distance behavior determined by the $\bm{8}$-baryons that form the  isospin doublets in the bottom row of either the $\frac{1}{2}^+$ octet in Fig.~\ref{fig:8barmult} or the $\frac{3}{2}^+$ octet. 
At long distance, they must approach the $\Xi_c \bar{D}_s^{(\ast)}$, $\Xi_c^{\prime (\ast)} \bar{D}_s^{(\ast)}$, or $\Omega_c \bar{D}^{(\ast)}$ threshold. 
Of the seven static-hadron-pair potentials with $(I,S) = \bigl( \frac{1}{2},-2 \bigr)$ listed in Table~\ref{tab:hadronpairpot}, the only ones that can connect to the low-energy adjoint-baryon potentials with $(I,S) = \bigl( \frac{1}{2},-2 \bigr)$ listed in Table~\ref{tab:8barpot} are $\bigl(\frac{1}{2}^+\bigr)\frac{1}{2}$, $\bigl(\frac{1}{2}^{+\prime}\bigr)\frac{1}{2}$, and $\bigl(\frac{3}{2}^+\bigr)\frac{3}{2} \big/ \frac{1}{2}$.
Whether any of these adjoint-baryon potentials cross below the static-hadron-pair thresholds before approaching them is determined by the strangeness splittings of the $\bm{8}$-baryons.
The strangeness splittings of $\bm{8}$-baryons can be calculated using lattice QCD with an $\bm{8}$ color source or lattice NRQCD with a heavy color-octet particle, but no such information is yet available.

\section{Hidden-charm Pentaquark Energies}
\label{sec:pentaquarkEnergies}

In this section, we use first-order perturbation theory in the spin splittings of charm hadrons to estimate the spin splittings for the hidden-charm pentaquarks.

\subsection{Spin Splittings}
\label{sec:SpinSplittings}

At first order in $1/m_Q$, the correction $\mathbf{V}_\mathrm{NLO}(\bm{r},\bm{S}_Q,\bm{S}_{\bar{Q}})$ to the B\nobreakdash-O potentials for hidden-heavy pentaquarks depends on the separation vector $\bm{r}$ and the spins $\bm{S}_Q$ and $\bm{S}_{\bar{Q}}$ of the heavy quark and heavy antiquark.
This potential includes short-distance terms that give spin splittings to $\bm{8}$-hadrons and long-distance terms that give spin splittings to a well-separated heavy baryon and heavy meson.
We focus on the long-distance term, which is particularly important as it splits each static-hadron-pair potential at large $r$ into several heavy-hadron-pair potentials.
It is also the dominant term in the case of an $S$-wave bound state near the static-hadron-pair threshold, because most of the probability is in the long-distance region beyond the range of the potential \cite{Braa04}.

At long distances, the light-QCD fields reduce to a  well-separated  $\bm{3}$-baryon and $\bm{\bar{3}}$-meson.
The potential $\mathbf{V}_\mathrm{NLO}(\bm{r},\bm{S}_Q,\bm{S}_{\bar{Q}})$ reduces to a constant potential $\mathbf{V}_\mathrm{SS}(\bm{S}_Q,\bm{S}_{\bar{Q}})$ that depends on the spins but not on $\bm{r}$. 
The long-distance spin-splitting potential was given for hidden-heavy tetraquarks in Ref.~\cite{Braa24b}.
For hidden-heavy pentaquarks, the long-distance spin-splitting potential is
\begin{equation}
\mathbf{V}_\mathrm{SS}(\bm{S}_Q,\bm{S}_{\bar{Q}}) 
 = C_\mathrm{B} \, \bm{j}_\mathrm{B} \cdot \bm{S}_Q + C_\mathrm{M} \, \bm{j}_\mathrm{M}\cdot \bm{S}_{\bar{Q}},
\label{VSS}
\end{equation}
where $C_\mathrm{B}$ and $C_\mathrm{M}$ are parameters of order $1/m_Q$ and $\bm{j}_\mathrm{B}$ and $\bm{j}_\mathrm{M}$ are the angular-momentum vectors of the well-separated $\bm{3}$-baryon and $\bm{\bar{3}}$-meson. 
The angular momentum $j_\mathrm{B}$ of the $\bm{3}$-baryon is an integer.
The angular momentum $j_\mathrm{M}$ of the $\bm{\bar{3}}$-meson is a half-integer.

We introduce the notation $(j_1,j_2) j_3$ for the angular-momentum multiplet with quantum number $j_3$ from the irreducible decomposition of the direct product of multiplets with quantum numbers $j_1$ and $j_2$.  
Implicit in the notation $(j_1,j_2) j_3$ are the triangle relations $\lvert j_1-j_2\rvert \le j_3 \le j_1+j_2$.
The triangle relations for $(j_1,j_2) j_3$ are invariant under permutations of $j_1$, $j_2$, and $j_3$.

The spin vectors of the heavy baryon and heavy meson are $\bm{J}_\mathrm{B} = \bm{j}_\mathrm{B} + \bm{S}_Q$ and $\bm{J}_\mathrm{M} = \bm{j}_\mathrm{M} + \bm{S}_{\bar{Q}}$.
The spin-splitting potential in Eq.~\eqref{VSS} is diagonal in the \emph{dihadron basis} $(J_\mathrm{B},J_\mathrm{M})$ labeled by the spins $J_\mathrm{B}$ and $J_\mathrm{M}$ of the heavy baryon and heavy meson.
Its eigenvalues are
\begin{multline}
E_\mathrm{SS}(J_\mathrm{B},J_\mathrm{M}) = \frac{1}{2j_\mathrm{B} + 1}  \biggl[ J_\mathrm{B} (J_\mathrm{B} +1) - j_\mathrm{B} (j_\mathrm{B} + 1) - \frac{3}{4} \biggr] \Delta_\mathrm{B}
\\
+ \frac{1}{2j_\mathrm{M} +1} \biggl[ J_\mathrm{M} (J_\mathrm{M} + 1) - j_\mathrm{M} (j_\mathrm{M} +1) - \frac{3}{4} \biggr]   \Delta_\mathrm{M} .
\label{VSS-diag}
\end{multline}
The angular momenta are constrained by the triangle relations for $(j_\mathrm{B},\frac{1}{2})J_\mathrm{B}$ and $(j_\mathrm{M},\frac{1}{2})J_\mathrm{M}$. 
If $j_\mathrm{B}=0$, the heavy-baryon spin-splitting term is zero. 
If $j_\mathrm{B} \ge 1$, the coefficient $\Delta_\mathrm{B}$ in Eq.~\eqref{VSS-diag} is the spin splitting of a heavy-baryon doublet: $C_\mathrm{B} = 2 \Delta_\mathrm{B} / (2j_\mathrm{B}+1)$.
The coefficient $\Delta_\mathrm{M}$ in Eq.~\eqref{VSS-diag}  is the spin splitting of a heavy-meson doublet: $C_\mathrm{M} = 2 \Delta_\mathrm{M} / (2j_\mathrm{M}+1)$.

A simple approximation to the spin splittings of the pentaquarks can be obtained by taking the spin-splitting potential $\mathbf{V}_\mathrm{SS}$ in Eq.~\eqref{VSS} into account using first-order perturbation theory.
The calculation of the spin splittings is described in Appendix~\ref{app:PentaquarkSplittingsAmplitudes}.
The angular momenta are constrained by the triangle relations for $(j,S_{Q \bar Q})J$.
The heavy-quark pair can be spin singlet ($S_{Q \bar Q}=0$) or spin triplet ($S_{Q \bar Q}=1$).
The ground state for $Q \bar{Q}$ in the B\nobreakdash-O potentials $(j{}^{\pi}) j \big/ \ldots \big/ \frac{1}{2}$ consists of states that form a degenerate heavy-quark spin symmetry (HQSS) multiplet.
If $j=\frac{1}{2}$, the HQSS multiplet is a triplet consisting of a spin-singlet state with $J=\frac{1}{2}$ and two spin-triplet states with $J=\frac{1}{2},\frac{3}{2}$.
If $j\geq \frac{3}{2}$, the HQSS multiplet is a quartet consisting of a spin-singlet state with $J=j$ and three spin-triplet states with $J=j-1,j,j+1$.
The pentaquarks in the HQSS multiplet all have the same parity $P=-\pi$.

We denote the block of the spin-splitting potential with total angular momentum $J$ by $\bigl(\mathbf{V}_\mathrm{SS} \bigr)^J$. 
The expectation values of the entries of $\bigl(\mathbf{V}_\mathrm{SS} \bigr)^J$ are given in Appendix~\ref{app:PentaquarkSplittingsAmplitudes}.
At first order in perturbation theory, the spin splittings of the pentaquarks with spin $J$ are given by the eigenvalues of  $\bigl(\mathbf{V}_\mathrm{SS} \bigr)^J$.
The spin-weighted average of the spin splittings of all the pentaquarks in an HQSS multiplet is zero.

\subsection{\texorpdfstring{$c\bar{c}$}{Hidden-charm} Pentaquarks}
\label{sec:nonstrangeEnergies}

We have identified the $c \bar{c}$ pentaquarks as bound states in the $\bigl(\frac{3}{2}^+\bigr)\frac{3}{2} \big/ \frac{1}{2}$ potentials that approach the $\Sigma_c^{(\ast)} \bar{D}^{(\ast)}$ threshold at large $r$.
The HQSS multiplet associated with the ground state for $c \bar{c}$ in these coupled potentials consists of one spin-singlet  state with $J^P= \frac{3}{2}^-$ and three spin-triplet  states with $J^P= \frac{1}{2}^-,\frac{3}{2}^-,\frac{5}{2}^-$. 
We approximate the spin splittings of the pentaquarks by first order perturbation theory in the spin-splitting term $\mathbf{V}_\mathrm{SS}$ in Eq.~\eqref{VSS}
with $\Delta_\mathrm{B}=66.5$~MeV and $\Delta_\mathrm{M}=142.1$~MeV.
The energies of the four pentaquarks are given in Appendix~\ref{app:PentaquarkSplittingsAmplitudes} as functions of the ground-state energy $\varepsilon_{3/2}$, $\Delta_\mathrm{B}$, and $\Delta_\mathrm{M}$.

\begin{table}
\centering
\begin{tabular}{lccccccc}
\toprule
$J^P$ & $\Delta  E$ [MeV] &  $\bigl( \frac{3}{2}^+, 0 \bigr)$   &  $\bigl( \frac{3}{2}^+, 1 \bigr)$ & $\Sigma_c \bar{D}$  &  $\Sigma_c^\ast \bar{D}$  &  $\Sigma_c \bar{D}^\ast$  &  $\Sigma_c^\ast \bar{D}^\ast$  \\
\midrule
$\frac{1}{2}^-$              & $-96.2$  &   \hphantom{+}0           &   $+1$   &   $-\frac{2}{3}$   &  \hphantom{+}0      &   $-\frac{2}{9}$        &   $+\frac{1}{9}$  \\
$\frac{3}{2}^-$              & $-45.1$  &    $+0.128$   &   $+0.872$   & \hphantom{+}0             &   $-0.611$     &   $-0.240$        &   $-0.149$   \\
$\frac{3}{2}^{- \prime}$ & \hphantom{0}$+6.6$ &  $-0.872$ &  $+0.128$   &  \hphantom{+}0              &   $+0.056$     &   $-0.649$        &   $+0.295$   \\
$\frac{5}{2}^-$              & $+57.7$ &   \hphantom{+}0            &   $+1$     &  \hphantom{+}0              &   \hphantom{+}0    &   \hphantom{+}0      &   $-1$	\\
\bottomrule
\end{tabular}
\caption{Lowest HQSS multiplet of $c \bar{c}$ pentaquarks with $j^\pi=\frac{3}{2}^+$ and $(I,S)= \bigl( \frac{1}{2}, 0 \bigr)$.
The energy $\Delta E$ for the pentaquark with quantum numbers $J^P$ is the spin splitting relative to the spin-weighted average of the four pentaquarks.
The columns labeled $\bigl( \frac{3}{2}^+, S_{Q \bar Q} \bigr)$ give the amplitude and probability for the heavy-quark-pair spin $S_{Q \bar Q}$.
The columns labeled by a charm-hadron pair give the amplitude and probability for that charm-hadron pair.  
We use the convention commonly used in tables of Clebsch-Gordan coefficients in which the probability is given by the absolute value of the entry and the amplitude is given by the square root of the probability multiplied by the sign of the entry.
For each pentaquark, the sum of the probabilities for the quantum numbers $(j^\pi, S_{Q \bar Q})$ is $1$, and the sum of the probabilities for the charm-hadron pairs is also $1$.}
\label{tab:nonstrange}
\end{table}

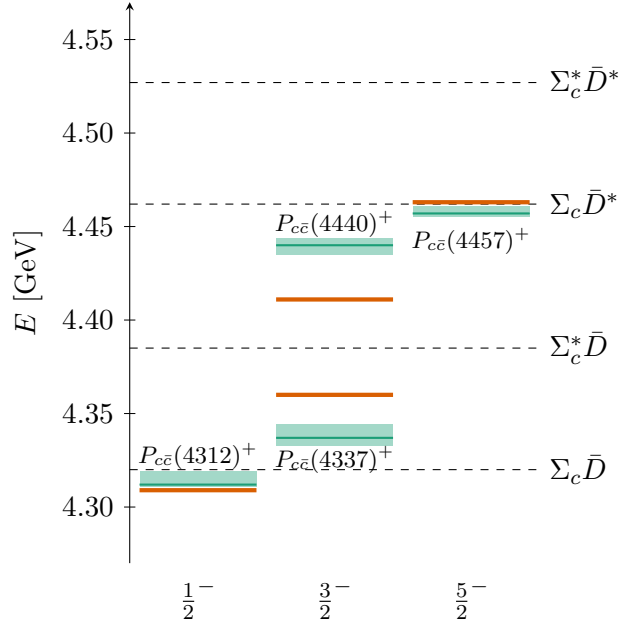
\begin{figure}
\tikzsetnextfilename{pentaquark-energy-levels-nonstrange}
\centering
\begin{tikzpicture}
\pgfmathsetmacro{\MSigma}{2.453}
\pgfmathsetmacro{\MSigmastar}{2.518}
\pgfmathsetmacro{\MD}{1.867}
\pgfmathsetmacro{\MDstar}{2.009}
\pgfplotsset{
	/pgfplots/error bars/draw error bar/.code 2 args={%
		\node[inner sep=0pt] (a) at #1 {};
		\node[inner sep=0pt] (b) at #2 {};
		\path[draw=none, fill, opacity=.4] ($(a) - (22pt,0)$) rectangle ($(b) + (22pt,0)$);
	}
}
\pgfplotstableread{
x	y
0	4.309
1	4.360
1	4.411
2	4.463
}{\nonstrangetheor}
\pgfplotstableread{
x 	y 	ey+ 	ey-
0	4.312	0.007	0.001
1	4.337	0.007	0.004
1	4.440	0.004	0.005
2	4.457	0.004	0.002
}{\nonstrangeexp}
\begin{axis}[
	width=7cm,
	height=9cm,
	only marks,
	black,
	clip mode=individual,
	ylabel={$E$},
	y unit=eV,
	y SI prefix=giga,
	axis y line=left,
	axis x line=bottom,
	ymin=4.27,
	ymax=4.57,
	xmin=-.5,
	xmax=2.5,
	x axis line style={draw=none},
	x tick style={draw=none},
	y tick style={black},
	xtick={0, 1, 2},
	xticklabels={$\frac{1}{2}^{-}$, $\frac{3}{2}^{-}$, $\frac{5}{2}^{-}$},
	ytick distance=0.05,
	yticklabel style={/pgf/number format/.cd,fixed,zerofill,precision=2},
	cycle list/Dark2
	]
	\begin{scope}[dashed]
		\draw (-.5,{\MSigma+\MD}) -- (2.5,{\MSigma+\MD})
		node[anchor=west] {$\Sigma_c\bar{D}$};
		\draw (-.5,{\MSigmastar+\MD}) -- (2.5,{\MSigmastar+\MD})
		node[anchor=west] {$\Sigma_c^\ast\bar{D}$};
		\draw (-.5,{\MSigma+\MDstar}) -- (2.5,{\MSigma+\MDstar})
		node[anchor=west] {$\Sigma_c\bar{D}^\ast$};
            		\draw (-.5,{\MSigmastar+\MDstar}) -- (2.5,{\MSigmastar+\MDstar})
		node[anchor=west] {$\Sigma_c^\ast\bar{D}^\ast$};
	\end{scope}
	\addplot+[mark=-,thick,mark size=22pt,
		error bars/.cd, y dir=both, y explicit]
		table[y error plus=ey+, y error minus=ey-] {\nonstrangeexp};
	\addplot+[mark=-,ultra thick,mark size=22pt] 
		table {\nonstrangetheor};
	\node[anchor=south] at (0, 4.315) {\footnotesize $P_{c\bar{c}}(4312)^+$};
	\node[anchor=north] at (1, 4.338) {\footnotesize $P_{c\bar{c}}(4337)^+$};
	\node[anchor=south, inner sep=.2pt] at (1, 4.444) {\footnotesize $P_{c\bar{c}}(4440)^+$};
	\node[anchor=north] at (2, 4.455) {\footnotesize $P_{c\bar{c}}(4457)^+$};
\end{axis}
\end{tikzpicture}
\caption{\label{fig:NonstrangeEnergyLevels}
Energy-level diagram for the  ground-state quartet of isospin-$\frac{1}{2}$ $c \bar{c}$ pentaquarks associated with the $\frac{3}{2}^+$ $\bm{8}$-baryon. 
The dashed horizontal lines are thresholds for a charm baryon and a charm meson. 
The shaded boxes are the error bands for the measured masses, with the central value highlighted by a thin line.
The thick bars are predictions given by the sum of a ground-state energy and spin splittings.
The ground-state energy is equal to the spin-weighted average of the central values of the observed masses assuming the predicted spins.
The spin splittings are from first order perturbation theory in the charm-hadron spin-splitting term.
}
\end{figure}

The spin splittings of the four pentaquarks states are given in Table~\ref{tab:nonstrange}.
Their spin-weighted average is zero.
The spin splitting of the $\frac{5}{2}^-$ pentaquark is equal to the spin splitting of the $\Sigma_c^\ast \bar{D}^\ast$ threshold with respect to the spin-weighted $\Sigma_c^{(\ast)} \bar{D}^{(\ast)}$ threshold.
The differences between the masses of the four observed pentaquarks listed in Table~\ref{tab:nonstrangepentaquarks} are close enough to the differences between the spin splittings of the four $\Sigma_c^{(\ast)} \bar{D}^{(\ast)}$ thresholds to be consistent with the perturbative approximation.  
We therefore identify $P_{c\bar{c}}(4312)^+$, $P_{c\bar{c}}(4337)^+$, $P_{c\bar{c}}(4440)^+$, and $P_{c\bar{c}}(4457)^+$ with the $\frac{1}{2}^-$, $\frac{3}{2}^-$,  $\frac{3}{2}^{-\prime}$, and $\frac{5}{2}^-$ pentaquarks in Table~\ref{tab:nonstrange}, respectively.
Our prediction $\frac{5}{2}^-$ for the quantum numbers of $P_{c\bar{c}}(4457)^+$ is different from almost all previous predictions in the literature \cite{Du19,Pav19,Xia19,Liu20,Gir21,Bur22,Den22,Li23,Xu23,Cly24,
Pen24,Bram25,Lia26}, agreeing instead with only a few existing studies \cite{Ani15,Shi21,Wan22}.
Given our predictions for the spin quantum numbers, the spin-weighted average of the central values of the masses of the observed $c\bar{c}$ pentaquarks in Table~\ref{tab:nonstrange} is $M_0 =4404.8$~MeV. 
We can predict the masses of the pentaquarks by adding $M_0$ to the spin splittings in Table~\ref{tab:nonstrange}.
The predicted masses are compared to the measured masses  in the energy-level diagram in Fig.~\ref{fig:NonstrangeEnergyLevels}.

Table~\ref{tab:nonstrange} also provides information about the nature of the four pentaquarks by giving amplitudes and probabilities for their components. 
The calculation of the amplitudes is described in Appendix~\ref{app:PentaquarkSplittingsAmplitudes}.
The amplitudes for the spin-singlet and spin-triplet components are given in the columns labeled $\bigl( \frac{3}{2}^+, 0 \bigr)$ and $\bigl( \frac{3}{2}^+, 1 \bigr)$.
The $\frac{1}{2}^-$ and $\frac{5}{2}^-$ pentaquarks are pure spin triplet, while the $\frac{3}{2}^-$  and $\frac{3}{2}^{-\prime}$ pentaquarks are linear combinations of spin-singlet and spin triplet.
The amplitudes for each of the constituent charm-hadron pairs are given in the columns labeled $\Sigma_c \bar{D}$,  $\Sigma_c^\ast \bar{D}$, $\Sigma_c \bar{D}^\ast$, and $\Sigma_c^\ast \bar{D}^\ast$. 
The $\frac{5}{2}^-$ pentaquark is pure $\Sigma_c^\ast \bar{D}^\ast$. 
The other three pentaquarks are superpositions of three charm-hadron pairs.
The $\Sigma_c^\ast \bar{D}$ component of the $\frac{3}{2}^{-\prime}$ pentaquark has the extremely small probability 0.0558. 
Part of the reason this probability is so small is that its expansion in powers of $\Delta_\mathrm{B}/\Delta_\mathrm{M}$ begins at second order with the term $(5/72) (\Delta_\mathrm{B}/\Delta_\mathrm{M})^2$. 

\subsection{Molecular Dominance with Ideal Mixing}
\label{sec:MoleculeDominance}

The proximity of the masses of the observed $c \bar{c}$ pentaquarks to thresholds for a charm baryon and charm meson has motivated their identifications as hadronic molecules consisting of charm hadrons.
The $P_{c\bar{c}}(4312)^+$ and $P_{c\bar{c}}(4380)^+$ have been often identified as molecules with constituents $\Sigma_c \bar{D}$ and $\Sigma_c^\ast \bar{D}$, while $P_{c\bar{c}}(4440)^+$ and $P_{c\bar{c}}(4457)^+$ have both been often identified as molecules with constituents $\Sigma_c \bar{D}^\ast$; see, for instance, Refs.~\cite{Du19,Pav19,Vol19,Xia19,Liu20,Cly24,Pen24} and references therein. 
Other references that considered the molecular nature of the $c\bar{c}$ pentaquarks include Ref.~\cite{Bur22}, where $P_{c\bar{c}}(4312)^+$, $P_{c\bar{c}}(4380)^+$, and $P_{c\bar{c}}(4440)^+$ are identified as molecules but a different nature is assigned to $P_{c\bar{c}}(4457)^+$, and Ref.~\cite{Xu23}, where $P_{c\bar{c}}(4312)^+$, $P_{c\bar{c}}(4380)^+$, and $P_{c\bar{c}}(4457)^+$ are identified as molecules but a different nature is assigned to $P_{c\bar{c}}(4440)^+$.

Within the general framework of the B\nobreakdash-O approximation, the molecular interpretation corresponds to near-threshold bound states in potentials that approach a static-hadron-pair threshold.
We refer to the assumption that there are near-threshold bound states in all the potentials that approach that threshold as \emph{molecular dominance}.
The B\nobreakdash-O potentials with flavor $(I,S) = \bigl(\frac{1}{2},0\bigr)$ that approach the $\Sigma_c^{(\ast)} \bar{D}^{(\ast)}$ threshold at large $r$ in Table~\ref{tab:hadronpairpot} are $\bigl(\frac{3}{2}^+\bigr)\frac{3}{2} \big/ \frac{1}{2}$ and $\bigl(\frac{1}{2}^{+\prime}\bigr)\frac{1}{2}$. 
We denote the ground-state energy of a $c \bar c$ pair in the $\bigl(\frac{3}{2}^+\bigr)\frac{3}{2} \big/ \frac{1}{2}$ potential by $\varepsilon_{3/2}$.
The corresponding HQSS multiplet consists of one $J^P=\frac{1}{2}^-$ state, two $\frac{3}{2}^-$ states, and one $\frac{5}{2}^-$ state.
We denote the ground-state energy of a $c \bar c$ pair in the $\bigl(\frac{1}{2}^{+\prime}\bigr)\frac{1}{2}$ potential by $\varepsilon_{1/2}$.
The corresponding HQSS multiplet consists of two $\frac{1}{2}^-$ states and one $\frac{3}{2}^-$ state.
If $\lvert \varepsilon_{1/2}-\varepsilon_{3/2}\rvert$ is small compared to the spin splittings $\Delta_\mathrm{M}$ and $\Delta_\mathrm{B}$, there is mixing between states with the same $J^P$ quantum numbers.
The energies of the seven pentaquarks can be  estimated using first-order perturbation theory in the spin-splitting term $\mathbf{V}_\mathrm{SS}$ in Eq.~\eqref{VSS} with $\Delta_\mathrm{B}=66.5$~MeV and $\Delta_\mathrm{M}=142.1$~MeV. 

\begin{table}
\centering
\begin{tabular}{lccccccccc}
\toprule
$J^P$ & $\Delta E$ [MeV] & $\bigl(\frac{1}{2}^+, 0\bigr)$ & $\bigl(\frac{1}{2}^+, 1\bigr)$ & $\bigl(\frac{3}{2}^+, 0\bigr)$ & $\bigl(\frac{3}{2}^+, 1\bigr)$ & $\Sigma_c \bar{D}$ & $\Sigma_c^\ast \bar{D}$ & $\Sigma_c \bar{D}^\ast$ & $\Sigma_c^\ast \bar{D}^\ast$ \\
\midrule
$\frac{1}{2}^-$                         & $-150.9$  &   $-\frac{1}{4}$ &   $+\frac{1}{12}$ & \hphantom{+}0                      & $-\frac{2}{3}$   & $+1$ & \hphantom{+}0 & \hphantom{+}0 & \hphantom{+}0 \\
$\frac{1}{2}^{- \prime}$            &    \hphantom{00}$-8.8$  &  $+\frac{1}{12}$ & $-\frac{25}{36}$ & \hphantom{+}0                      & $-\frac{2}{9}$    & \hphantom{+}0 & \hphantom{+}0 & $+1$ & \hphantom{+}0 \\
$\frac{1}{2}^{- \prime \prime}$ & \hphantom{0}$+57.7$  & $-\frac{2}{3}$ &         $-\frac{2}{9}$ & \hphantom{+}0                      & $+\frac{1}{9}$    & \hphantom{+}0 & \hphantom{+}0 & \hphantom{+}0 & $+1$ \\
$\frac{3}{2}^-$                          & \hphantom{0}$-84.4$  & \hphantom{+}0                        &      $+\frac{1}{3}$ & $-\frac{1}{4}$   & $-\frac{5}{12}$ & \hphantom{+}0 & $+1$ & \hphantom{+}0 & \hphantom{+}0 \\
$\frac{3}{2}^{- \prime}$            &    \hphantom{00}$-8.8$  & \hphantom{+}0                        &      $-\frac{1}{9}$  & $+\frac{1}{3}$  & $-\frac{5}{9}$   & \hphantom{+}0 & \hphantom{+}0 & $+1$ & \hphantom{+}0 \\
$\frac{3}{2}^{- \prime \prime}$ & \hphantom{0}$+57.7$  & \hphantom{+}0                        &      $-\frac{5}{9}$ & $-\frac{5}{12}$ & $-\frac{1}{36}$ & \hphantom{+}0 & \hphantom{+}0 & \hphantom{+}0 & $+1$ \\
$\frac{5}{2}^-$                         & \hphantom{0}$+57.7$  & \hphantom{+}0                        & \hphantom{+}0                          & \hphantom{+}0                        & $-1$                & \hphantom{+}0 & \hphantom{+}0 & \hphantom{+}0 & $+1$ \\ 
\bottomrule
\end{tabular}
\caption{Lowest HQSS multiplet of $c \bar{c}$ pentaquarks with $(I,S)= \bigl( \frac{1}{2}, 0 \bigr)$ in the case of molecular dominance with ideal mixing between the $j^\pi=\frac{1}{2}^+$ and $j^\pi=\frac{3}{2}^+$ potentials.
The energy $\Delta E$ for the pentaquark with quantum numbers $J^P$ is the spin splitting relative to the spin-weighted average of the $\Sigma_c \bar{D}$,  $\Sigma_c^\ast \bar{D}$,  $\Sigma_c \bar{D}^\ast$, and  $\Sigma_c^\ast \bar{D}^\ast$ thresholds.
The columns labeled $(j^\pi, S_{Q \bar Q})$ give the amplitude and probability for the light-QCD quantum numbers $j^\pi$ and heavy-quark-pair spin $S_{Q \bar Q}$.
The columns labeled by a charm-hadron pair give the amplitude and probability for that charm-hadron pair.
The probability is the absolute value of the entry and the amplitude is the square root of the probability multiplied by the sign of the entry.
For each pentaquark, the sum of the probabilities for the quantum numbers $(j^\pi, S_{Q \bar Q})$ is $1$.
 }
\label{tab:molecule}
\end{table}

We will consider only the case of \emph{ideal mixing} in which the ground-state energies in the $\bigl(\frac{3}{2}^+\bigr)\frac{3}{2} \big/ \frac{1}{2}$ and $\bigl(\frac{1}{2}^{+\prime}\bigr)\frac{1}{2}$ potentials are tuned to have the same values $\varepsilon_{3/2} = \varepsilon_{1/2}$.
The energies of the seven pentaquarks are given in Appendix~\ref{app:PentaquarkSplittingsAmplitudes} as functions of $\varepsilon_{3/2}$, $\Delta_\mathrm{B}$, and $\Delta_\mathrm{M}$.
The spectrum can be determined most easily by realizing that the baryon-meson pairs $\Sigma_c \bar{D}$,  $\Sigma_c^\ast \bar{D}$,  $\Sigma_c \bar{D}^\ast$, and  $\Sigma_c^\ast \bar{D}^\ast$ are all eigenstates of the spin-splitting term $\mathbf{V}_\mathrm{SS}$ in Eq.~\eqref{VSS}.
The bound states are therefore baryon-meson molecules with these constituents.
The $\Sigma_c \bar{D}$ molecule has quantum numbers $\frac{1}{2}^-$ and is identified with $P_{c\bar{c}}(4312)^+$.
The $\Sigma_c^\ast \bar{D}$ molecule has quantum numbers $\frac{3}{2}^-$ and we identify it with $P_{c\bar{c}}(4337)^+$.
The two $\Sigma_c \bar{D}^\ast$ molecules have quantum numbers $\frac{1}{2}^{-\prime}$ and $\frac{3}{2}^{-\prime}$ and  the two states are identified with either $P_{c\bar{c}}(4440)^+$ and $P_{c\bar{c}}(4457)^+$ (Scenario 1) or $P_{c\bar{c}}(4457)^+$ and $P_{c\bar{c}}(4440)^+$ (Scenario 2).
The three $\Sigma_c^\ast \bar{D}^\ast$ molecules have quantum numbers $\frac{1}{2}^{-\prime\prime}$, $\frac{3}{2}^{-\prime\prime}$, and $\frac{5}{2}^-$.
They are identified as pentaquarks with energies near the $\Sigma_c^\ast \bar{D}^\ast$ threshold at 4526.7~MeV  that have not yet been observed.
The spin splittings of the seven pentaquark states are given in Table~\ref{tab:molecule}. 
Their spin-weighted average is zero.
Note that the four distinct spin splittings coincide with those of the four $\Sigma_c^{(\ast)} \bar{D}^{(\ast)}$ thresholds.

Since we have only four candidates for the seven pentaquark states and we have two options for the quantum numbers of $P_{c\bar{c}}(4337)^+$ and $P_{c\bar{c}}(4440)^+$, we cannot calculate the spin-weighted average of the pentaquark masses.
We therefore assume instead that the spin-weighted mass of the seven pentaquarks is equal to the spin-weighted $\Sigma_c^{(\ast)} \bar{D}^{(\ast)}$ threshold, which is $M_0 = 4469.8$~MeV. 
We can predict the masses of the pentaquarks by adding $M_0$ to the spin splittings in Table~\ref{tab:nonstrange}. 
The predicted masses are compared to the measured masses in the energy-level diagram in Fig.~\ref{fig:NonstrangeEnergyLevels}.

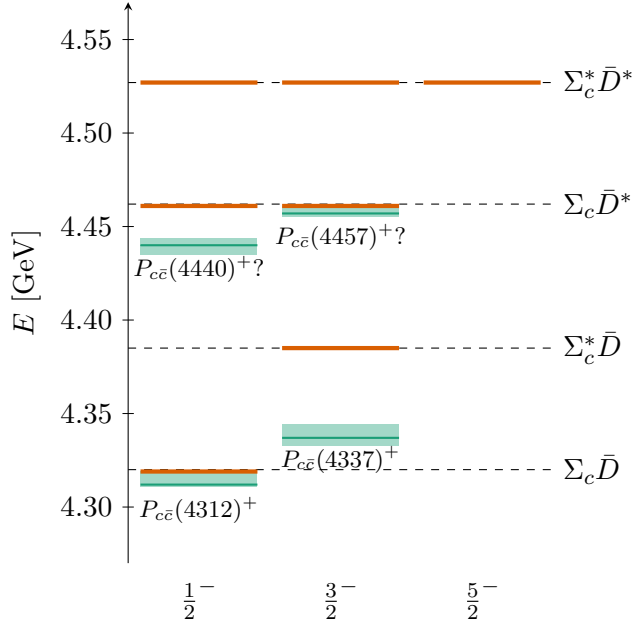
\begin{figure}
\tikzsetnextfilename{pentaquark-energy-levels-nonstrange-molecular-dominance}
\centering
\begin{tikzpicture}
\pgfmathsetmacro{\MSigma}{2.453}
\pgfmathsetmacro{\MSigmastar}{2.518}
\pgfmathsetmacro{\MD}{1.867}
\pgfmathsetmacro{\MDstar}{2.009}
\pgfplotsset{
	/pgfplots/error bars/draw error bar/.code 2 args={%
		\node[inner sep=0pt] (a) at #1 {};
		\node[inner sep=0pt] (b) at #2 {};
		\path[draw=none, fill, opacity=.4] ($(a) - (22pt,0)$) rectangle ($(b) + (22pt,0)$);
	}
}
\pgfplotstableread{
x	y
0	4.319
0	4.461
0	4.527
1	4.385
1	4.461
1	4.527
2	4.527
}{\moleculartheor}
\pgfplotstableread{
x 	y 	ey+ 	ey-
0	4.312	0.007	0.001
1	4.337	0.007	0.004
0	4.440	0.004	0.005
1	4.457	0.004	0.002
}{\molecularexp}
\begin{axis}[
	width=7.2cm,
	height=9cm,
	only marks,
	black,
	clip mode=individual,
	ylabel={$E$},
	y unit=eV,
	y SI prefix=giga,
	axis y line=left,
	axis x line=bottom,
	ymin=4.27,
	ymax=4.57,
	xmin=-.5,
	xmax=2.5,
	x axis line style={draw=none},
	x tick style={draw=none},
	y tick style={black},
	xtick={0, 1, 2},
	xticklabels={$\frac{1}{2}^{-}$, $\frac{3}{2}^{-}$, $\frac{5}{2}^{-}$},
	ytick distance=0.05,
	yticklabel style={/pgf/number format/.cd,fixed,zerofill,precision=2},
	cycle list/Dark2
	]
	\begin{scope}[dashed]
		\draw (-.5,{\MSigma+\MD}) -- (2.5,{\MSigma+\MD})
		node[anchor=west] {$\Sigma_c\bar{D}$};
		\draw (-.5,{\MSigmastar+\MD}) -- (2.5,{\MSigmastar+\MD})
		node[anchor=west] {$\Sigma_c^\ast\bar{D}$};
		\draw (-.5,{\MSigma+\MDstar}) -- (2.5,{\MSigma+\MDstar})						node[anchor=west] {$\Sigma_c\bar{D}^\ast$};
            		\draw (-.5,{\MSigmastar+\MDstar}) -- (2.5,{\MSigmastar+\MDstar})					node[anchor=west] {$\Sigma_c^\ast\bar{D}^\ast$};
	\end{scope}
	\addplot+[mark=-,thick,mark size=22pt,
		error bars/.cd, y dir=both, y explicit]
		table[y error plus=ey+, y error minus=ey-] {\molecularexp};
	\addplot+[mark=-,ultra thick,mark size=22pt] 
		table {\moleculartheor};
	\node[anchor=north] at (0, 4.311) {\footnotesize $P_{c\bar{c}}(4312)^+$};
	\node[anchor=north] at (1, 4.338) {\footnotesize $P_{c\bar{c}}(4337)^+$};
	\node[anchor=north,inner sep=.3pt] at (0, 4.435) {\footnotesize $P_{c\bar{c}}(4440)^+?$};
	\node[anchor=north] at (1, 4.457) {\footnotesize $P_{c\bar{c}}(4457)^+?$};
\end{axis}
\end{tikzpicture}
\caption{\label{fig:MoleculeNonstrangeEnergyLevels}
Energy-level diagram for the isospin-$\frac{1}{2}$ $c \bar{c}$ pentaquarks associated with the $\frac{3}{2}^+$ and $\frac{1}{2}^{+\prime}$ $\bm{8}$-baryons under the assumption of molecular dominance with ideal mixing.
The dashed horizontal lines are the thresholds for a charm baryon and a charm meson. 
The shaded boxes are the error bars for the measured masses, with the central value highlighted by a thin line.
The quantum numbers of $P_{c\bar{c}}(4440)^+$ and $P_{c\bar{c}}(4457)^+$ were chosen to be $\frac{1}{2}^-$ and $\frac{3}{2}^-$ (Scenario 1), but the opposite assignment $\frac{3}{2}^-$ and $\frac{1}{2}^-$ (Scenario 2) is also possible.
The blue bars are the predictions of molecular dominance with ideal mixing, assuming the ground-state energy is equal to the spin-weighted average of the thresholds for $\Sigma_c \bar{D}$, $\Sigma_c^\ast \bar{D}$,  $\Sigma_c \bar{D}^\ast$, and $\Sigma_c^\ast \bar{D}^\ast$.
}
\end{figure}

We now compare the energy levels of the $c \bar{c}$ pentaquarks from molecular dominance with ideal mixing in Fig.~\ref{fig:MoleculeNonstrangeEnergyLevels} with those in Fig.~\ref{fig:NonstrangeEnergyLevels} from the B\nobreakdash-O approximation with the $\bigl(\frac{3}{2}^+\bigr)\frac{3}{2} \big/ \frac{1}{2}$ potentials only. 
The quality of the fits are comparable.
The fit for molecular dominance could be improved by removing the assumption of ideal mixing.
One qualitative difference between the B\nobreakdash-O approximation and molecular dominance is in the assignment of quantum numbers.
Both approaches predict the quantum numbers of $P_{c\bar{c}}(4312)^+$ and $P_{c\bar{c}}(4337)^+$ to be $\frac{1}{2}^-$ and $\frac{3}{2}^-$.
The B\nobreakdash-O prediction for the quantum numbers of $P_{c\bar{c}}(4440)^+$ and $P_{c\bar{c}}(4457)^+$ is $\frac{3}{2}^-$ and $\frac{5}{2}^-$.
The molecular dominance predictions are either $\frac{1}{2}^-$ and $\frac{3}{2}^-$ in Scenario 1 or $\frac{3}{2}^-$ and $\frac{1}{2}^-$ in Scenario 2.
If the quantum numbers of $P_{c\bar{c}}(4457)^+$ were determined to be $\frac{5}{2}^-$, molecular dominance would be excluded.
The other qualitative difference between the two approaches is that molecular dominance predicts the existence of three additional pentaquarks with energies near the $\Sigma_c^\ast \bar{D}^\ast$ threshold that have not been observed.

Table~\ref{tab:molecule} also provides information about the nature of the seven pentaquarks by giving amplitudes and probabilities for their components.
The calculation of the amplitudes is described in Appendix~\ref{app:PentaquarkSplittingsAmplitudes}.
Six of the seven pentaquarks have three nonzero components in the $\bm{8}$-hadron basis $(j^\pi, S_{Q \bar Q})$. 
The exception is the $\frac{5}{2}^-$ pentaquark, which is pure $\bigl(\frac{3}{2}^+, 1 \bigr)$.
In the dihadron basis  $(J_\mathrm{B},J_\mathrm{M})$, the quantum numbers $\bigl(\frac{1}{2},0 \bigr)$, $\bigl(\frac{3}{2},0 \bigr)$, $\bigl(\frac{1}{2}, 1 \bigr)$ and $\bigl(\frac{3}{2}, 1 \bigr)$ correspond to the charm-hadron pairs $\Sigma_c \bar{D}$,  $\Sigma_c^\ast \bar{D}$, $\Sigma_c \bar{D}^\ast$, and $\Sigma_c^\ast \bar{D}^\ast$, respectively.
For each pentaquark, there is only one nonzero amplitude and its value has been chosen to be $+1$.
The pentaquark with amplitude $+1$ for $\mathrm{B\,M}$ can be interpreted as a molecule composed of that baryon-meson pair. 
The spin splittings of each of the seven pentaquarks is equal to the spin splitting of that baryon meson pair, as illustrated in Fig.~\ref{fig:MoleculeNonstrangeEnergyLevels}.

In Ref.~\cite{Bram25}, the B\nobreakdash-O approximation was applied to the $c \bar c$ pentaquarks under the assumption of molecular dominance. 
That study disregards the constraint that the $\bigl(\frac{3}{2}^+\bigr)\frac{3}{2} \big/ \frac{1}{2}$ and $\bigl(\frac{1}{2}^{+\prime}\bigr)\frac{1}{2}$ potentials at short distance must approach the repulsive color-octet potential offset by the energy of a ground-state $\bm{8}$-baryon and a same-parity orbitally excited $\bm{8}$-baryon, respectively.
The energy of the same-parity orbitally excited $\bm{8}$-baryon is expected to be larger by about 600~MeV.
The ground states in the $\bigl(\frac{3}{2}^+\bigr)\frac{3}{2} \big/ \frac{1}{2}$ and $\bigl(\frac{1}{2}^{+\prime}\bigr)\frac{1}{2}$ potentials correspond to a total of $4+3=7$ pentaquarks.
If the spin-splitting potential is treated as a first-order perturbation, the energies of the seven pentaquarks are determined by the ground-state energies $\varepsilon_{3/2}$ and $\varepsilon_{1/2}$ and the known energy splittings $\Delta_\mathrm{B}$ and $\Delta_\mathrm{M}$. 
In Ref.~\cite{Bram25}, the four lowest-energy pentaquarks were identified with $P_{c\bar{c}}(4312)^+$, $P_{c\bar{c}}(4380)^+$, $P_{c\bar{c}}(4440)^+$, and $P_{c\bar{c}}(4457)^+$.
Note that the second pentaquark was identified with $P_{c\bar{c}}(4380)^+$ rather than $P_{c\bar{c}}(4337)^+$.
The parameters $\varepsilon_{1/2}$ and $\varepsilon_{3/2}$  were determined by minimizing the $\chi^2$ for the four observed pentaquark energies with error bars and the predicted energies of the four lowest-energy pentaquarks.
Results were given for two scenarios in which both $\varepsilon_{1/2}$ and $\varepsilon_{3/2}$ are negative and the lower of the two energies is either $\varepsilon_{1/2}$ or $\varepsilon_{3/2}$.
If $\varepsilon_{1/2} < \varepsilon_{3/2}$, the energies relative to the spin-weighted $\Sigma_c^{(\ast)}\bar{D}^{(\ast)}$ threshold are $\varepsilon_{1/2} = -23$~MeV and $\varepsilon_{3/2} = -1$~MeV.
If $\varepsilon_{1/2} > \varepsilon_{3/2}$, the energies are $\varepsilon_{1/2} = -0.5$~MeV and $\varepsilon_{3/2} = -14$~MeV.
In both scenarios, the difference $\varepsilon_{1/2} - \varepsilon_{3/2}$ is small compared to the spin splittings $\Delta_\mathrm{B}$ and $\Delta_\mathrm{M}$, so the results are qualitatively similar to those for molecular dominance with ideal mixing in Fig.~\ref{fig:MoleculeNonstrangeEnergyLevels}.
The quantum numbers for $P_{c\bar{c}}(4312)^+$ and $P_{c\bar{c}}(4380)^+$ were predicted to be $\frac{1}{2}^-$ and $\frac{3}{2}^-$. 
The quantum numbers for $P_{c\bar{c}}(4440)^+$ and $P_{c\bar{c}}(4457)^+$ were predicted to be either $\frac{1}{2}^-$ and $\frac{3}{2}^-$ (Scenario 1) if $\varepsilon_{1/2} < \varepsilon_{3/2}$ or $\frac{3}{2}^-$ and $\frac{1}{2}^-$ (Scenario 2) if $\varepsilon_{1/2} > \varepsilon_{3/2}$.
Three additional pentaquarks with $J^P = \frac{1}{2}^-$, $\frac{3}{2}^-$, and $\frac{5}{2}^-$ were predicted near the $\Sigma_c^\ast \bar{D}^\ast$ threshold. The energies of the three pentaquarks near each threshold increase with increasing $J$ in Scenario 1 and they decrease with increasing $J$ in Scenario 2.

\subsection{\texorpdfstring{$c\bar{c}s$}{Strange Hidden-charm} Pentaquarks}
\label{sec:strangeEnergies}

We have identified the $c \bar{c}s$ pentaquarks as bound states or resonances in the $\bigl(\frac{1}{2}^{+\prime}\bigr)\frac{1}{2}$ potential that approaches the $\Xi_c \bar{D}^{(\ast)}$ threshold at large $r$.
The HQSS multiplet associated with the ground state for $c \bar{c}$ in this potential consists of one spin-singlet  state with $J^P= \frac{1}{2}^-$ and two spin-triplet  states with $J^P= \frac{1}{2}^{-\prime}$ and $\frac{3}{2}^-$.
We approximate the spin splittings of the pentaquarks by first order perturbation theory in the spin-splitting term $\mathbf{V}_\mathrm{SS}$ in Eq.~\eqref{VSS} with $\Delta_\mathrm{B}=0$ and $\Delta_\mathrm{M}=142.1$~MeV.  
The energies of the three pentaquarks are given in Appendix~\ref{app:PentaquarkSplittingsAmplitudes} as functions of the ground-state energy $\varepsilon_{1/2}$ and $\Delta_\mathrm{M}$.

\begin{table}
\centering
\begin{tabular}{lccccc}
\toprule
$J^P$ & $\Delta E$ [MeV] & $\bigl( \frac{1}{2}^+, 0 \bigr)$ & $\bigl( \frac{1}{2}^+, 1 \bigr)$ & $\Xi_c \bar{D}$ & $\Xi_c \bar{D}^\ast$ \\
\midrule
$\frac{1}{2}^-$              & $-106.6$& $-\frac{1}{4}$        & $-\frac{3}{4}$ & $+1$        & $\hphantom{+}0$ \\
$\frac{1}{2}^{- \prime}$ & \hphantom{0}$+35.5$ & $-\frac{3}{4}$ & $+\frac{1}{4}$  & $\hphantom{+}0$  &  $+1$   \\
$\frac{3}{2}^-$              & \hphantom{0}$+35.5$  &  $\hphantom{+}0$                 & $-1$  & $\hphantom{+}0$  & $+1$                       \\
\bottomrule
\end{tabular}
\caption{Lowest HQSS multiplet of $c \bar{c}s$ pentaquarks with $j^\pi=\frac{1}{2}^+$ and $(I,S) =(0,-1)$.
The energy $\Delta E$ for the pentaquark with quantum numbers $J^P$ is the spin splitting relative to the spin-weighted average of the three pentaquarks.
The columns labeled $\bigl( \frac{1}{2}^+, S_{Q \bar Q} \bigr)$ give the amplitude and probability for the heavy-quark-pair spin $S_{Q \bar Q}$.
The column labeled by a charm-hadron pair gives the amplitude and probability for that charm-hadron pair. 
The probability is the absolute value of the entry and the amplitude is the square root of the probability multiplied by the sign of the entry.
For each pentaquark, the sum of the probabilities for the quantum numbers $(j^\pi, S_{Q \bar Q})$ is $1$.
 }
\label{tab:strange}
\end{table}

\begin{figure}
\tikzsetnextfilename{pentaquark-energy-levels-strange}
\centering
\begin{tikzpicture}
\pgfmathsetmacro{\MXi}{2.469}
\pgfmathsetmacro{\MD}{1.867}
\pgfmathsetmacro{\MDstar}{2.009}
\pgfplotsset{
	/pgfplots/error bars/draw error bar/.code 2 args={%
		\node[inner sep=0pt] (a) at #1 {};
		\node[inner sep=0pt] (b) at #2 {};
		\path[draw=none, fill, opacity=.4] ($(a) - (22pt,0)$) rectangle ($(b) + (22pt,0)$);
	}
}
\pgfplotstableread{
x	y
0	4.322
0	4.464
1	4.464
}{\strangetheor}
\pgfplotstableread{
x 	y 	ey+ 	ey-
0	4.338	0.001	0.001
1	4.459	0.006	0.003
}{\strangeexp}
\begin{axis}[
	width=5.7cm,
	height=6cm,
	only marks,
	black,
	clip mode=individual,
	ylabel={$E$},
	y unit=eV,
	y SI prefix=giga,
	axis y line=left,
	axis x line=bottom,
	ymin=4.29,
	ymax=4.49,
	xmin=-.5,
	xmax=1.5,
	x axis line style={draw=none},
	x tick style={draw=none},
	y tick style={black},
	xtick={0, 1, 2},
	xticklabels={$\frac{1}{2}^{-}$, $\frac{3}{2}^{-}$},
	ytick distance=0.05,
	yticklabel style={/pgf/number format/.cd,fixed,zerofill,precision=2},
	cycle list/Dark2
	]
	\begin{scope}[dashed]
		\draw (-.5,{\MXi+\MD}) -- (1.5,{\MXi+\MD})
		node[anchor=west] {$\Xi_c\bar{D}$};
		\draw (-.5,{\MXi+\MDstar}) -- (1.5,{\MXi+\MDstar})
		node[anchor=west] {$\Xi_c\bar{D}^\ast$};
	\end{scope}
	\addplot+[mark=-,thick,mark size=22pt,
		error bars/.cd, y dir=both, y explicit]
		table[y error plus=ey+, y error minus=ey-] {\strangeexp};
	\addplot+[mark=-,ultra thick,mark size=22pt] 
		table {\strangetheor};
	\node[anchor=south] at (0, 4.339) {\footnotesize $P_{c\bar{c}s}(4338)^0$};
	\node[anchor=north] at (1, 4.456) {\footnotesize $P_{c\bar{c}s}(4459)^0?$};
\end{axis}
\end{tikzpicture}
\caption{\label{fig:StrangeEnergyLevels}
Energy-level diagram for the ground-state triplet of isospin-0 $c \bar{c}s$ pentaquarks associated with the $\frac{1}{2}^{+\prime}$ $\bm{8}$-baryon. 
The dashed horizontal lines are the thresholds for a charm baryon and a charm meson.
The shaded boxes are the error bands for the measured masses, with the central value highlighted by a thin line.
For illustration purposes, we assign the quantum numbers $\frac{3}{2}^-$ to $P_{c\bar{c}s}(4459)^0$, but the assignment $\frac{1}{2}^-$ is also possible.
The thick bars are predictions given by the sum of a ground-state energy and spin splittings.
The ground-state energy is equal to the spin-weighted average of the central values of the masses of the two observed pentaquarks and the third pentaquark assuming it has the same mass as $P_{c\bar{c}s}(4459)^0$.
The spin splittings are from first order perturbation theory in the charm-hadron spin-splitting term.
}
\end{figure}
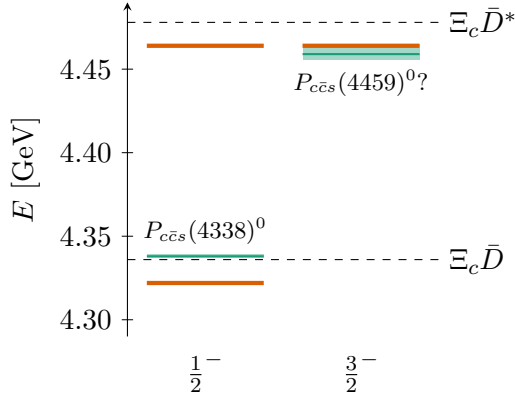

The spin splittings of the three pentaquarks are given in Table~\ref{tab:strange}. 
Their spin-weighted average is zero.
The predicted spin splitting between the $\frac{1}{2}^{-\prime}$ or $\frac{3}{2}^-$ pentaquark and the $\frac{1}{2}^-$ pentaquark is 142.1~MeV. 
We consider this close enough to the difference $120.6^{+5.6}_{-3.2}$~MeV  between the masses of the two observed $c \bar{c}s$ pentaquarks in Table~\ref{tab:strangepentaquarks} to be consistent with our perturbative approximation.  
We therefore identify $P_{c\bar{c}s}(4338)^0$ with the $\frac{1}{2}^-$ pentaquark and we identify $P_{c\bar{c}s}(4459)^0$ with either the $\frac{1}{2}^{-\prime}$ pentaquark or the $\frac{3}{2}^-$ pentaquark.
The quantum numbers $\frac{1}{2}^-$ of $P_{c\bar{c}s}(4338)^0$ agree with its  experimentally determined quantum numbers.
The quantum numbers of $P_{c\bar{c}s}(4459)^0$ are predicted to be either $\frac{1}{2}^-$ or $\frac{3}{2}^-$.
We predict that there  is another $c\bar{c}s$ pentaquark nearby in energy with the other quantum numbers.
Given the observed quantum numbers of $P_{c\bar{c}s}(4338)^0$, our prediction $J^P=\frac{1}{2}^-$ or $\frac{3}{2}^-$ for the quantum numbers of $P_{c\bar{c}s}(4459)^0$, and the assumption that the mass of the third pentaquark is similar to that of $P_{c\bar{c}s}(4459)^0$, the spin-weighted average of the central values of the masses of the the three pentaquarks is $M_0 = 4428.7$~MeV. 
We can predict the masses of the pentaquarks by adding $M_0$  to the spin splittings in Table~\ref{tab:strange}. 
The predicted masses are compared to the measured masses in the energy-level diagram in Fig.~\ref{fig:StrangeEnergyLevels}.

Table~\ref{tab:strange} also provides information about the nature of the three pentaquarks by giving amplitudes and probabilities for their components.
The calculation of the amplitudes is described in Appendix~\ref{app:PentaquarkSplittingsAmplitudes}.
The amplitudes and probabilities for the spin-singlet and spin-triplet components are given in the columns labeled $\bigl( \frac{1}{2}^+, 0 \bigr)$ and $\bigl( \frac{1}{2}^+, 1 \bigr)$.
The $\frac{3}{2}^-$ pentaquark is pure spin triplet while the $\frac{1}{2}^-$ and $\frac{1}{2}^{-\prime}$ pentaquarks are linear combinations of spin singlet and spin triplet.
The amplitudes and probabilities of the constituent charm-hadron pairs are given in the columns labeled $\Xi_c \bar{D}$ and $\Xi_c \bar{D}^\ast$. 
The $\frac{1}{2}^-$ pentaquark is pure $\Xi_c \bar{D}$ and the $\frac{1}{2}^{-\prime}$ and $\frac{3}{2}^-$ pentaquarks are pure $\Xi_c \bar{D}^\ast$.

\section{Hidden-charm Pentaquark Decays}
\label{sec:Decays}

In this section, we present model-independent predictions for ratios of decay rates of hidden-charm pentaquarks.

\subsection{Pentaquark Decays}
\label{sec:PentaquarkDecays}

There are decays of pentaquarks whose rates can be calculated analytically up to a multiplicative factor.
The cancellation of that factor in ratios of decay rates gives model-independent predictions for the ratios.

\subsubsection{Model-Independent Ratios}
\label{sec:ModelIndependentRatios}

The possible decay modes of a hidden-charm pentaquark include light hadrons, charmonium plus light hadrons, and charm-hadron pairs. 
The decay into light hadrons requires annihilation of the $c \bar{c}$ pair. 
The inclusive decay rate into light hadrons has a perturbative QCD suppression by at least two powers of the running coupling constant $\alpha_s(m_c)$.  
The decay into charmonium plus light hadrons proceeds through the transition of the $c \bar c$ pair into charmonium and the transition of the light-QCD state into the light hadron.
The semi-inclusive decay rate into charmonium plus light hadrons can be calculated using the methods of BOEFT developed for quarkonium hybrids in Ref.~\cite{Bra22}.
Decays into charm-hadron pairs can proceed through transitions between B\nobreakdash-O potentials.
These decays could produce large partial widths unless they are suppressed.
Most of the observed pentaquarks have surprisingly narrow decay widths.
It is therefore important to identify a suppression mechanism for each of the kinematically allowed decays into charm-hadron pairs for every pentaquark with a narrow width.

The spins of the heavy quark and the heavy antiquark are conserved  in the heavy-quark limit.
Heavy-quark spin symmetry (HQSS) can be used to simplify the $T$-matrix elements for two-body decays of a pentaquark. 
We focus on pentaquarks belonging to an HQSS multiplet for a ground state whose wavefunction is dominated by $S$-wave components with $L_{Q \bar Q} = 0$.
The light-QCD state in the pentaquark at short distances is an adjoint baryon with quantum numbers $j^\pi$.
The pentaquark can be decomposed into states $\vert (j^\pi,S_{Q \bar Q}) J \rangle$ labeled by the $\bm{8}$-hadron basis. $\mathrm{A}$ 2-body final state that consists of a heavy baryon and a heavy meson
can be decomposed into states  $\vert ( (j^\prime{}^{\pi^\prime}, L_\mathrm{BM} ) L,S_{Q \bar Q}^\prime ) J^\prime \rangle$ labeled by their total light-QCD quantum numbers $j^\prime{}^{\pi^\prime}$, their orbital angular momentum  $L_\mathrm{BM}$, the B\nobreakdash-O angular momentum $L$, the spin $S_{Q \bar Q}^\prime$ of the heavy-quark pair, and the total angular momentum $J^\prime$.
Conservation of total angular momentum implies $J^\prime = J$.
Heavy-quark spin symmetry implies that $S_{Q \bar Q}^\prime = S_{Q \bar Q}$ and that  the $T$-matrix element does not depend on $S_{Q \bar Q}$.
The $T$-matrix element therefore  can be expressed in terms of a reduced $T$-matrix element between light-QCD states:
\begin{equation}
\bigl\langle  \bigl( ( j^\prime{}^{\pi^\prime} , L_\mathrm{BM} ) L,S_{Q \bar Q}^\prime \bigr)J^\prime \big\vert T \big\vert (j^\pi,S_{Q \bar Q})J \bigr\rangle  = \delta_{J^\prime,J}\, \delta_{S_{Q \bar Q}^\prime, S_{Q \bar Q}}\,  \bigl\langle  ( j^\prime{}^{\pi^\prime} , L_\mathrm{BM} ) L \big\Vert T \big\Vert j^\pi \bigr\rangle.
\label{reducedTmatrixelement}
\end{equation}
The reduced $T$-matrix element depends on four angular momenta.
Conservation of parity requires $\pi^\prime (-1)^{L_\mathrm{BM}} = \pi$.

The exact conservation of total angular momentum $J$ and the conservation of the heavy-quark-pair spin $S_{Q \bar Q}$ in the heavy-quark limit implies that the B\nobreakdash-O angular momentum $L$ is also conserved  in the heavy-quark limit.
Conservation of the B\nobreakdash-O angular momentum implies that the reduced $T$-matrix element in Eq.~\eqref{reducedTmatrixelement} is nonzero only if $L=j$: 
\begin{equation}
\bigl\langle  ( j^\prime{}^{\pi^\prime} , L_\mathrm{BM} ) L \big\Vert T \big\Vert j^\pi \bigr\rangle
=  \delta_{L,j}\,  \bigl\langle  ( j^\prime{}^{\pi^\prime} , L_\mathrm{BM} ) j \big\Vert T \big\Vert j^\pi \bigr\rangle.
\label{reducedT-Lconservation}
\end{equation}
Thus the reduced $T$-matrix element depends on only three angular momenta and they satisfy the triangle relations for $( j^\prime , L_\mathrm{BM}) j $.

Model-independent ratios of decay rates of hidden-heavy hadrons into pairs of heavy hadrons were derived in Ref.~\cite{Braa24a}.
There are model-independent ratios if the $T$-matrix elements can be expressed in terms of a single reduced $T$-matrix element.
The square of  the reduced $T$-matrix element then cancels out of a ratio of decay rates.
There are model-independent ratios if the triangle relations and the parity constraint imply that the reduced $T$-matrix element is nonzero for only one pair of $j^\prime$ and $L_\mathrm{BM}$.

\subsubsection{Relative Velocities}
\label{sec:RelativeVelocities}

The decay rate of a pentaquark can be expressed as the product of the square of the $T$-matrix element and a phase-space factor.
The two-body phase-space factor depends on the relative velocity of the decay products $\mathrm{A}$ and $\mathrm{B}$, which can be defined as the velocity of one of the products in the rest frame of the other.
For the decay of a pentaquark with mass $M_\mathrm{P}$ into  two hadrons with masses $M_\mathrm{A}$ and $M_\mathrm{B}$, the relativistic expression for the relative velocity is
\begin{equation}
v_\mathrm{P,AB}= \sqrt{\lambda(M_\mathrm{P}^2,M_\mathrm{A}^2,M_\mathrm{B}^2)} \Big/ \bigl(M_\mathrm{P}^2 - M_\mathrm{A}^2 - M_\mathrm{B}^2 \bigr) ,
\label{vPBMrel}
\end{equation}
where $\lambda(x,y,z) = x^2+y^2+z^2-2(xy+yz+zx)$.
If the energy $M_\mathrm{P}-M_\mathrm{A}-M_\mathrm{B}$ of the pentaquark relative to the threshold is much smaller than both $M_\mathrm{A}$ and $M_\mathrm{B}$, then $v_\mathrm{P,AB}  \ll 1$ and it can be approximated by the nonrelativistic expression
\begin{equation}
v_\mathrm{P,AB} \approx \sqrt{2 (M_\mathrm{P} - M_\mathrm{A} - M_\mathrm{B}) / \mu_\mathrm{AB}}\, ,
\label{vPBM}
\end{equation}
where $\mu_\mathrm{AB} = M_\mathrm{A} M_\mathrm{B} / (M_\mathrm{A} + M_\mathrm{B})$ is the reduced mass of the decay products.

For the decay of a pentaquark $\mathrm{P}$ into a baryon $\mathrm{B}$ and a meson $\mathrm{M}$ with orbital angular momentum $L_\mathrm{BM}$, the reduced $T$-matrix element has a factor of $v^{L_\mathrm{BM}}$, where $v$ is the relative velocity of the baryon and meson.
We will assume the most important kinematic effects associated with the spin splittings of $\mathrm{P}$, $\mathrm{B}$, and $\mathrm{M}$ can be taken into account by replacing $v$ by the velocity $v_\mathrm{P,\mathrm{BM}}$ in Eq.~\eqref{vPBMrel}.
The phase-space factor in the decay rate provides an additional factor of $v$.
Thus the decay rate has a factor $v_\mathrm{P,BM}^{2 L_\mathrm{BM} + 1}$.

In the case of decays of the pentaquark into a heavy baryon and a heavy meson, there may also be approximate model-independent ratios because the heavy baryon and the heavy meson are always at least somewhat nonrelativistic.
The factor $v^{L_\mathrm{BM}}$ in the reduced $T$-matrix element favors decay channels with the smallest value of $L_\mathrm{BM}$.
There are approximate model-independent ratios if the triangle relations and the parity constraint for the smallest $L_\mathrm{BM}$ allow only a single reduced $T$-matrix element.

\subsubsection{Charmonium Plus a Light Baryon}
\label{sec:Quarkoniumlighthadron}

A hidden-charm pentaquark can decay into charmonium plus a single light baryon. 
We consider a pentaquark in the HQSS multiplet for an adjoint baryon with light-QCD quantum numbers $j^\pi$.
The decay requires the transition of the $c \bar c$ pair in the pentaquark into charmonium and  the transition of the light-QCD state into the light baryon.
For simplicity, we consider only the $S$-wave charmonium states $J/\psi$ and $\eta_c$.
They are the spin-triplet ($S_{Q \bar Q}=1$) and spin-singlet ($S_{Q \bar Q}=0$) members of an HQSS multiplet with quantum numbers $1^{--}$ and $0^{-+}$.

The pentaquark can be decomposed into $\bm{8}$-hadron basis states $\vert (j^\pi,S_{Q \bar Q}) J \rangle$.
We take the light baryon $h$ to have light-QCD quantum numbers $j_h^{\pi_h}$ and orbital angular momentum $L_h$.
The 2-body final state can be decomposed into states  $\vert ( (j_h^{\pi_h}, L_h) L,S_{Q \bar Q}^\prime ) J^\prime \rangle$ labeled also by the B\nobreakdash-O angular momentum $L$, the spin $S_{Q \bar Q}^\prime$ of the heavy-quark pair, and the total angular momentum $J^\prime$.
Conservation of total angular momentum implies $J^\prime = J$.
Heavy-quark spin symmetry implies that $S_{Q \bar Q}^\prime = S_{Q \bar Q}$ and that  the $T$-matrix element does not depend on $S_{Q \bar Q}$.
Conservation of B\nobreakdash-O angular momentum implies that $L=j$.
The $T$-matrix element therefore  can be expressed in terms of a reduced $T$-matrix element between light-QCD states:
\begin{equation}
\bigl\langle  \bigl( ( j_h^{\pi_h}, L_h ) L, S_{Q \bar Q}^\prime)J^\prime \big\vert T \big\vert (j^\pi,S_{Q \bar Q})J \bigr\rangle  = \delta_{J^\prime,J}\, \delta_{L,j}\, \delta_{S_{Q \bar Q}^\prime, S_{Q \bar Q}}\,  \bigl\langle ( j_h^{\pi_h}, L_h ) j \big\Vert T \big\Vert j^\pi \bigr\rangle.
\label{reducedT-charmoniumbaryon}
\end{equation}
The reduced $T$-matrix element depends on  three angular momenta that satisfy the triangle relations for $( j_h,L_h ) j $.
Conservation of parity requires $\pi_h (-1)^{L_h} = \pi$.
If the mass of the light hadron is large enough that kinematics require it to have a small relative velocity $v$, the reduced $T$-matrix element scales like $v^{L_h}$.

There are model-independent ratios of decay rates if the triangle relations and the parity constraint allow only one value of $L_h$.
We consider specifically the case where the light baryon is the proton or the $\Lambda$, both of which have $j_h^{\pi_h}= \frac{1}{2}^+$.
If $\pi = +$, the parity constraint requires $L_h$ to be even.
If $j^\pi=  \frac{1}{2}^+$, the only allowed value of $L_h$ is zero.
If $j^\pi=  \frac{3}{2}^+$, the only allowed value of $L_h$ is 2.
In both these cases, there are model-independent ratios of decay rates.

If the pentaquark is pure spin triplet, it can decay into $J/\psi\, h$ but not $\eta_c\, h$.
If it is a linear combination of spin singlet and spin triplet, it can decay into both  $\eta_c\, h$ and  $J/\psi\, h$.
The model-independent branching ratio for a pentaquark $\mathrm{P}$ to decay into $\eta_c\, h$ and $J/\psi\, h$ is
\begin{equation}
\frac{\Gamma[\mathrm{P}\to\eta_c\, h]}{\Gamma[\mathrm{P} \to J/\psi\, h]} 
= \frac{P_{\mathrm{P},( j^\pi ,0)}}{P_{\mathrm{P},( j^\pi ,1)}} 
\left(\frac{v_{\mathrm{P},\eta_c\, h}}{v_{\mathrm{P},J/\psi\, h}} \right)^{2L_h+1},
\label{BReta/psi}
\end{equation}
where $P_{\mathrm{P},(j^\pi,S_{Q \bar Q})}$ is the probability for the $(j^\pi,S_{Q \bar Q})$ component of $\mathrm{P}$ and $v_\mathrm{P,AB}$ is the relative velocity defined in Eq.~\eqref{vPBMrel}.

\subsubsection{Constituent Charm-Hadron Pairs}
\label{sec:Constituents}

If a B\nobreakdash-O potential that approaches a static-hadron-pair threshold at large $r$ has a bound state with energy very close to the threshold, its wave function will be predominantly $S$-wave and it will extend far beyond the range of the potential. When the spin-splitting potential $\mathbf{V}_\mathrm{SS}$ in Eq.~\eqref{VSS} is taken into account, the bound state will be split into an HQSS multiplet of pentaquarks $\mathrm{P}$. The static-hadron-pair threshold will be split into the thresholds for an HQSS multiplet of heavy baryons $\mathrm{B}$ and an HQSS multiplet of heavy mesons $\mathrm{M}$. In the heavy-quark limit, each pentaquark $\mathrm{P}$ can be expressed as a linear combination of the baryon-meson pairs $\mathrm{B\,M}$ with amplitudes $A_{\mathrm{P},\mathrm{BM}}$. The pentaquark $\mathrm{P}$ can decay into its constituents $\mathrm{B\,M}$ if the decay is kinematically allowed, which requires their masses to satisfy $M_\mathrm{P} > M_\mathrm{B} +M_\mathrm{M}$. The decay can proceed through an $S$-wave transition if $A_{\mathrm{P},\mathrm{BM}}$ is nonzero. If $A_{\mathrm{P},\mathrm{BM}}=0$, the pentaquark may still decay into $\mathrm{B\,M}$ through a $D$-wave transition with a rate suppressed by $v_{\mathrm{P},\mathrm{BM}}^4$.

The rates for decays into constituents through an $S$-wave channel can be calculated using first-order perturbation theory in the spin-splitting potential $\mathbf{V}_\mathrm{SS}(\bm{S}_Q,\bm{S}_{\bar{Q}})$ in Eq.~\eqref{VSS}.
The decay rate can be expressed as the product of the square of the $T$-matrix element and a phase-space factor.
The phase-space factor is proportional to the relative velocity $v_\mathrm{P,BM}$ of the decay products in Eq.~\eqref{vPBMrel}.
The $T$-matrix element for the decay of a pentaquark $\mathrm{P}$ into a baryon $\mathrm{B}$ and a meson $\mathrm{M}$ at first order in $\mathbf{V}_\mathrm{SS}$ can be expressed as
\begin{equation}
\bigl\langle \mathrm{B\,M} \big\vert T \big\vert \mathrm{P} \bigr\rangle
= E_\mathrm{SS}(J_\mathrm{B},J_\mathrm{M}) \,
\bigl\langle \mathrm{B\,M} \big\vert \mathrm{P} \bigr\rangle,
\label{decayamplitude}
\end{equation}
where we used the fact that the $\mathrm{B\,M}$ scattering state is an eigenstate of $\mathbf{V}_\mathrm{SS}$ with the eigenvalue in Eq.~\eqref{VSS-diag}.
The inner product $\bigl\langle \mathrm{B\,M} \big\vert \mathrm{P} \bigr\rangle$ at zeroth order in $\mathbf{V}_\mathrm{SS}$ can be expressed as the product of  the amplitude $A_{\mathrm{P},\mathrm{B}\mathrm{M}}$ for the pentaquark to have constituents $\mathrm{B\,M}$ and the radial integral of the product of the bound-state wavefunction for the pentaquark and a spherical Bessel function for the baryon-meson state.
If the relative velocity $v_{\mathrm{P},\mathrm{BM}}$ is small, that radial integral is approximately the same for all baryons in the HQSS multiplet of $\mathrm{B}$ and all mesons in the HQSS multiplet of $\mathrm{M}$.
Its square therefore nearly cancels in ratios of decay rates.  
If the pentaquarks $\mathrm{P}$ and $\mathrm{P}^\prime$ have energies above the $\mathrm{B\,M}$ and $\mathrm{B}^\prime \mathrm{M}^\prime$ thresholds, the ratio of their decay rates can be approximated by
\begin{equation}
\frac{\Gamma[\mathrm{P} \to \mathrm{B}\, \mathrm{M}]}{\Gamma[\mathrm{P}^\prime \to \mathrm{B}^\prime \mathrm{M}^\prime]} = 
\frac{P_{\mathrm{P},\mathrm{B}\, \mathrm{M}}}{P_{\mathrm{P}^\prime,\mathrm{B}^\prime \mathrm{M}^\prime}} \,
\left( \frac{E_\mathrm{SS}(J_\mathrm{B},J_\mathrm{M})}{E_\mathrm{SS}(J_{\mathrm{B}^\prime},J_{\mathrm{M}^\prime})} \right)^2 \,
\frac{v_{\mathrm{P},\mathrm{BM}}}{v_{\mathrm{P}^\prime,\mathrm{B}^\prime \mathrm{M}^\prime}} ,
\label{decayrates}
\end{equation}
where $P_{\mathrm{P},\mathrm{B}\, \mathrm{M}}=\lvert A_{\mathrm{P},\mathrm{B}\, \mathrm{M}}\rvert^2$ is the probability for $\mathrm{P}$ to have constituents $\mathrm{B\,M}$.  

\subsubsection{Lower-Energy Charm-Hadron Pairs}
\label{LowerEnergyCharmHadrons}

A hidden-charm pentaquark can decay into a charm baryon and a charm meson whose threshold is lower than that of the constituents of the pentaquark.
The pentaquark $\mathrm{P}$  belongs to the HQSS multiplet of a $j^\pi$ adjoint baryon and has quantum numbers $J^P$.
The charm baryon $\mathrm{B}$ and a charm meson $\mathrm{M}$ belong to the HQSS multiplets of a $\bm{3}$-baryon and a $\bm{\bar{3}}$-meson with light-QCD quantum numbers ${j_\mathrm{B}^\prime}^{\pi_\mathrm{B}^\prime}$ and ${j_\mathrm{M}^\prime}^{\pi_\mathrm{M}^\prime}$.
They have spin and parity quantum numbers $J_\mathrm{B}^{P_\mathrm{B}}$ and $J_\mathrm{M}^{P_\mathrm{M}}$.
The pentaquark is an $S$-wave bound state and the baryon-meson pair is a scattering state with  orbital angular momentum $L_\mathrm{BM}$.

The heavy-quark-pair spin $S_{Q \bar Q}$  and the B\nobreakdash-O angular momentum $L$ are both conserved in the heavy-quark limit.
HQSS can be exploited to express the $T$-matrix elements for all the decays $\mathrm{P} \to \mathrm{B\,M}$ in terms of reduced $T$-matrix elements between light-QCD states, as in Eq.~\eqref{reducedTmatrixelement}. 
Conservation of B\nobreakdash-O angular momentum implies that $L=j$.
In the case of decays into lower-energy charm-hadron pairs, the reduced $T$-matrix elements have the form
\begin{equation}
\bigl\langle  \bigl( ({j_\mathrm{B}^\prime}^{\pi_\mathrm{B}^\prime}, {j_\mathrm{M}^\prime}^{\pi_\mathrm{M}^\prime}) j^\prime, L_\mathrm{BM} \bigr) L \big\Vert T \big\Vert j^\pi\bigr\rangle 
 =   \delta_{L,j}\,  \bigl\langle  \bigl( ({j_\mathrm{B}^\prime}^{\pi_\mathrm{B}^\prime}, {j_\mathrm{M}^\prime}^{\pi_\mathrm{M}^\prime})  j^\prime, L_\mathrm{BM} \bigr) j \big\Vert T \big\Vert j^\pi\bigr\rangle .
\label{reducedT}
\end{equation}
The angular momenta satisfy the triangle relations for $(j_\mathrm{B}^\prime,j_\mathrm{M}^\prime )j^\prime$ and $(j^\prime, L_\mathrm{BM} \bigr) j$. 
The parity constraint is $\pi = \pi_\mathrm{B}^\prime\,  \pi_\mathrm{B}^\prime\,  (-1)^{L_\mathrm{BM}}$.
The charm baryon and the charm meson are nonrelativistic and the reduced $T$-matrix elements scales with their relative velocity $v$ as $v^{L_\mathrm{BM} }$.
The leading term in the expansion of the $T$-matrix element in powers of $v$ is the one with the smallest possible value $L_\mathrm{min}$ of $L_\mathrm{BM}$.

There are approximate model-independent ratios of the decay rates for $\mathrm{P} \to \mathrm{B\,M}$ if the triangle relations and the parity constraint for the reduced $T$ matrix element in Eq.~\eqref{reducedT} with $L_\mathrm{BM} = L_\mathrm{min}$ allow only a single value of $j^\prime$.
The model-independent ratios are derived in Appendix~\ref{app:PentaquarkDecays}. 
The factors in the decay rate for $\mathrm{P} \to \mathrm{B\,M}$ that depend on angular momenta and the relative velocity are the product of  $v_\mathrm{P,BM}^{2L_\mathrm{min}+1}$ and a number determined by angular-momentum algebra and by the amplitudes for $\mathrm{P}$ in the $\bm{8}$-hadron basis $(j^\pi,S_{Q \bar Q})$.
The ratio of the decay rates for $\mathrm{P^\prime} \to \mathrm{B^\prime\,M^\prime}$ and $\mathrm{P} \to \mathrm{B\,M}$ is the ratio of two such expressions.
Since the parity constraint requires $L_\mathrm{min}$ to be either even or odd, the next-to-minimal term in the expansion in powers of $v$ has orbital angular momentum $L_\mathrm{min}+2$.
The leading corrections to the model-independent ratio are therefore suppressed by $v_\mathrm{P,BM}^4$ or $v_\mathrm{P^\prime,B^\prime\,M^\prime}^4$. 
In the special case $L_\mathrm{min} = 0$, the model-independent ratios simplify dramatically.
The factors in the decay rate for $\mathrm{P} \to \mathrm{B\,M}$ that depend on angular momenta and the relative velocity are the product of  $v_\mathrm{P,BM}$ and a number determined by angular-momentum algebra.

\subsection{\texorpdfstring{$c\bar{c}$}{Hidden-charm} Pentaquarks}
\label{sec:nonstrangePentaquarks}

The observed $c \bar{c}$ pentaquarks were discovered through their decays into $J/\psi\, p$.
Their decay widths are listed in Table~\ref{tab:nonstrangepentaquarks}.  
An identification of these pentaquarks  should explain why they have narrow widths.

\subsubsection{Charmonium Plus Proton}
\label{sec:nonstrangeQuarkoniumlighthadron}

We have identified $P_{c \bar{c}}(4312)$, $P_{c \bar{c}}(4337)$, $P_{c \bar{c}}(4440)$, and $P_{c \bar{c}}(4457)$ as the members of the HQSS quartet associated with the ground state for $c \bar c$ in the coupled $\bigl(\frac{3}{2}^+\bigr) \frac{3}{2} \big/ \frac{1}{2}$ potentials with flavor $(I,S) = \bigl(\frac{1}{2},0\bigr)$.
The discovery decay mode $J/\psi\, p$ requires the transition of the $c \bar c$ pair in the pentaquark into $J/\psi$ and the transition of the light-QCD state into the proton.
Table~\ref{tab:nonstrange} shows that all four pentaquarks are predicted to have substantial probabilities for the spin-triplet component $\bigl(\frac{3}{2}^+,1 \bigr)$.
This is compatible with the transition of the $c \bar c$ pair into the spin-triplet charmonium state $J/\psi$.
The  proton has isospin $\frac{1}{2}$ and quantum numbers $\frac{1}{2}^+$.
The light-QCD state in the  pentaquark $P_{c \bar c}^+$ has isospin $\frac{1}{2}$ and quantum numbers $\frac{3}{2}^+$. 
The parity constraint requires the orbital angular momentum $L_p$ of the proton to be even.
The transition from the light-QCD state to the proton must therefore proceed through a $D$-wave channel.
The $D$-wave factor $v^4$ ranges from 0.24 for $P_{c \bar{c}}(4312)^+ \to J/\psi\, p$ to 0.48  for  $P_{c \bar{c}}(4457)^+ \to \eta_c\, p$, so it does not provide significant suppression.
We conclude that our identifications of the four $c \bar{c}$ pentaquarks are compatible with their discovery decay mode.  

HQSS and conservation of B\nobreakdash-O angular momentum can be used to express the $T$-matrix elements for the decays of $P_{c \bar{c}}^+$ into $J/\psi\, p$ and $\eta_c\, p$  in terms of  a single reduced $T$-matrix element of the form $\bigl\langle \bigl(\frac{1}{2}^+,2\bigr)\frac{3}{2} \big\Vert T \big\Vert  \frac{3}{2}^+ \bigr\rangle$ with proton orbital angular momentum $L_p=2$.
The branching ratios of $P_{c \bar{c}}^+$ into $\eta_c\, p$ and $J/\psi\, p$ can therefore be predicted using the model-independent ratios in Eq.~\eqref{BReta/psi}.
The $\frac{1}{2}^-$ and $\frac{5}{2}^-$ pentaquarks are pure spin triplets, so $P_{c \bar{c}}(4312)^+$ and $P_{c \bar{c}}(4457)^+$ can decay into $J/\psi\, p$ but not into $\eta_c\, p$.
The two $\frac{3}{2}^-$ pentaquarks are linear combinations of spin singlet and spin triplet, so $P_{c \bar{c}}(4337)^+$ and $P_{c \bar{c}}(4440)^+$ can decay into both $\eta_c\, p$ and $J/\psi\, p$.
Their decay rates into $\eta_c\, p$ and $J/\psi\, p$ are proportional to the $\bigl( \frac{3}{2}^+, 0 \bigr)$ and $\bigl( \frac{3}{2}^+, 1 \bigr)$ probabilities from Table~\ref{tab:nonstrange}.
They are also proportional to the fifth power of the relative velocity in Eq.~\eqref{vPBMrel}.
The predictions for the branching ratios are 
\begin{subequations}
\begin{align}
\Gamma[P_{c \bar{c}}(4312)^+ \to \eta_c\, p]/\Gamma[P_{c \bar{c}}(4312)^+ \to J/\psi\, p] &= 0,
\\
\Gamma[P_{c \bar{c}}(4337)^+ \to \eta_c\, p]/\Gamma[P_{c \bar{c}}(4337)^+ \to J/\psi\, p] &= 0.15\,  (1.094)^5 \approx  0.23, 
\\
\Gamma[P_{c \bar{c}}(4440)^+ \to \eta_c\, p]/\Gamma[P_{c \bar{c}}(4440)^+ \to J/\psi\, p] &= 6.8\,  (1.062)^5 \approx  9.2,
\\
\Gamma[P_{c \bar{c}}(4457)^+ \to \eta_c\, p]/\Gamma[P_{c \bar{c}}(4457)^+ \to J/\psi\, p] &= 0. 
\end{align}
\label{etac/Jpsi:BO}%
\end{subequations} 
A real number in parentheses raised to the fifth power is a ratio of relative velocities.
The central values of the pentaquark masses in Table~\ref{tab:nonstrangepentaquarks} were used to calculate the relative velocities.
The large branching ratio for $P_{c \bar{c}}(4440)^+$ suggests that observation of the decay mode $\eta_c\, p$ could be easiest for this pentaquark.

\subsubsection{Constituent Charm-Hadron Pairs}
\label{sec:nonstrangeConstituents}

If a $c \bar{c}$ pentaquark $\mathrm{P}$ is a molecule whose constituents include a charm baryon $\mathrm{B}$ and a charm meson $\mathrm{M}$ whose threshold is below the mass of $\mathrm{P}$,  decays into $\mathrm{B\,M}$  could give the pentaquark a large width.
The possible constituents of the observed $c \bar{c}$ pentaquarks are $\Sigma_c\bar{D}$, $\Sigma_c^\ast\bar{D}$, $\Sigma_c\bar{D}^\ast$, and $\Sigma_c^\ast\bar{D}^\ast$, and their thresholds are given in Table~\ref{tab:nonstrangepairs}.
The only decays that are kinematically allowed are the decays of $P_{c \bar{c}}(4337)$, $P_{c \bar{c}}(4440)$, and $P_{c \bar{c}}(4457)$ into $\Sigma_c\bar{D}$ and $P_{c \bar{c}}(4440)$ and $P_{c \bar{c}}(4457)$ into $\Sigma_c^\ast\bar{D}$.
We have identified $P_{c \bar{c}}(4312)$, $P_{c \bar{c}}(4337)$, $P_{c \bar{c}}(4440)$, and $P_{c \bar{c}}(4457)$ with the $\frac{1}{2}^-$, $\frac{3}{2}^-$, $\frac{3}{2}^{-\prime}$, and $\frac{5}{2}^-$ pentaquarks.

The probabilities  for each of the four pentaquarks  to have each of the four charm-hadron pairs as its constituents are given by Table~\ref{tab:nonstrange}.
The zeroes in Table~\ref{tab:nonstrange} for the $\Sigma_c\bar{D}$ probabilities of $P_{c \bar{c}}(4337)$, $P_{c \bar{c}}(4440)$, and $P_{c \bar{c}}(4457)$ indicate that their decay rates into  $\Sigma_c\bar{D}$ are suppressed by a $D$-wave factor $v^4$ that ranges from $9.1\times10^{-4}$  for $P_{c \bar{c}}(4337)$ to 0.048  for $P_{c \bar{c}}(4457)$.
The zero in Table~\ref{tab:nonstrange} for the $\Sigma_c^\ast\bar{D}$ probability of $P_{c \bar{c}}(4457)$ indicates that its decay rate into $\Sigma_c^\ast\bar{D}$ is suppressed by a $D$-wave factor $v^4=0.015$. 
The only decay that can proceed through an $S$-wave channel is $P_{c\bar{c}}(4440) \to \Sigma_c^\ast\bar{D}$.
 However its decay rate is suppressed by the small probability 0.056 for the $\Sigma_c^\ast\bar{D}$ component of $P_{c\bar{c}}(4440)$ in Table~\ref{tab:nonstrange}.
This small probability is a remarkable prediction of the B\nobreakdash-O framework.
Thus the rates for all  the decays into constituent charm-hadron pairs that are kinematically allowed are suppressed either by a small $D$-wave factor or by a small probability for the constituents.

\subsubsection{Lower-Energy Charm-Hadron Pairs}
\label{sec:nonstrangeHeavyhadronpairs}

We next consider decays of the $c \bar{c}$ pentaquarks into $\Lambda_c\bar{D}$ and $\Lambda_c\bar{D}^\ast$, whose thresholds in Table~\ref{tab:nonstrangepairs} are below those of the constituents $\Sigma_c^{(\ast)}\bar{D}^{(\ast)}$ of the pentaquarks.
The decay of a pentaquark into $\Lambda_c \bar{D}$ or $\Lambda_c \bar{D}^\ast$ can proceed through a transition between the $\bigl(\frac{3}{2}^+\bigr) \frac{1}{2}$ potential that approaches the $\Sigma_c^{(\ast)} \bar{D}^{(\ast)}$ threshold and the $\bigl(\frac{1}{2}^+\bigr) \frac{1}{2}$ potential that approaches the $\Lambda_c \bar{D}^{(\ast)}$ threshold.
The parity constraint requires the orbital angular momentum $L_\mathrm{BM}$ of the baryon-meson pair to be even.
The transition of the light-QCD state from $j^\pi=\frac{3}{2}^+$ to $\frac{1}{2}^+$ must proceed through a $D$-wave channel.
The decay rates of the pentaquarks into $\Lambda_c\bar{D}$ and $\Lambda_c\bar{D^\ast}$ therefore have a $D$-wave suppression factor. 
The $D$-wave factor $v^4$ ranges from $9.6 \times 10^{-4}$  for $P_{c \bar{c}}(4312) \to  \Lambda_c\, \bar{D}^\ast$ to 0.17  for  $P_{c \bar{c}}(4457) \to  \Lambda_c\, \bar{D}$.

There are approximate model-independent ratios for the decays of the pentaquarks into $\Lambda_c\, \bar{D}$ and $\Lambda_c\, \bar{D}^\ast$.
The charm baryon is an HQSS singlet whose light-QCD state is the $0^+$ $\bm{3}$-baryon.
The charm mesons $\bar{D}$ and $\bar{D}^\ast$ are an HQSS doublet whose light-QCD states are the $\frac{1}{2}^+$ $\bm{\bar{3}}$-meson.
The reduced matrix element in Eq.~\eqref{reducedT} with the minimal orbital angular momentum is $\bigl\langle  \bigl( \bigl(0^+,\frac{1}{2}^+\bigr) j^\prime, L_\mathrm{min} \bigr) \frac{3}{2} \big\Vert T \big\Vert \frac{3}{2}^+ \bigr\rangle$.
The triangle relation for $\bigl( 0,\frac{1}{2} \bigr) j^\prime$ requires $j^\prime =\frac{1}{2}$. 
Since the parity constraint requires $L_\mathrm{min}$  to be even, the triangle relation for $(j^\prime, L_\mathrm{min}) \frac{3}{2}$ implies that $L_\mathrm{min} = 2$.
The factors in the decay rate for $\mathrm{P} \to \mathrm{B\,M}$ are determined in Appendix~\ref{app:PentaquarkDecays}.
They are the product of  $v_\mathrm{P,BM}^5$ and a number determined by angular-momentum algebra and by the amplitudes for the $\bigl(\frac{1}{2}^+,S_{Q \bar Q}\bigr)$ components of $\mathrm{P}$, which are given in Table~\ref{tab:nonstrange}.
The model-independent ratio of the decay rates for $\mathrm{P} \to \mathrm{B\,M}$ and $\mathrm{P^\prime} \to \mathrm{B^\prime\,M^\prime}$ is the ratio of two such factors.

To specify all the model-independent ratios for decays into $\Lambda_c\, \bar{D}$ and $\Lambda_c\, \bar{D}^\ast$, we give the ratios of their decay rates to that for $P_{c\bar{c}}(4312) \to \Lambda_c\, \bar{D}^\ast$: 
\begin{subequations}
\begin{align}
\Gamma[P_{c\bar{c}}(4312) \to \Lambda_c\, \bar{D}]/\Gamma[P_{c\bar{c}}(4312) \to  \Lambda_c\, \bar{D}^\ast] &= 0, 
\label{nonstrangeBRatio4312}
\\
\Gamma[P_{c\bar{c}}(4337) \to  \Lambda_c\, \bar{D}]/\Gamma[P_{c\bar{c}}(4312) \to \Lambda_c\, \bar{D}^\ast] &= 0.0334\,  (3.034)^5 \approx 8.6, \label{nonstrangeGammaRatio4337}
\\
\Gamma[P_{c\bar{c}}(4440) \to  \Lambda_c\, \bar{D}]/\Gamma[P_{c\bar{c}}(4312) \to \Lambda_c\, \bar{D}^\ast] &=0.367\,  (3.587)^5 \approx  220 ,  
\\
\Gamma[P_{c\bar{c}}(4457) \to  \Lambda_c\, \bar{D}]/\Gamma[P_{c\bar{c}}(4312) \to \Lambda_c\, \bar{D}^\ast] &= \tfrac{2}{5}\,  (3.659)^5 \approx  260 ,
\\
\Gamma[P_{c\bar{c}}(4337) \to  \Lambda_c\, \bar{D}^\ast]/\Gamma[P_{c\bar{c}}(4312) \to \Lambda_c\, \bar{D}^\ast] &= 0.967\,  (1.552)^5 \approx  8.7, \label{nonstrangeGammaRatio4337*}
\\
\Gamma[P_{c\bar{c}}(4440) \to  \Lambda_c\, \bar{D}^\ast]/\Gamma[P_{c\bar{c}}(4312) \to \Lambda_c\, \bar{D}^\ast] &=0.633\,  (2.716)^5 \approx  94, 
\\
\Gamma[P_{c\bar{c}}(4457) \to  \Lambda_c\, \bar{D}^\ast]/\Gamma[P_{c\bar{c}}(4312) \to \Lambda_c\, \bar{D}^\ast] &= \tfrac{3}{5}\,  (2.844)^5 \approx  110 . 
\end{align}
\label{nonstrangeGammaRatios(*)}%
\end{subequations}
A real number in parentheses raised to the fifth power is a ratio of relative velocities.
The nonzero ratios in Eqs.~\eqref{nonstrangeGammaRatios(*)} are all large, because the $D$-wave factor for $P_{c\bar{c}}(4312) \to  \Lambda_c\, \bar{D}^\ast$ is much smaller than those for the other decays.

The ratios in Eqs.~\eqref{nonstrangeGammaRatios(*)} do not take into account the uncertainties in the pentaquark masses.
The largest error from the pentaquark masses comes from the denominators, because the $P_{c\bar{c}}(4312)$ mass is close to the $\Lambda_c\, \bar{D}^\ast$ threshold. 
The energy $M_\mathrm{P} - M_\mathrm{B} -M_\mathrm{M}$ of $P_{c\bar{c}}(4312)$ relative to the $\Lambda_c\, \bar{D}^\ast$ threshold is $16.9^{+6.8}_{-0.9}$ MeV. 
If the mass of $P_{c\bar{c}}(4312)$ is increased by its upper error bar or decreased by its lower error bar, the factor $v_\mathrm{P,BM}^5$ is changed from its central value by the multiplicative factor $1^{+1.29}_{-0.13}$.
The effect of the uncertainty in the  $P_{c\bar{c}}(4312)$ mass can be estimated by dividing each of the ratios in Eqs.~\eqref{nonstrangeGammaRatios(*)} by the factor $1^{+1.29}_{-0.13}$.

The ratios for decays into both $\Lambda_c\, \bar{D}$ and $\Lambda_c\, \bar{D}^\ast$ can be obtained by adding the appropriate ratios in Eqs.~\eqref{nonstrangeGammaRatios(*)}.
The predictions to two digits for the partial width into  $\Lambda_c\, \bar{D}$ and $\Lambda_c\, \bar{D}^\ast$ of $P_{c\bar{c}}(4337)$, $P_{c\bar{c}}(4440)$, and $P_{c\bar{c}}(4457)$ divided by that for $P_{c\bar{c}}(4312)$ are 17, 310, and 370. 
The measured total widths of the four pentaquarks are given in Table~\ref{tab:nonstrangepentaquarks} and they are all comparable to within the large errors.
The predicted widths of $P_{c \bar{c}}(4312)$ and $P_{c \bar{c}}(4337)$ into  $\Lambda_c\, \bar{D}$ and $\Lambda_c\, \bar{D}^\ast$ are more than 1 and 2 orders of magnitude smaller than those of $P_{c \bar{c}}(4440)$ and $P_{c \bar{c}}(4457)$.
Thus only $P_{c\bar{c}}(4440)$ and $P_{c\bar{c}}(4457)$ could have decays into $\Lambda_c\, \bar{D}$ and $\Lambda_c\, \bar{D}^\ast$ that contribute significantly to their total widths.  

Model-independent branching ratios of the four pentaquarks into $\Lambda_c\, \bar{D}$ and $\Lambda_c\, \bar{D}^\ast$ can be obtained by dividing the appropriate ratios in Eqs.~\eqref{nonstrangeGammaRatios(*)}:
\begin{subequations}
\begin{align}
\Gamma[P_{c\bar{c}}(4312) \to \Lambda_c\, \bar{D}]/\Gamma[P_{c\bar{c}}(4312) \to  \Lambda_c\, \bar{D}^\ast] &= 0, 
\\
\Gamma[P_{c\bar{c}}(4337) \to  \Lambda_c\, \bar{D}]/\Gamma[P_{c\bar{c}}(4337) \to \Lambda_c\, \bar{D}^\ast] &= 0.0346\,  (1.955)^5 \approx  0.99, 
\label{nonstrangeBRatio4337}
\\
\Gamma[P_{c\bar{c}}(4440) \to  \Lambda_c\, \bar{D}]/\Gamma[P_{c\bar{c}}(4440) \to \Lambda_c\, \bar{D}^\ast] &=0.579\,  (1.320)^5 \approx  2.3,  
\\
\Gamma[P_{c\bar{c}}(4457) \to  \Lambda_c\, \bar{D}]/\Gamma[P_{c\bar{c}}(4457) \to \Lambda_c\, \bar{D}^\ast] &=\tfrac{2}{3}\,  (1.287)^5 \approx  2.4.  
\end{align}
\label{nonstrangeBRatios}%
\end{subequations} 
The branching ratios are all comparable with the exception of that for $P_{c\bar{c}}(4312)$, which is zero.

\subsection{Molecular Dominance with Ideal Mixing}
\label{sec:MoleculeDominanceDecay}

In 2019, Voloshin pointed out that if the $c \bar{c}$ pentaquarks are interpreted as charm-hadron molecules, heavy-quark spin symmetry (HQSS) can be used to derive the ratios of their decay rates into various final states \cite{Vol19}.
In the general B\nobreakdash-O framework, charm-hadron  molecules correspond to bound states in the $\bigl(\frac{3}{2}^+\bigr)\frac{3}{2} \big/ \frac{1}{2}$ and $\bigl(\frac{1}{2}^{+\prime}\bigr)\frac{1}{2}$ potentials that approach the $\Sigma_c^{(\ast)} \bar{D}^{(\ast)}$ threshold at large $r$.
We refer to the case of bound states in both the $\bigl(\frac{3}{2}^+\bigr)\frac{3}{2} \big/ \frac{1}{2}$ and $\bigl(\frac{1}{2}^{+\prime}\bigr)\frac{1}{2}$ potentials as \emph{molecular dominance}.
We refer to the simple case in which the energies $\varepsilon_{3/2} $ and $\varepsilon_{1/2}$ of the ground states for $c \bar c$ in the $\bigl(\frac{3}{2}^+\bigr)\frac{3}{2} \big/ \frac{1}{2}$ and $\bigl(\frac{1}{2}^{+\prime}\bigr)\frac{1}{2}$ potentials are equal as \emph{ideal mixing}.
In the simple case of molecular dominance with ideal mixing, the spin splittings of the seven pentaquarks in the ground-state HQSS multiplets are given in Table~\ref{tab:molecule}. 
Each of the seven pentaquarks is predicted to be at a charm-hadron-pair threshold, as can be seen in the energy-level diagram in Fig.~\ref{fig:MoleculeNonstrangeEnergyLevels}.
We refer to the three pentaquarks near the $\Sigma_c^\ast \bar{D}^\ast$ threshold with quantum numbers $\frac{1}{2}^{\prime\prime -}$, $\frac{3}{2}^{\prime -}$, and $\frac{5}{2}^-$  by the same name $P_{c \bar{c}}(4530)$, because all three pentaquarks are predicted to have masses near the $\Sigma_c^\ast \bar{D}^\ast$ threshold at 4526.7~MeV.
In the simple case of molecular dominance with ideal mixing, Voloshin derived analytic results for some ratios of decay rates.

\subsubsection{Charmonium Plus Proton}
\label{sec:moleculeQuarkoniumlighthadron}

HQSS and conservation of B\nobreakdash-O angular momentum can be used to express the $T$-matrix elements for the decays of $P_{c \bar{c}}^+$ into $J/\psi\, p$ and $\eta_c\, p$  in terms of two reduced $T$-matrix elements of the form $\bigl\langle \bigl(\frac{1}{2}^+,2\bigr)\frac{3}{2} \big\Vert T \big\Vert  \frac{3}{2}^+ \bigr\rangle$ and $\bigl\langle \bigl(\frac{1}{2}^+,0\bigr)\frac{1}{2} \big\Vert T \big\Vert  \frac{1}{2}^+ \bigr\rangle$ with proton orbital angular momentum $L_p=2$ and and 0, respectively.
The mass of the proton is large enough that it is somewhat nonrelativistic, so the reduced $T$-matrix element scales with its relative velocity as $v^{L_p}$.
The square of the reduced $T$-matrix element for $L_p=2$ is suppressed compared to that for  $L_p=0$ by a factor of $v^4$ that ranges from 0.24  for $P_{c \bar{c}}(4312) \to J/\psi\, p$ to 0.53 for $P_{c \bar{c}}(4530) \to \eta_c\, p$.
In the velocity for $P_{c \bar{c}}(4530)$, we have assumed its mass is equal to the $\Sigma_c^\ast \bar{D}^\ast$ threshold.
These  $D$-wave suppression factors are not very small.
If $D$-wave amplitudes are ignored anyway, the branching ratios of the pentaquarks into $\eta_c\, p$ and $J/\psi\, p$ can be predicted using Eqs.~\eqref{BReta/psi}. 
The decay rates into $\eta_c\, p$ and $J/\psi\, p$ are proportional to the $\bigl(\frac{1}{2}^+, 0\bigr)$ and $\bigl(\frac{1}{2}^+, 1\bigr)$ probabilities in Table~\ref{tab:molecule} and to the relative velocities defined in Eq.~\eqref{vPBMrel}.
The predictions for the branching ratios  are 
\begin{subequations}
\begin{align}
\Gamma[P_{c \bar{c}}(4312)^+ \to \eta_c\, p]/\Gamma[P_{c \bar{c}}(4312)^+ \to J/\psi\, p] &= 3\, (1.10) \approx  3.3, 
\label{etac/Jpsi:4312}
\\
\Gamma[P_{c \bar{c}}(4337)^+ \to \eta_c\, p]/\Gamma[P_{c \bar{c}}(4337)^+ \to J/\psi\, p] &=  0, 
\label{etac/Jpsi:4337}
\\
\Gamma[P_{c \bar{c}}(4440)^+ \to \eta_c\, p]/\Gamma[P_{c \bar{c}}(4440)^+ \to J/\psi\, p] &= \bigl(  \tfrac{3}{25},~ 0 \bigr) (1.06) \approx  (0.13, ~ 0), 
\label{etac/Jpsi:4440a}
\\
\Gamma[P_{c \bar{c}}(4457)^+ \to \eta_c\, p]/\Gamma[P_{c \bar{c}}(4457)^+ \to J/\psi\, p] &= \bigl(  0,~ \tfrac{3}{25}  \bigr) ,
\label{etac/Jpsi:4457a}
\\
\Gamma[P_{c \bar{c}}(4530)^+ \to \eta_c\, p]/\Gamma[P_{c \bar{c}}(4530)^+ \to J/\psi\, p] &=  (  3, ~ 0, ~ 0 ) (1.05) \approx  (3.1, ~0, ~0).
\label{etac/Jpsi:4457}
\end{align}
\label{etac/Jpsi:Voloshin}%
\end{subequations}
A single real number in parentheses  is a ratio of relative velocities.
The two branching ratios for $P_{c \bar{c}}(4440)^+$ in Eq.~\eqref{etac/Jpsi:4440a} and $P_{c \bar{c}}(4457)^+$ in Eq.~\eqref{etac/Jpsi:4457a} assume their quantum numbers are $\frac{1}{2}^{-\prime}$ and $\frac{3}{2}^{-\prime}$ (Scenario 1) or $\frac{3}{2}^{-\prime}$ and $\frac{1}{2}^{-\prime}$  (Scenario 2).
The three branching ratios for $P_{c \bar{c}}(4530)^+$ are for the three states near the $\Sigma_c^\ast \bar{D}^\ast$ threshold with quantum numbers $\frac{1}{2}^{-\prime\prime}$, $\frac{3}{2}^{-\prime\prime}$, and $\frac{5}{2}^-$.
The analytic factors 3 in Eq.~\eqref{etac/Jpsi:4312} and $\frac{3}{25}$ in Eq.~\eqref{etac/Jpsi:4440a} have been derived previously by Voloshin \cite{Vol19}.
The predictions for the four lowest-energy pentaquarks in Eqs.~\eqref{etac/Jpsi:4312}-\eqref{etac/Jpsi:4440a} are very different from those from the B\nobreakdash-O approximation in Eqs.~\eqref{etac/Jpsi:BO}.
Given that the $D$-wave suppression factors are not very small, the physical relevance of the ratios  in Eqs.~\eqref{etac/Jpsi:Voloshin} is questionable.

The semi-inclusive decay rates of exotic hidden-heavy hadrons into quarkonium plus light hadrons can be calculated using BOEFT \cite{Bra22}. 
The decay rates for $P_{c \bar{c}}(4312)^+ \to J/\psi+X$ and $P_{c \bar{c}}(4312)^+ \to \eta_c+X$ were calculated  in Ref.~\cite{Bram25}.
For the four lowest-energy pentaquarks, the predicted decay rates ranged from about 3~MeV to about 30 MeV.
These partial widths are comparable to the total widths of the observed pentaquarks  listed in Table~\ref{tab:nonstrangepentaquarks}.

\subsubsection{Constituent Charm-Hadron Pairs}
\label{sec:moleculeConstituents}

If a $c \bar{c}$ pentaquark is a molecule whose constituents include a charm baryon and a charm meson whose threshold is below the pentaquark mass,  decays into these constituents could give the pentaquark a large width.
The possible constituents of the observed $c \bar{c}$ pentaquarks are $\Sigma_c\bar{D}$, $\Sigma_c^\ast\bar{D}$, $\Sigma_c\bar{D}^\ast$, and $\Sigma_c^\ast\bar{D}^\ast$, and their thresholds are given in Table~\ref{tab:nonstrangepairs}.  
The probabilities for each of the seven pentaquarks to have each of the four pairs of constituents are given in Table~\ref{tab:molecule}.
Each pentaquark is a molecule with a single pair of charm-hadron constituents and zero probability for constituents with a lower-energy threshold.
The decay rate of the molecule into a charm-hadron pair with a lower threshold has a $D$-wave factor $v^4$.
Of the four  observed pentaquarks in the energy-level diagram in Fig.~\ref{fig:MoleculeNonstrangeEnergyLevels}, the only one whose mass is above the threshold for its constituents is $P_{c \bar{c}}(4337)$.
Its mass is about 16~MeV above the threshold, so it can decay by an $S$-wave transition into $\Sigma_c\bar{D}$.
Each of the other three observed pentaquarks in Fig.~\ref{fig:MoleculeNonstrangeEnergyLevels} has a mass whose central value is below the threshold for its constituents, but its decay width is large enough that its line shape extends above the threshold.
It can decay into its constituents by an $S$-wave transition from the part of the line shape that is above the threshold.
Thus all four observed pentaquarks may have large partial widths from $S$-wave decays into their constituents.

\subsubsection{Lower-Energy Charm-Hadron Pairs}
\label{sec:moleculeHeavyhadronpairs}

We next consider the decays of the pentaquarks into $\Lambda_c\bar{D}$ and $\Lambda_c\bar{D}^\ast$, whose thresholds in Table~\ref{tab:nonstrangepairs} are below those of the constituents $\Sigma_c^{(\ast)}\bar{D}^{(\ast)}$ of the pentaquarks.
The decay of a pentaquark into $\Lambda_c \bar{D}$ or $\Lambda_c \bar{D}^\ast$ can proceed through an $S$-wave transition between the $\frac{1}{2}^+$ light-QCD state in the pentaquark and the $\frac{1}{2}^+$ light-QCD state in the charm-hadron pair.
It can also proceed through a $D$-wave transition between the $\frac{3}{2}^+$ light-QCD state in the pentaquark and the $\frac{1}{2}^+$ light-QCD state in the charm-hadron pair.

There are model-independent ratios of the decay rates of the pentaquarks into $\Lambda_c\, \bar{D}$ and $\Lambda_c\, \bar{D}^\ast$.
The charm baryon is an HQSS singlet whose light-QCD state is the $0^+$ $\bm{3}$-baryon.
The charm mesons $\bar{D}$ and $\bar{D}^\ast$ are an HQSS doublet whose light-QCD states are the $\frac{1}{2}^+$ $\bm{\bar{3}}$-meson.
The reduced matrix elements in Eq.~\eqref{reducedT} for $j^\pi = \frac{1}{2}^+$ and $\frac{3}{2}^+$ with the minimal orbital angular momentum are $\bigl\langle  \bigl( \bigl(0^+,\frac{1}{2}^+\bigr) j^\prime, L_\mathrm{min} \bigr) \frac{1}{2} \big\Vert T \big\Vert \frac{1}{2}^+ \bigr\rangle$ and $\bigl\langle  \bigl( \bigl(0^+,\frac{1}{2}^+\bigr) j^\prime, L_\mathrm{min} \bigr) \frac{3}{2} \big\Vert T \big\Vert \frac{3}{2}^+ \bigr\rangle$.
The triangle relation for $\bigl( 0,\frac{1}{2} \bigr) j^\prime$ requires $j^\prime =\frac{1}{2}$.
The triangle relation for $(j^\prime, L_\mathrm{min}) \frac{1}{2}$ implies that $L_\mathrm{min} =0$ or 1.
The triangle relation for $(j^\prime, L_\mathrm{min}) \frac{3}{2}$ implies that $L_\mathrm{min} =1$ or 2.
The parity constraint requires $L_\mathrm{min}$ to be even.
Thus the minimal  orbital angular momentum is $L_\mathrm{min} = 0$.
The decays proceed predominantly through the $\bigl( \frac{1}{2}^+,0 \bigr)$ and $\bigl( \frac{1}{2}^+,1 \bigr)$ components of the pentaquark, whose amplitudes are given in Table~\ref{tab:molecule}.
The factors in the decay rate for $\mathrm{P} \to \mathrm{B\,M}$  that depend on angular momenta and the relative velocity are the product of $v_\mathrm{P,BM}$ and a number determined by angular-momentum algebra and by the amplitudes for the $\bigl(\frac{1}{2}^+,S_{Q \bar Q}\bigr)$ components of $\mathrm{P}$.
The model-independent ratio of the decay rates for $\mathrm{P} \to \mathrm{B\,M}$ and $\mathrm{P^\prime} \to \mathrm{B^\prime\,M^\prime}$ is the ratio of two such factors.

To specify all the model-independent ratios of the decay rates of the pentaquarks into $\Lambda_c\, \bar{D}$ and $\Lambda_c\, \bar{D}^\ast$, we give the ratios of the decay rates to that for $P_{c\bar{c}}(4312) \to \Lambda_c\, \bar{D}^\ast$: 
\begin{subequations}
\begin{align}
\Gamma[P_{c\bar{c}}(4337) \to  \Lambda_c\, \bar{D}]/\Gamma[P_{c\bar{c}}(4312) \to \Lambda_c\, \bar{D}^\ast] &= 0,   
\label{4337/4312:LambdacD}
\\
\Gamma[P_{c\bar{c}}(4440) \to  \Lambda_c\, \bar{D}]/\Gamma[P_{c\bar{c}}(4312) \to \Lambda_c\, \bar{D}^\ast] &= (1,~ 0) \,  (3.59) \approx (3.6,~0)\, ,
\label{4440/4312:LambdacD}
\\
\Gamma[P_{c\bar{c}}(4457) \to  \Lambda_c\, \bar{D}]/\Gamma[P_{c\bar{c}}(4312) \to \Lambda_c\, \bar{D}^\ast] &= (0,~ 1) \,  (3.66) \approx (0,~ 3.7) \, ,
\label{4457/4312:LambdacD}
\\
\Gamma[P_{c \bar{c}}(4530) \to  \Lambda_c\, \bar{D}]/\Gamma[P_{c\bar{c}}(4312) \to \Lambda_c\, \bar{D}^\ast] &= (2,~ 0,~ 0) (3.92) \approx (7.8,0, ~ 0 ),
\label{4530/4312:LambdacD} \\
\Gamma[P_{c\bar{c}}(4337) \to  \Lambda_c\, \bar{D}^\ast]/\Gamma[P_{c\bar{c}}(4312) \to \Lambda_c\, \bar{D}^\ast] &= 1\, (1.55) \approx 1.6,
\label{4337/4312:LambdacDstar}
\\
\Gamma[P_{c\bar{c}}(4440) \to  \Lambda_c\, \bar{D}^\ast]/\Gamma[P_{c\bar{c}}(4312) \to \Lambda_c\, \bar{D}^\ast] &=\bigl( \tfrac{4}{3},  ~\tfrac{1}{3}\bigr) (2.72) \approx (3.6,~0.91), 
\label{4440/4312:LambdacDstar}
\\
\Gamma[P_{c\bar{c}}(4457) \to  \Lambda_c\, \bar{D}^\ast]/\Gamma[P_{c\bar{c}}(4312) \to \Lambda_c\, \bar{D}^\ast] &= \bigl( \tfrac{1}{3},  ~\tfrac{4}{3}\bigr) (2.84) \approx (0.95,~ 3.8), 
\label{4457/4312:LambdacDstar}
\\
\Gamma[P_{c \bar{c}}(4530) \to  \Lambda_c\, \bar{D}^\ast]/\Gamma[P_{c\bar{c}}(4312) \to \Lambda_c\, \bar{D}^\ast] &=\bigl( \tfrac{2}{3}, ~ \tfrac{5}{3}, ~ 0 \bigr) (3.27) \approx (2.2,~5.5 , ~0 ).
\label{4530/4312:LambdacDstar}
\end{align}
\label{moleculeBRatios(*)}%
\end{subequations} 
A single real number in parentheses is a ratio of relative velocities.
The two ratios for $P_{c \bar{c}}(4440)$ in Eqs.~\eqref{4440/4312:LambdacD} and \eqref{4440/4312:LambdacDstar} and for $P_{c \bar{c}}(4457)$ in Eqs.~\eqref{4457/4312:LambdacD} and \eqref{4457/4312:LambdacDstar} assume their quantum numbers are $\frac{1}{2}^{-\prime}$ and $\frac{3}{2}^{-\prime}$ (Scenario 1) or $\frac{3}{2}^{-\prime}$ and $\frac{1}{2}^{-\prime}$  (Scenario 2).
The three ratios for $P_{c \bar{c}}(4530)$ in Eqs.~\eqref{4530/4312:LambdacD} and \eqref{4530/4312:LambdacDstar} are for the three pentaquarks near the $\Sigma_c^\ast \bar{D}^\ast$ threshold with quantum numbers $\frac{1}{2}^{-\prime\prime}$ , $\frac{3}{2}^{-\prime\prime}$, and $\frac{5}{2}$.
There are six zeros among the model-independent ratios in Eqs.~\eqref{moleculeBRatios(*)}.
These  zeros indicate that the decay  in the numerator proceeds through a $D$-wave transition with a rate suppressed by $v^4$.

The ratios in Eqs.~\eqref{nonstrangeGammaRatios(*)} do not take into account the uncertainties in the pentaquark masses.
The largest error from the pentaquark masses comes from the denominators, because the $P_{c\bar{c}}(4312)$ mass is close to the $\Lambda_c\, \bar{D}^\ast$ threshold. 
If the mass of $P_{c\bar{c}}(4312)$ is increased by its upper error bar or decreased by its lower error bar, the relative velocity $v_\mathrm{P,BM}$ is changed from its central value by the multiplicative factor $1^{+0.18}_{-0.03}$.
The effect of the uncertainty in the  $P_{c\bar{c}}(4312)$ mass can be estimated by dividing each of the ratios in Eqs.~\eqref{nonstrangeGammaRatios(*)} by the factor $1^{+0.18}_{-0.03}$.

The ratios for decays into both $\Lambda_c\, \bar{D}$ and $\Lambda_c\, \bar{D}^\ast$ can be obtained by adding the appropriate ratios in Eqs.~\eqref{moleculeBRatios(*)}.
The prediction for the partial width into $\Lambda_c\, \bar{D}$ and $\Lambda_c\, \bar{D}^\ast$ for $P_{c \bar{c}}(4337)$ divided by  that for $P_{c \bar{c}}(4312)$ is 1.6. 
The predictions for $P_{c \bar{c}}(4440)$ and $P_{c \bar{c}}(4457)$ are 7.2  and 0.91  (Scenario 1) or 0.95  and 7.5 (Scenario 2).
The three ratios are very different from the ratios 17, 310, and 370 predicted by the B\nobreakdash-O approximation  in Eqs.~\eqref{moleculeBRatios(*)}.
The biggest difference between the predictions of molecular dominance for decays into $\Lambda_c\, \bar{D}$ and $\Lambda_c\, \bar{D}^\ast$ and those for the B\nobreakdash-O approximation is that they proceed through $S$-wave transitions instead of $D$-wave transitions.
The resulting partial widths of the observed pentaquarks into $\Lambda_c\, \bar{D}$ and $\Lambda_c\, \bar{D}^\ast$ are likely to be too large to accommodate the small total widths in Table~\ref{tab:nonstrange}.

Approximate model-independent branching ratios into $\Lambda_c\, \bar{D}$ and $\Lambda_c\, \bar{D}^\ast$ can be obtained for six of the seven pentaquarks by dividing the appropriate ratios from Eq.~\eqref{moleculeBRatios(*)}:
\begin{subequations}
\begin{align}
\Gamma[P_{c\bar{c}}(4312) \to \Lambda_c\, \bar{D}]/\Gamma[P_{c\bar{c}}(4312) \to  \Lambda_c\, \bar{D}^\ast] &= 0,  
\\
\Gamma[P_{c\bar{c}}(4337) \to  \Lambda_c\, \bar{D}]/\Gamma[P_{c\bar{c}}(4337) \to \Lambda_c\, \bar{D}^\ast] &=0,  
\\
\Gamma[P_{c\bar{c}}(4440) \to  \Lambda_c\, \bar{D}]/\Gamma[P_{c\bar{c}}(4440) \to \Lambda_c\, \bar{D}^\ast] &= \bigl( \tfrac{3}{4},~ 0 \bigr)(1.32) \approx  (0.99, ~ 0), 
\label{moleculeGammaRatio4440}  
\\
\Gamma[P_{c\bar{c}}(4457) \to  \Lambda_c\, \bar{D}]/\Gamma[P_{c\bar{c}}(4457) \to \Lambda_c\, \bar{D}^\ast] &= \bigl( 0,~ \tfrac{3}{4} \bigr)(1.29) \approx  (0,~ 0.97),
\label{moleculeGammaRatio4457} 
\\
\Gamma[P_{c \bar{c}}(4530) \to  \Lambda_c\, \bar{D}]/\Gamma[P_{c \bar{c}}(4530) \to \Lambda_c\, \bar{D}^\ast] &= ( 3, ~ 0) (1.20) \approx (3.6,~ 0).  \label{moleculeGammaRatio4530}
\end{align}
\label{moleculeGammaRatios}%
\end{subequations} 
The two branching ratios for $P_{c \bar{c}}(4440)$ in Eq.~\eqref{moleculeGammaRatio4440} and for $P_{c \bar{c}}(4457)$ in Eq.~\eqref{moleculeGammaRatio4440} assume their quantum numbers are  $\frac{1}{2}^{-\prime}$ and $\frac{3}{2}^{-\prime}$ (Scenario 1) or $\frac{3}{2}^{-\prime}$ and $\frac{1}{2}^{-\prime}$  (Scenario 2).
The two branching ratios for $P_{c \bar{c}}(4530)$ in Eq.~\eqref{moleculeGammaRatio4530} are for the two states near the $\Sigma_c^\ast \bar{D}^\ast$ threshold with quantum numbers $\frac{1}{2}^{-\prime\prime}$ and $\frac{3}{2}^{-\prime\prime}$.
The analytic factor $\frac{3}{4}$ in Eq.~\eqref{moleculeGammaRatio4440} has been derived previously by Voloshin \cite{Vol19}.

The 7th pentaquark is the $\frac{5}{2}^-$ pentaquark near the $\Sigma_c^\ast \bar{D}^\ast$ threshold.
The dominant $T$-matrix elements for its decays into $\Lambda_c\, \bar{D}$ and $\Lambda_c\, \bar{D}^\ast$ can be expressed in terms of the single reduced $T$-matrix element $\bigl\langle \bigl(\frac{1}{2}^+,2\bigr)\frac{3}{2} \big\Vert T \big\Vert  \frac{3}{2}^+ \bigr\rangle$ with minimal orbital angular momentum $L_\mathrm{min} =2 $.
These decays therefore proceed through $D$-wave transitions.
The approximate model-independent branching ratio for the decays of the $\frac{5}{2}^-$ pentaquark into $\Lambda_c\, \bar{D}$ and $\Lambda_c\, \bar{D}^\ast$  is
\begin{equation}
\Gamma[P_{c \bar{c}}(4530) \to  \Lambda_c\, \bar{D}]/\Gamma[P_{c \bar{c}}(4530) \to \Lambda_c\, \bar{D}^\ast] = (2/3) (1.197)^5\approx 1.6.
\label{BR[5/2->D,D*]}
\end{equation}
In the velocity ratio, we have assumed the mass of $P_{c \bar{c}}(4530)$ is equal to the $\Sigma_c^\ast \bar{D}^\ast$ threshold.
The decay rates of the $\frac{5}{2}^-$ pentaquark into $\Lambda_c\, \bar{D}$ and into $\Lambda_c\, \bar{D}^\ast$ should be very small compared to the nonzero decay rates in the ratios in Eqs.~\eqref{moleculeGammaRatios}, because they have $D$-wave suppression factors $v^4$.

 The ratios of the decay rates of the pentaquarks into $\Lambda_c\, \bar{D}$ and $\Lambda_c\, \bar{D}^\ast$ have been calculated in Ref.~\cite{Bram25} for two models with molecular dominance: Scenario 1 with ground-state energies $\varepsilon_{1/2} = -23$~MeV and $\varepsilon_{3/2} = -1$~MeV and Scenario 2 with $\varepsilon_{1/2} = -0.5$~MeV and $\varepsilon_{3/2} = -14$~MeV.
In both scenarios, the difference $\varepsilon_{1/2} - \varepsilon_{3/2}$ is small compared to the spin splittings $\Delta_\mathrm{B}$ and $\Delta_\mathrm{M}$,
so the predicted ratios are qualitatively similar to those for molecular dominance with ideal mixing in Eqs.~\eqref{moleculeBRatios(*)}.
The decay rates into $\Lambda_c\, \bar{D}$ are predicted to be zero for $P_{c\bar{c}}(4312)$ and for all three pentaquarks with $J^P = \frac{3}{2}^-$ in both Scenarios 1 and 2 as well as in the limit of ideal mixing.
It is convenient to consider ratios of the  decay rates  to that for $P_{c\bar{c}}(4312) \to \Lambda_c\, \bar{D}^\ast$ as in Eqs.~\eqref{moleculeBRatios(*)}.
For both scenarios, the pattern of the nonzero prefactors of the velocity ratio is qualitatively similar to that in the limit of ideal mixing.
The smallest nonzero prefactor is for the decay of the $\frac{3}{2}^{-\prime}$ pentaquark into $\Lambda_c\, \bar{D}^\ast$: 0.326  in Scenario 1 and 0.333  in Scenario 2 compared to $\frac{1}{3}$ in the limit of ideal mixing.
The largest prefactor is for the decay of the $\frac{1}{2}^{-\prime\prime}$ pentaquark into $\Lambda_c\, \bar{D}$: 1.53  in Scenario 1 and 2.34  in Scenario 2 compared to $2$ in the limit of ideal mixing.

\subsection{\texorpdfstring{$c\bar{c}s$}{Strange Hidden-charm} Pentaquarks}
\label{sec:strangeDecays}

The observed $c \bar{c}s$ pentaquarks were discovered through their decays into $J/\psi\, \Lambda$.
Their decay widths are listed in Table~\ref{tab:strangepentaquarks}.  
An identification of these pentaquarks should explain why $P_{c\bar{c}s}(4338)^0$ and $P_{c\bar{c}s}(4459)^0$ have narrow widths. 

\subsubsection{Charmonium Plus Lambda}
\label{sec:strangeQuarkoniumlighthadron}

We have identified $P_{c\bar{c}s}(4338)^0$ and $P_{c\bar{c}s}(4459)^0$ as two of the members of an HQSS triplet for the ground state of $c \bar c$ in the $\bigl(\frac{1}{2}^+\bigr) \frac{1}{2}$ potential with flavor $(I,S) = (0,-1)$. 
Some of the properties of these three pentaquarks are listed in Table~\ref{tab:strange}.
The discovery decay mode $J/\psi\, \Lambda$ requires the transition of the $c \bar c$ pair in the pentaquark  into $J/\psi$ and the transition of the light-QCD state into $\Lambda$.
The $J/\psi$ is a spin-triplet charmonium state.
Table~\ref{tab:strange} shows that all three pentaquarks are predicted to have substantial probabilities for the spin-triplet component labeled $\bigl(\frac{1}{2}^+,1 \bigr)$.
This is compatible with the transition of the $c \bar c$ pair into $J/\psi$. 
The $\Lambda$ is the isospin-0 member of a SU(3) octet and its quantum numbers are $\frac{1}{2}^+$.
The  light-QCD state in the pentaquark $P_{c\bar{c}s}^0$ has isospin zero and quantum numbers $\frac{1}{2}^+$. 
The transition to $\Lambda$ can therefore proceed through an $S$-wave channel.
We conclude that our identifications of the $c\bar{c}s$ pentaquarks are compatible with their discovery decay mode.
 
HQSS and conservation of B\nobreakdash-O angular momentum  can be used to express the $T$-matrix elements for the decays of $P_{c \bar{c} s}^0$ into $J/\psi\, \Lambda$ and $\eta_c\, \Lambda$  in terms of  a single reduced $T$-matrix element of the form $\bigl\langle \bigl(\frac{1}{2}^+,0\bigr)\frac{1}{2} \big\Vert T \big\Vert  \frac{1}{2}^+ \bigr\rangle$ with $\Lambda$ orbital angular momentum $L_\Lambda = 0$.
The branching ratio of $P_{c \bar{c} s}^0$ into $\eta_c\, \Lambda$ and $J/\psi\, \Lambda$ can be predicted using the model-independent ratio in Eq.~\eqref{BReta/psi}.
The $\frac{3}{2}^-$ state is pure heavy-quark-pair spin triplet, so it can decay into $J/\psi\, \Lambda$ but not into $\eta_c\, \Lambda$.
The $\frac{1}{2}^-$ and $\frac{1}{2}^{-\prime}$  pentaquarks are linear combinations of spin singlet and spin triplet, so they can decay into both $J/\psi\, \Lambda$ and  $\eta_c\, \Lambda$.
Their decay widths are proportional to the spin-singlet and spin-triplet probabilities given by Table~\ref{tab:strange}.
They are also proportional to the relative velocity in Eq.~\eqref{vPBMrel}.
The predictions for their branching ratios into $\eta_c\, \Lambda$ and $J/\psi\, \Lambda$ are 
\begin{subequations}
\begin{align}
\Gamma[P_{c \bar{c} s}(4338)^0 \to \eta_c\, \Lambda]/\Gamma[P_{c \bar{c}s}(4338)^0 \to J/\psi\, \Lambda] &= \tfrac{1}{3}\, (1.28)= 0.43, 
\\
\Gamma[P_{c \bar{c} s}(4459)^0 \to \eta_c\, \Lambda]/\Gamma[P_{c \bar{c}s}(4459)^0 \to J/\psi\, \Lambda] &= (3, ~0)\, (1.13) = (3.4,~0).
\end{align}
\end{subequations}
The two branching ratios for $P_{c \bar{c}s}(4459)^0$ correspond to the two options $\frac{1}{2}^{-\prime}$ and $\frac{3}{2}^-$ for its quantum numbers.

\subsubsection{Constituent Charm-Hadron Pairs}
\label{sec:srangeConstituents}

If a $c \bar{c}s$ pentaquark is a molecule whose constituents include a charm baryon and a charm meson whose threshold is below the pentaquark mass,  decays into these constituents could give the pentaquark a large width.
The possible constituents of the observed $c \bar{c}s$ pentaquarks are $\Xi_c \bar{D}$ and  $\Xi_c \bar{D}^\ast$, and their thresholds are given in Table~\ref{tab:strangepairs}.
We have identified $P_{c \bar{c}s}(4338)$ with the $\frac{1}{2}^-$ pentaquark, whose constituents are $\Xi_c \bar{D}$ as indicated in Table~\ref{tab:strange}.
We have identified $P_{c \bar{c}s}(4459)$ with either the $\frac{1}{2}^{-\prime}$ or the $\frac{3}{2}^-$ pentaquark whose constituents are both predicted to be $\Xi_c \bar{D}^\ast$, as indicated in Table~\ref{tab:strange}.
The central value of the mass of $P_{c\bar{c}s}(4338)^0$ is above the $\Xi_c \bar{D}$ threshold by about 2~MeV. 
The central value of the mass of $P_{c\bar{c}s}(4459)^0$ is below the $\Xi_c \bar{D}^\ast$ threshold by about 19~MeV.
However the widths of the two pentaquarks are large enough that both have line shapes that extend well beyond the threshold.
Thus $P_{c\bar{c}s}(4338)$ can decay by $S$-wave into $\Xi_c \bar{D}$ through the part of the line shape that is above the $\Xi_c \bar{D}$  threshold. 
Similarly $P_{c\bar{c}s}(4459)$ can decay by $S$-wave into $\Xi_c \bar{D}^\ast$  through the part of the line shape that is above the $\Xi_c \bar{D}^\ast$ threshold.
There may be some suppression from the fraction of the line shape that is above the threshold. 
There may also be some suppression from the mismatch between the flavors of the baryon-meson pair at large $r$ and the adjoint baryon at small $r$.
The probability for the flavor $\Xi_c \bar{D}^{(\ast)}$ is 1 at large $r$ but it could be as small as $\frac{2}{3}$ at small $r$.
Another source of suppression is the factor of relative velocity from the phase space.
These sources of suppression may not be sufficient to explain the narrow widths of $P_{c \bar{c} s}(4338)$ and $P_{c \bar{c} s}(4459)$.

\subsubsection{Lower-Energy Charm-Hadron Pairs}
\label{sec:strangeHeavyhadronpairs}

We next consider the  decays of the $c\bar{c}s$ pentaquarks into $\Lambda_c\bar{D}_s$ and $\Lambda_c\bar{D}^\ast_s$, whose thresholds are given in Table~\ref{tab:strangepairs}.
The mass of $P_{c\bar{c}s}(4338)^0$ is above the $\Lambda_c\bar{D}_s$ threshold and the mass of $P_{c\bar{c}s}(4459)$ is above  both the $\Lambda_c\bar{D}_s$ and $\Lambda_c\bar{D}_s^\ast$ thresholds.
The decay of a pentaquark into $\Lambda_c \bar{D}_s$ or $\Lambda_c \bar{D}_s^\ast$ can proceed through a transition between the $\bigl(\frac{1}{2}^{+ \prime}\bigr) \frac{1}{2}$ potential that approaches the $\Xi_c \bar{D}^{(\ast)}$ threshold and the $\bigl(\frac{1}{2}^+\bigr) \frac{1}{2}$ potential that approaches the $\Lambda_c \bar{D}_s^{(\ast)}$ threshold.
The transition must proceed through an $S$-wave channel.

There are approximate model-independent ratios of the decay rates of the pentaquarks into $\Lambda_c\, \bar{D}_s$ and $\Lambda_c\, \bar{D}_s^\ast$.
The charm baryon $\Lambda_c$ is an HQSS singlet whose light-QCD state is the $0^+$ $\bm{3}$-baryon.
The charm mesons $\bar{D}_s$ and $\bar{D}^\ast_s$ are an HQSS doublet whose light-QCD states are the $\frac{1}{2}^+$ $\bm{\bar{3}}$-meson.
The reduced matrix element in Eq.~\eqref{reducedT} with the minimal orbital angular momentum is $\bigl\langle  \bigl( \bigl(0^+,\frac{1}{2}^+\bigr) j^\prime, L_\mathrm{min} \bigr) \frac{1}{2} \big\Vert T \big\Vert \frac{1}{2}^+ \bigr\rangle$.
The triangle relation for $\bigl( 0,\frac{1}{2} \bigr) j^\prime$ requires $j^\prime = \frac{1}{2}$.
The parity constraint implies that $L_\mathrm{min}$ is even.
The triangle relation for $(j^\prime, L_\mathrm{min}) \frac{1}{2}$ implies that $L_\mathrm{min} =0$ or 1.
Thus the minimal orbital angular momentum is $L_\mathrm{min} = 0$.
The factors in the decay rate for $\mathrm{P} \to \mathrm{B\,M}$ that depend on angular momenta and the relative velocity are particularly simple in the case $L_\mathrm{min} = 0$ and are given in Appendix~\ref{app:PentaquarkDecays}.
They are the product of $v_\mathrm{P,BM}$ and the probability for the constituents of $\mathrm{P}$ to have the same spins $J_B$ and $J_M$ as the heavy baryon $\mathrm{B}$ and heavy meson $\mathrm{M}$, which are given in Table~\ref{tab:strange}.

The probability factors in the decay rates of the $c \bar c s$ pentaquark $\mathrm{P}$ into $\Lambda_c\, \bar{D}_s$ and $\Lambda_c\, \bar{D}_s^\ast$ are identical to the probabilities for the components $\Xi_c\, \bar{D}$ and $\Xi_c\, \bar{D}^\ast$ of $\mathrm{P}$ in Table~\ref{tab:strange}.
The decay rate for $P_{c \bar{c} s} (4459) \to \Lambda_c\, \bar{D}_s$ is predicted to be zero regardless  of whether $P_{c \bar{c} s} (4459)$ is identified with the $\frac{1}{2}^{-\prime}$ pentaquark or the $\frac{3}{2}^-$ pentaquark. 
The only decay rates that are predicted to be nonzero are those for $P_{c \bar{c} s} (4338) \to \Lambda_c\, \bar{D}_s$ and $P_{c \bar{c} s} (4459) \to \Lambda_c\, \bar{D}_s^\ast$.
The probability factors from Table~\ref{tab:strange} are 1, so the prediction for the ratio of these decay rates is the ratio of the velocity factors defined in Eq.~\eqref{vPBMrel}:
\begin{equation}
\Gamma\bigl[P_{c \bar{c} s}(4459)^0  \to  \Lambda_c^+\, D_s^{\ast  -}\bigr] / \Gamma\bigl[P_{c \bar{c} s}(4338)^0  \to \Lambda_c^+\, D_s^-\bigr] = 0.85^{+0.04}_{-0.02}. 
\label{Gamma4459/4338}
\end{equation} 
The error bars come mostly from the uncertainty in the mass of $P_{c\bar{c}s}(4459)^0$.
The ratio of the measured widths of $P_{c\bar{c}s}(4459)^0$ and $P_{c\bar{c}s}(4338)^0$ in Table~\ref{tab:strangepentaquarks} is $2.5^{+1.6}_{-1.4}$. 
This is compatible within errors with the prediction in Eq.~\eqref{Gamma4459/4338} because of the large uncertainty in the decay width of $P_{c\bar{c}s}(4459)^0$.

Since the decays $P_{c \bar{c} s} (4338) \to \Lambda_c\, \bar{D}_s$ and $P_{c \bar{c} s} (4459) \to \Lambda_c\, \bar{D}_s^\ast$ can proceed through $S$-wave transitions, their rates may be too large to accommodate the narrow observed widths of  $P_{c \bar{c} s} (4338)$ and $P_{c \bar{c} s} (4459)$.
The narrow widths could arise from SU(3)-flavor suppression.
At large $r$, the flavor of the $\bigl(\frac{1}{2}^{+\prime}\bigr)\frac{1}{2}$ potential is $\Xi_c\bar{D}^{(\ast)}$, which is orthogonal to $\Lambda_c\bar{D}_s^{(\ast)}$. 
At small $r$, the flavor of the $\bigl(\frac{1}{2}^{+\prime}\bigr)\frac{1}{2}$ potential is determined by the mixing between the SU(3)-singlet $\bm{8}$-baryon and the isospin-0 $\bm{8}$-baryon at the center of the SU(3) octet.
The probability for the flavor to be $\Lambda_c\bar{D}_s^{(\ast)}$ is between 0 and $\frac{1}{3}$.
Since the $c \bar{c} s$ pentaquarks are $S$-wave bound states with energies near the $\Xi_c \bar{D}^{(\ast)}$ threshold, their wavefunctions extend far beyond the range of the potentials.
If their decays into $\Lambda_c\, \bar{D}_s$ and $\Lambda_c\, \bar{D}_s^\ast$ are dominated by the long-distance region where the flavor of the constituents  is orthogonal to $\Lambda_c \bar{D}_s^{(\ast)}$, those decays could have significant flavor suppression.

\section{Summary}
\label{sec:Summary}

A specific pattern for the exotic hidden-heavy hadrons was proposed in Ref.~\cite{Braa24b}.
They were identified as members of HQSS multiplets associated with energy levels in diabatic B\nobreakdash-O potentials whose dependence on $r$ has a specific form motivated by the continuity of the spectrum of QCD in the presence of static color sources as functions of $r$.
The potentials are repulsive color-Coulomb potentials offset by the energy of an $\bm{8}$-hadron at small $r$, they cross below a triplet-hadron-pair threshold at intermediate $r$, and they approach that threshold at large $r$.
Only those potentials associated with the lowest energy $\bm{8}$-hadrons
cross below the static-hadron-pair threshold before approaching it, and these are the only potentials that may support bound states.
Since the energy levels in the potentials are necessarily close to the static-hadron-pair threshold, the energies of the hidden-heavy hadrons are close to heavy-hadron-pair thresholds.
In Ref.~\cite{Alas25}, we applied this proposal to hidden-heavy pentaquarks.
In this paper, we described the application to hidden-charm pentaquarks in much greater detail.

We considered in detail the two cases corresponding to the hidden-heavy pentaquarks that have been discovered.
One of the two cases is isospin-$\frac{1}{2}$ $c \bar{c}$ pentaquarks.
The relevant B\nobreakdash-O potentials with flavor $(I,S) = \bigl( \frac{1}{2}, 0 \bigr)$ have B\nobreakdash-O  quantum numbers $\bigl( \frac{3}{2}^+ \bigr) \frac{3}{2} \big/ \frac{1}{2}$.
At small $r$, they are repulsive color-Coulomb potentials offset by the energy of the $\bm{8}$-baryon with flavor $\bigl( \frac{1}{2}, 0 \bigr)$ and light-QCD quantum numbers $j^\pi = \frac{3}{2}^+$. 
At large $r$, they approach the triplet-hadron-pair threshold associated with $\Sigma_c^{(\ast)} \bar{D}^{(\ast)}$ threshold. 
The HQSS multiplet for the ground state in these potentials consists of four pentaquarks that we identify as $P_{c\bar{c}}(4312)^+$, $P_{c\bar{c}}(4337)^+$, $P_{c\bar{c}}(4440)^+$, and $P_{c\bar{c}}(4457)^+$.
Their spin splittings were estimated by treating the term in the Hamiltonian that produces the spin splittings of charm baryons and charm mesons as a first-order perturbation. 
These spin splittings for the four pentaquarks are given in Table~\ref{tab:nonstrange} and their energy-level diagram is shown in Fig.~\ref{fig:NonstrangeEnergyLevels}.
We predict the $J^P$ quantum numbers of $P_{c\bar{c}}(4312)^+$, $P_{c\bar{c}}(4337)^+$, $P_{c\bar{c}}(4440)^+$, and $P_{c\bar{c}}(4457)^+$ to be $\frac{1}{2}^-$, $\frac{3}{2}^-$, $\frac{3}{2}^-$, and $\frac{5}{2}^-$, respectively.
The surprising prediction that $P_{c\bar{c}}(4457)^+$ has quantum numbers $\frac{5}{2}^-$ agrees with only a few previous studies \cite{Ani15,Shi21,Wan22}.
The vast majority of the existing predictions in the literature are $\frac{1}{2}^-$ or $\frac{3}{2}^-$ \cite{Du19,Pav19,Xia19,Liu20,Gir21,Bur22,Den22,Li23,Xu23,Cly24,
Pen24,Bram25,Lia26}.
The determination of the $P_{c\bar{c}}(4457)^+$ quantum numbers to be $\frac{5}{2}^-$ would therefore provide strong support for the pattern of exotic hidden-heavy hadrons proposed in Ref.~\cite{Braa24b}.

The amplitudes and probabilities for the four $c \bar c$ pentaquarks in the $\bm{8}$-hadron basis and in the dihadron basis are given  in Table~\ref{tab:nonstrange}.
They were used to determine approximate model-independent ratios of decay rates of the pentaquarks.
We showed that our identifications of the four isospin-$\frac{1}{2}$ $c \bar c$ pentaquarks are compatible with their narrow widths.
The only decay into constituent  charm-hadron pairs that can proceed through an $S$-wave transition is $P_{c\bar{c}}(4440) \to \Sigma_c^\ast \bar{D}$, but its rate is suppressed by the small probability 0.056 for these constituents.
This small probability is a remarkable prediction of the B\nobreakdash-O framework.
The four pentaquarks all have $D$-wave decays into the lower-energy charm-hadron pairs $\Lambda_c \bar{D}$ and $\Lambda_c \bar{D}^\ast$, so the rates for these decays are suppressed by the $D$-wave factor $v^4$, where $v$ is the relative velocity of the decay products.
This $D$-wave suppression is another remarkable prediction of the B\nobreakdash-O framework.

The assumption of molecular dominance is that if there is a bound state in a B\nobreakdash-O potential that approaches a specific triplet-hadron-pair threshold at large $r$, then there are bound states in all the B\nobreakdash-O potentials that approach that threshold.
In the case of isospin-$\frac{1}{2}$ $c \bar c$ pentaquarks, molecular dominance requires both the $\bigl(\frac{3}{2}^+\bigr)\frac{3}{2} \big/ \frac{1}{2}$ and $\bigl( \frac{1}{2}^{+\prime} \bigr) \frac{1}{2}$ potentials to support bound states.
At small $r$, the $\bigl(\frac{3}{2}^+\bigr)\frac{3}{2} \big/ \frac{1}{2}$ potentials approach the repulsive color-Coulomb potential offset by the energy of the ground-state $\bm{8}$-baryon with quantum numbers $j^\pi=\frac{3}{2}^+$. 
At small $r$, the $\bigl(\frac{1}{2}^{+\prime}\bigr)\frac{1}{2}$ potential approaches the repulsive color-Coulomb potential offset by the energy of the same-parity orbitally-excited  $\bm{8}$-baryon with quantum numbers $j^\pi=\frac{1}{2}^{+\prime}$.
The existence of $c \bar c$ pentaquarks implies that the energy of the $\frac{3}{2}^+$ $\bm{8}$-baryon must be low enough for the $\bigl(\frac{3}{2}^+\bigr)\frac{3}{2} \big/ \frac{1}{2}$ potentials to support bound states.
Since the $\frac{1}{2}^{+\prime}$ $\bm{8}$-baryon is a same-parity orbital excitation, its energy is unlikely to be low enough for the $\bigl( \frac{1}{2}^{+\prime} \bigr) \frac{1}{2}$ potential to support bound states.

Molecular dominance implies that the ground states in the $\bigl(\frac{3}{2}^+\bigr)\frac{3}{2} \big/ \frac{1}{2}$ and $\bigl( \frac{1}{2}^{+\prime} \bigr) \frac{1}{2}$ potentials consist of seven pentaquarks.
Ideal mixing is the additional assumption that the energies of the ground states in those two potentials are equal. 
The spin splittings of the seven pentaquarks in the case of molecular dominance with ideal mixing were estimated by treating the spin splittings of charm hadrons as a first-order perturbation.  
The spin splittings for the seven pentaquarks are given in Table~\ref{tab:molecule} and their energy-level diagram is shown in Fig.~\ref{fig:MoleculeNonstrangeEnergyLevels}.
The $P_{c\bar{c}}(4312)^+$ and $P_{c\bar{c}}(4337)^+$ are both predicted to be near the $\Sigma_c \bar{D}$ threshold with the same quantum numbers $\frac{1}{2}^-$ and $\frac{3}{2}^-$ as in the B\nobreakdash-O approximation.
The $P_{c\bar{c}}(4440)^+$ and $P_{c\bar{c}}(4457)^+$ are both predicted to be near the $\Sigma_c \bar{D}^\ast$ threshold with quantum numbers either $\frac{1}{2}^-$ and $\frac{3}{2}^-$ (Scenario 1) or $\frac{3}{2}^-$ and $\frac{1}{2}^-$ (Scenario 2).
The prediction for $P_{c\bar{c}}(4457)^+$ is different from the prediction $\frac{5}{2}^-$ from the B\nobreakdash-O approximation.
Three additional pentaquarks are predicted to be near the $\Sigma_c^\ast \bar{D}^\ast$ threshold with quantum numbers $\frac{1}{2}^-$, $\frac{3}{2}^-$, and $\frac{5}{2}^-$, but no such pentaquarks have been observed.
The amplitudes and probabilities for the seven $c \bar c$ pentaquarks in the $\bm{8}$-hadron basis and in the dihadron basis are given in Table~\ref{tab:nonstrange}.
They were used to derive approximate model-independent ratios of decay rates of the pentaquarks. 
The predictions differ significantly from those of the B\nobreakdash-O approximation.
Another difference between the predictions of molecular dominance and the B\nobreakdash-O approximation is that the decays into $\Lambda_c\, \bar{D}$ and $\Lambda_c\, \bar{D}^\ast$ proceed through $S$-wave transitions instead of $D$-wave transitions.
The resulting partial widths into $\Lambda_c\, \bar{D}$ and $\Lambda_c\, \bar{D}^\ast$ are likely to be too large to accommodate the small total widths of the observed pentaquarks in Table~\ref{tab:nonstrange}.

The other case we considered in detail is isospin-0 $c \bar{c}s$ pentaquarks.
The relevant B\nobreakdash-O potential with flavor $(I,S) = \bigl( 0, -1 \bigr)$ has B\nobreakdash-O  quantum numbers $\bigl( \frac{1}{2}^{+\prime} \bigr) \frac{1}{2}$.
At small $r$, it is a repulsive color-Coulomb potential offset by the energy of the $\bm{8}$-baryon with light-QCD quantum numbers $\frac{1}{2}^{+\prime}$.
At large $r$, it approaches the $\Xi_c \bar{D}^{(\ast)}$ threshold.
The spin splittings of the pentaquarks were estimated by treating the spin splittings of charm baryons and charm mesons as a first-order perturbation.
The spin splittings of the three pentaquarks are given in Table~\ref{tab:strange} and their energy-level diagram is shown in Fig.~\ref{fig:StrangeEnergyLevels}.
We identify $P_{c\bar{c}s}(4338)^0$ with the $\frac{1}{2}^-$ pentaquark near the $\Xi_c \bar{D}$ threshold.
Its $J^P$ quantum numbers  are in agreement with experiment. 
We identify $P_{c\bar{c}s}(4459)^0$ with one of two pentaquarks near the $\Xi_c \bar{D}^\ast$ threshold whose quantum numbers are $\frac{1}{2}^{- \prime}$ and $\frac{3}{2}^-$.
We did not identify any convincing mechanism for the suppression of the decays of these pentaquarks into the constituents $\Xi_c \bar{D}$ and $\Xi_c \bar{D}^\ast$.

The amplitudes and probabilities for the three  $c \bar c s$ pentaquarks in the $\bm{8}$-hadron basis and in the dihadron basis are given in Table~\ref{tab:strange}.
They were used to determine approximate model-independent ratios of decay rates of the pentaquarks.
The $P_{c\bar{c}s}(4338)^0$ and $P_{c\bar{c}s}(4459)^0$ have $S$-wave decays into the lower-energy charm-hadron pairs $\Lambda_c \bar{D}_s$ and $\Lambda_c \bar{D}^\ast_s$, respectively.
The small partial widths for these decays could come from SU(3)-flavor suppression.
At large $r$, the flavor of the   $\bigl( \frac{1}{2}^{+\prime} \bigr) \frac{1}{2}$ potential is orthogonal to that of these charm-hadron pairs. 
At small $r$, the flavor of the potential is dominated by the $\bm{8}$-baryon with quantum numbers $\frac{1}{2}^{+\prime}$.
The $c \bar{c} s$ pentaquarks are $S$-wave bound states with energies near the $\Xi_c \bar{D}^{(\ast)}$ threshold and with wavefunctions that extend far beyond the range of the potentials.
If their decays into $\Lambda_c\, \bar{D}_s$ and $\Lambda_c\, \bar{D}_s^\ast$ are dominated by the long-distance region where the flavor of the constituents  is orthogonal to $\Lambda_c \bar{D}_s^{(\ast)}$, those decays could have significant flavor suppression.

The proposal for the pattern of  exotic hidden-heavy hadrons in Ref.~\cite{Braa24b} implies that it is largely determined by the spectrum of $\bm{8}$-hadrons in QCD.
The existence of tetraquarks near charm-meson-pair thresholds implies that the energies of $\bm{8}$-mesons are near the critical energies for the ground state in the appropriate B\nobreakdash-O  potentials to be at the static-meson-pair thresholds \cite{Braa24b}.
The existence of $(I,S) = \bigl( \frac{1}{2}, 0 \bigr)$ pentaquarks near the  $\Sigma_c \bar{D}$ and $\Sigma_c \bar{D}^{(\ast)}$ thresholds implies that the energy of the $j^\pi = \frac{3}{2}^+$ $\bm{8}$-baryon is near the critical energy for the ground state in the $\bigl(\frac{3}{2}^+\bigr)\frac{3}{2} \big/ \frac{1}{2}$ potentials to be at the  $\Sigma_c^{(\ast)} \bar{D}^{(\ast)}$ threshold.
The existence of $(I,S) = (0,-1)$ pentaquarks near the  $\Xi_c \bar{D}$ and  $\Xi_c \bar{D}^\ast$ thresholds implies that the energy of the $j^\pi = \frac{1}{2}^+$ $\bm{8}$-baryon is near the critical energy for the ground state in the  $\bigl( \frac{1}{2}^{+\prime} \bigr) \frac{1}{2}$ potentials to be at the  $\Xi_c \bar{D}^{(\ast)}$ threshold.
More accurate estimates of the  $\bm{8}$-hadron energies can be obtained by solving Schr\"odinger equations for the hidden-heavy tetraquarks and pentaquarks that take into account the spin splittings of heavy baryons and heavy mesons nonperturbatively.
The spectrum of $\bm{8}$-hadrons in QCD has been predicted using various constituent models \cite{ATLAS19review}.
The discoveries of the exotic hidden-heavy hadrons provides a powerful incentive for definitive calculations of the spectrum of $\bm{8}$-hadrons using lattice QCD with a static $\bm{8}$ source or lattice NRQCD with a heavy color-octet particle.

Definitive predictions of the properties of exotic hidden-heavy hadrons will require lattice QCD calculations of the relevant B\nobreakdash-O potentials.
Until those potentials are available, the best one can do is develop models for the potentials consistent with the constraints from BOEFT.
Simple models for the B\nobreakdash-O potentials for isospin-0 tetraquarks have been developed and solutions of the Schr\"odinger equation have been used to calculate their spin splittings nonperturbatively \cite{Bram26,Alas26}. Improved models for the B\nobreakdash-O potentials should give more accurate predictions for the energies of the HQSS partners of those exotic hidden-heavy hadrons that have been discovered.
The experimental discoveries of additional hidden-heavy tetraquarks and pentaquarks consistent with those predictions would confirm that the pattern of the exotic hidden-heavy hadrons in QCD is finally understood.

\acknowledgments
This work was supported in part by the U.S. Department of Energy under grant DE-SC0011726.
R. B. acknowledges sponsorship from the Alexander von Humboldt Foundation.
This work contributes to the goals of the US DOE ExoHad Topical Collaboration, Contract DE-SC0023598.

\appendix

\section{Pentaquark Spin Splittings and Amplitudes}
\label{app:PentaquarkSplittingsAmplitudes}

In this Appendix, we describe the calculations of the spin splittings in the HQSS multiplets for the $c\bar{c}$ pentaquarks and the $c\bar{c}s$ pentaquarks. We also describe the calculations of the amplitudes of the pentaquarks in the $\bm{8}$-hadron basis and in the dihadron basis.

\subsection{General Case}
\label{app:GeneralSpinSplittings}

Pentaquarks can be identified as bound states in diabatic B\nobreakdash-O  potentials that at large $r$ approach the threshold for a $\bm{3}$-baryon and a $\bm{\bar{3}}$-meson. 
We take their light-QCD angular momenta to be the integer $j_\mathrm{B}$ and the half-integer $j_\mathrm{M}$ and we take their parities to be $\pi_\mathrm{B}$ and $\pi_\mathrm{M}$. 
The diabatic  B\nobreakdash-O potentials that approach the threshold at large $r$ are among a set of coupled potentials $(j^\pi) j \big/ \ldots \big/ \frac{1}{2}$ that at small $r$ approach the repulsive color-Coulomb potential offset by the energy of an $\bm{8}$-baryon with quantum numbers $j^\pi$.
The parity of the $\bm{8}$-baryon is $\pi = \pi_\mathrm{B} \pi_\mathrm{M}$ and its spin can range from $\lvert j_\mathrm{B} - j_\mathrm{M}\rvert$ to $j_\mathrm{B} + j_\mathrm{M}$.

There could be bound states in any of the coupled potentials $(j^\pi) j \big/ \ldots \big/ \frac{1}{2}$.
We denote the energy of the ground state in the coupled $(j^\pi) j \big/ \ldots \big/ \frac{1}{2}$ potentials by $ \varepsilon_{j^\pi} $.
The static Hamiltonian $\mathbf{H}_0$ is obtained by truncating the diabatic potential at zeroth order in $1/m_Q$.
It is diagonal in the \emph{$\bm{8}$-hadron basis} labeled by $(j^\pi,S_{Q \bar Q})$, where $S_{Q \bar Q}$ is the total spin of the heavy-quark pair.
The static Hamiltonian restricted to the ground states is
\begin{equation}
\bigl( \mathbf{H}_0 \bigr)_{(j^\pi,S_{Q \bar Q}),((j^\pi)^\prime,S_{Q \bar Q}^\prime)}
= \varepsilon_{j^\pi} \, \delta(j^\pi,(j^\pi)^\prime) \, \delta(S_{Q \bar Q},S_{Q \bar Q}^\prime ).
\label{H0}
\end{equation}

The spin-dependent terms in the diabatic potential at first order in $1/m_Q$ reduce at long distances to the spin-splitting term $\mathbf{V}_\mathrm{SS}$ in Eq.~\eqref{VSS}.
The spin-splitting term is diagonal in the \emph{dihadron basis} labeled by $(J_\mathrm{B},J_\mathrm{M})$, where  $(J_\mathrm{B}$ and $J_\mathrm{M})$ are the spins of a heavy baryon and a heavy meson. Its eigenvalues $E_\mathrm{SS}(J_\mathrm{B},J_\mathrm{M})$ are given in Eq.~\eqref{VSS-diag}.
The change of basis from the angular momenta in the $\bm{8}$-hadron basis to those in the dihadron basis  is given by the recoupling formula
\begin{multline}
\vket{\bigl( (j_\mathrm{B}, \frac{1}{2}) J_\mathrm{B},  (j_\mathrm{M}, \frac{1}{2}) J_\mathrm{M} \bigr)}{J}{M_J}
= (-1) \sqrt{\tilde{J}_\mathrm{B} \tilde{J}_\mathrm{M}}
\\
\times 
\sum_{j,S_{Q \bar Q}} \sqrt{\tilde{j} \tilde{S}_{Q \bar Q} }
\begin{Bmatrix}
j_\mathrm{B}          & \frac{1}{2}                     & J_\mathrm{B} \\
j_\mathrm{M}         & \frac{1}{2}                     & J_\mathrm{M} \\
j &  S_{Q \bar Q} & J \end{Bmatrix}
\vket{\bigl( (j_\mathrm{B},j_\mathrm{M})j, (\frac{1}{2},\frac{1}{2})S_{Q \bar Q} \bigr)}{J}{M_J} ,
\label{penta-di}
\end{multline}
where we have introduced the shorthand $\tilde{J} = 2 J+ 1$ and the expression with nine angular momenta inside curly brackets is a Wigner $9j$ symbol.
The entries $\frac{1}{2}$ in the Wigner $9j$ symbol are the spins of the heavy quark and antiquark. 
The factor of $(-1)$ takes into account the change in order of fermionic operators associated with the $\bm{\bar{3}}$-meson and the heavy quark.
The inverse of the recoupling formula in Eq.~\eqref{penta-di} is
\begin{multline}
\vket{\bigl( (j_\mathrm{B},j_\mathrm{M})j, (\frac{1}{2},\frac{1}{2})S_{Q \bar Q} \bigr)}{J}{M_J}
= (-1)  \sqrt{\tilde{j}\,  \tilde{S}_{Q \bar Q} } 
\\
\times 
\sum_{J_\mathrm{B},J_\mathrm{M}} \sqrt{\tilde{J}_\mathrm{B} \tilde{J}_\mathrm{M}}
\begin{Bmatrix}
j_\mathrm{B}          & \frac{1}{2}                     & J_\mathrm{B} \\
j_\mathrm{M}         & \frac{1}{2}                     & J_\mathrm{M} \\
j &  S_{Q \bar Q} & J \end{Bmatrix}
\vket{\bigl( (j_\mathrm{B}, \frac{1}{2}) J_\mathrm{B},  (j_\mathrm{M}, \frac{1}{2}) J_\mathrm{M} \bigr)}{J}{M_J} .
\label{di-penta}
\end{multline}

The energies of the pentaquarks can be approximated by first-order perturbation theory in the spin-splitting potential $\mathbf{V}_\mathrm{SS}$ in Eq.~\eqref{VSS}.
The calculation of the matrix elements of $\mathbf{V}_\mathrm{SS}$ in the $\bm{8}$-hadron basis $(j^\pi,S_{Q \bar Q})$ involves lengthy angular-momentum algebra that has been described extensively in Refs.~\cite{Braa24a} and \cite{Alas26}. 
The matrix elements are simpler for states that are dominated by $S$-wave channels with $L_{Q \bar Q}=0$.
We expect this to be the case for the ground states in the diabatic B\nobreakdash-O potentials. 
Conservation of the total angular momentum $J$ ensures that $\mathbf{V}_\mathrm{SS}$ is block diagonal in $J$.
In the block with pentaquark spin $J$, the matrix elements of $\mathbf{V}_\mathrm{SS}$ in the $\bm{8}$-hadron basis are \cite{Alas26} 
\begin{multline}
\bigl( \mathbf{V}_\mathrm{SS} \bigr)^{J}_{(j,S_{Q \bar Q}),(j^\prime,S_{Q \bar Q}^\prime)}
=  \sqrt{\tilde{j}\, \tilde{S}_{Q \bar Q}\, \tilde{j}^\prime\, \tilde{S}_{Q \bar Q}^\prime}
\\
\times
\sum_{J_\mathrm{B} J_\mathrm{M}} \tilde{J}_\mathrm{B} \tilde{J}_\mathrm{M}
\begin{Bmatrix}
j_\mathrm{B}          & \frac{1}{2}                     & J_\mathrm{B} \\
j_\mathrm{M}         & \frac{1}{2}                     & J_\mathrm{M} \\
j &  S_{Q \bar Q} & J \end{Bmatrix}
\begin{Bmatrix}
j_\mathrm{B}          & \frac{1}{2}                     & J_\mathrm{B} \\
j_\mathrm{M}         & \frac{1}{2}                     & J_\mathrm{M} \\
j^\prime &  S_{Q \bar Q} ^\prime & J \end{Bmatrix}
E_\mathrm{SS}(J_\mathrm{B},J_\mathrm{M}),
\label{<V1SS>^J}
\end{multline} 
where $E_\mathrm{SS}(J_\mathrm{B},J_\mathrm{M})$ is the spin splitting in Eq.~\eqref{VSS-diag}.

In the absence of spin splittings, the pentaquarks corresponding to the ground state of a $c \bar c$ pair  in the coupled  $(j^\pi) j \big/ \ldots \big/ \frac{1}{2}$ potentials have degenerate energies $\varepsilon_{j^\pi}$.
They form a heavy-quark spin symmetry (HQSS) multiplet with parity $P=-\pi$.
The angular momenta in a pentaquark with spin $J$ are constrained by the triangle relations for $(j,S_{Q \bar Q})J$.
If $j=\frac{1}{2}$, the HQSS multiplet is a triplet consisting of a spin-singlet ($S_{Q \bar Q}=0$) state with $J=\frac{1}{2}$ and two spin-triplet ($S_{Q \bar Q}=1$) states with $J=\frac{1}{2},\frac{3}{2}$.
If $j\geq \frac{3}{2}$, the HQSS multiplet is a quartet consisting of a spin-singlet state with $J=j$ and three spin-triplet states with $J=j-1,j,j+1$.

At  first-order perturbation theory in $\mathbf{V}_\mathrm{SS}$, the energies of the pentaquarks with spin $J$ are the eigenvalues of the sum of the static Hamiltonian $\mathbf{H}_0 $ in Eq.~\eqref{H0} and the $J$ block $(\mathbf{V}_\mathrm{SS})^J$ of the spin-splitting potential in Eq.~\eqref{<V1SS>^J}.
They are functions of the ground-state energies $\varepsilon_{j^\pi}$, the heavy-meson spin splitting $\Delta_\mathrm{M}$, and the heavy-baryon spin splitting $\Delta_\mathrm{B}$.
If there are bound states in only the coupled potentials $(j^\pi) j \big/ \ldots \big/ \frac{1}{2}$, the static Hamiltonian $\mathbf{H}_0$ is a unit matrix multiplied by $\varepsilon_{j^\pi}$.
The spin splittings of the pentaquarks with spin $J$ are the eigenvalues of the $J$ block $(\mathbf{V}_\mathrm{SS})^J$ of the spin-splitting potential in Eq.~\eqref{<V1SS>^J} with $j^\prime = j$. 

The amplitudes for each of the pentaquarks in the $\bm{8}$-hadron basis or in the dihadron basis can be obtained as a byproduct of the calculation of their energies.
The amplitudes $A^\mathrm{P}_{(j^\pi,S_{Q \bar Q})}$ for a pentaquark $\mathrm{P}$ with spin $J$ in the $\bm{8}$-hadron basis are given by the normalized eigenvectors of the $J$ block of $\mathbf{H}_0 + \mathbf{V}_\mathrm{SS}$.
If there are bound states in only the coupled potentials $(j^\pi) j \big/ \ldots \big/ \frac{1}{2}$, the amplitudes are functions of $\Delta_\mathrm{B}/\Delta_\mathrm{M}$.
The amplitudes for a pentaquark $\mathrm{P}$ in the dihadron basis can be obtained from those in the $\bm{8}$-hadron basis by using Eq.~\eqref{penta-di} to recouple the angular momenta.
Given the amplitudes $A^\mathrm{P}_{(j^\pi,S_{Q \bar Q})}$ for a pentaquark $\mathrm{P}$ with spin $J$  in the $\bm{8}$-hadron basis, the amplitudes in the dihadron basis are
\begin{equation}
A^\mathrm{P}_{(J_B,J_M)}= (-1)\, \sqrt{ \tilde{J}_\mathrm{B} \tilde{J}_\mathrm{M} }\,
 \sum_{j,S_{Q \bar Q} } \sqrt{\tilde{j}  \tilde{S}_{Q \bar Q}} 
\begin{Bmatrix}
j_\mathrm{B}         & \frac{1}{2}                     & J_\mathrm{B} \\
j_\mathrm{M}          & \frac{1}{2}                     & J_\mathrm{M} \\
j &  S_{Q \bar Q} & J \end{Bmatrix}
A^\mathrm{P}_{(j^\pi,S_{Q \bar Q})} .
\label{Ap-Adh}
\end{equation}

\subsection{Nonstrange Hidden-charm Pentaquarks}
\label{app:NonstrangeSpinSplittings}

We identify the observed $c \bar c$ pentaquarks as members of an HQSS multiplet corresponding to the ground state for a $c \bar c$ pair in the $\bigl( \frac{3}{2}^+ \bigr) \frac{3}{2} \big/ \frac{1}{2}$ potentials with $(I,S) = \bigl( \frac{1}{2}, 0 \bigr)$ that approach the static-hadron-pair threshold corresponding  to $\Sigma_c^{(\ast)} \bar D^{(\ast)}$ at large $r$.
We denote the energy of the ground state with respect to the spin-weighted $\Sigma_c^{(\ast)} \bar D^{(\ast)}$ threshold by $\varepsilon_{3/2}$.
The HQSS multiplet consists of a spin-singlet state with $J^P=\frac{3}{2}^-$ and spin-triplet states $\frac{1}{2}^-$, $\frac{3}{2}^-$, and $\frac{5}{2}^-$.
The static Hamiltonian $\mathbf{H}_0$ restricted to the subspace spanned by the ground states is a diagonal  $4\times 4$ matrix whose diagonal entries are $\varepsilon_{3/2}$.

We can approximate the energies of the four pentaquarks using first-order perturbation theory in the spin-splitting potential $\mathbf{V}_\mathrm{SS}$ in Eq.~\eqref{VSS}.
The entries of the matrix for the $J$ block of $\mathbf{V}_\mathrm{SS}$ are given by Eq.~\eqref{<V1SS>^J} with $j = \frac{3}{2}$, $j_\mathrm{B} = 1$, and $j_\mathrm{M} = \frac{1}{2}$. 
If a pentaquark has $J^P =\frac{1}{2}^-$ or $\frac{5}{2}^-$, the triangle relations for $(j,S_{Q \bar Q})J$ allow only spin triplet.
The spin splitting of the pentaquark is the entry of the $J$ block of $\mathbf{V}_\mathrm{SS}$ with $S_{Q \bar Q} = S_{Q \bar Q}^\prime=1$, which is linear in $\Delta_\mathrm{B}$ and $\Delta_\mathrm{M}$.
If $J^P=\frac{3}{2}^-$, the triangle relations allow both spin singlet and spin triplet.
The energies of the $\frac{3}{2}^-$ and $\frac{3}{2}^{-\prime}$ pentaquarks must be calculated using degenerate perturbation theory in $\mathbf{V}_\mathrm{SS}$.
The $J=\frac{3}{2}$ block of $\mathbf{V}_\mathrm{SS}$ is the $2\times2$ matrix given by Eq.~\eqref{<V1SS>^J} with $S_{Q \bar Q}= 0,1$ and $S_{Q \bar Q}^\prime = 0,1$.
The spin splittings of the two $\frac{3}{2}^-$ pentaquarks are the eigenvalues of that matrix.
The energies of the four pentaquarks are
\begin{subequations}
\begin{align}
E\bigl(\tfrac{1}{2}^-\bigr) &= \varepsilon_{3/2} - \tfrac{5}{12} \Delta_\mathrm{M} - \tfrac{5}{9} \Delta_\mathrm{B} ,
\\
E\bigl(\tfrac{3}{2}^-\bigr) &= \varepsilon_{3/2} - \tfrac{1}{12}\Delta_\mathrm{M} - \tfrac{1}{9} \Delta_\mathrm{B} 
- \tfrac{1}{9} \sqrt{9 \Delta_\mathrm{M}^2 - 21\Delta_\mathrm{M} \Delta_\mathrm{B}  + 16  \Delta_\mathrm{B}^2},
\\
E\bigl(\tfrac{3}{2}^{-\prime}\bigr) &= \varepsilon_{3/2} - \tfrac{1}{12}\Delta_\mathrm{M} - \tfrac{1}{9} \Delta_\mathrm{B} 
+ \tfrac{1}{9}  \sqrt{ 9\Delta_\mathrm{M}^2  - 21  \Delta_\mathrm{M} \Delta_\mathrm{B} + 16 \Delta_\mathrm{B}^2},
\\
E\bigl(\tfrac{5}{2}^-\bigr) &= \varepsilon_{3/2} + \tfrac{1}{4} \Delta_\mathrm{M} + \tfrac{1}{3} \Delta_\mathrm{B} .
\end{align}
\label{E^J-nonstrange}%
\end{subequations}
Their spin-weighted average is zero.
The numerical values of the spin splittings for $\Delta_\mathrm{M}=142.1$~MeV and $\Delta_\mathrm{B}=66.5$~MeV are given in Table~\ref{tab:nonstrange}.

The amplitudes $A^\mathrm{P}_{(3/2^+,S_{Q \bar Q})}$ for a pentaquark $\mathrm{P}$ with spin $J$ in the $\bm{8}$-hadron basis are the normalized eigenvectors of the $J$  block of the spin-splitting potential $( \mathbf{V}_\mathrm{SS})^J$ in Eq.~\eqref{<V1SS>^J} with $j = j^\prime = \frac{3}{2}$.
The $\frac{1}{2}^-$ and $\frac{5}{2}^-$ pentaquarks are both pure $\bigl( \frac{3}{2}^+, 1 \bigr)$.
The amplitudes for the $\frac{3}{2}^-$ and $\frac{3}{2}^{-\prime}$ pentaquarks are the linear combinations of $\bigl(\frac{3}{2}^+,0\bigr)$ and $\bigl(\frac{3}{2}^+,1\bigr)$ given by the normalized eigenvectors of the $2 \times 2$  matrix for the $J=\frac{3}{2}$  block of the spin-splitting potential in Eq.~\eqref{<V1SS>^J}.
Their numerical values are given in Table~\ref{tab:nonstrange}.

Given the amplitudes $A^\mathrm{P}_{(3/2^+,S_{Q \bar Q})}$ for the pentaquark $\mathrm{P}$ in the $\bm{8}$-hadron basis  in Table~\ref{tab:nonstrange}, its amplitudes $A^\mathrm{P}_{(J_B,J_M)}$ in the dihadron basis can be obtained by using Eq.~\eqref{Ap-Adh} with the sum over $j$  restricted to $j=\frac{3}{2}$.
The amplitudes in the dihadron basis for $J^P = \frac{1}{2}^-$ and $\frac{5}{2}^-$ are given analytically and the amplitudes for $\frac{3}{2}^-$ and $\frac{3}{2}^{-\prime}$ are given numerically in Table~\ref{tab:nonstrange}.

\subsection{Molecular Dominance}
\label{app:MolDomSpinSplittings}

The assumption of molecular dominance is that if some of the B\nobreakdash-O  potentials that approach a static-hadron-pair threshold at large $r$ have bound states, then they all do.
Since the $\bigl( \frac{3}{2}^+ \bigr) \frac{3}{2} \big/ \frac{1}{2}$ potentials that approach the static-hadron-pair threshold corresponding to $\Sigma_c^{(\ast)} \bar D^{(\ast)}$ have bound states, molecular dominance implies that the $\bigl( \frac{1}{2}^{+\prime} \bigr) \frac{1}{2}$ potential must also have bound states.
The ground-state for a $c \bar c$ pair in the $\bigl( \frac{3}{2}^+ \bigr) \frac{3}{2} \big/ \frac{1}{2}$ potentials consists of a spin-singlet state 
$J^P=\frac{3}{2}^-$ and spin-triplet states $\frac{1}{2}^-$, $\frac{3}{2}^-$, and $\frac{5}{2}^-$.
The ground-state for a $c \bar c$ pair in the $\bigl( \frac{1}{2}^+ \bigr) \frac{1}{2}$ potential consists of a spin-singlet state $J^P=\frac{1}{2}^-$ and spin-triplet states $\frac{1}{2}^-$ and$\frac{3}{2}^-$.
We denote the energies of the ground states in the $\bigl( \frac{3}{2}^+ \bigr) \frac{3}{2} \big/ \frac{1}{2}$ and $\bigl( \frac{1}{2}^{+\prime} \bigr)  \frac{1}{2}$  potentials with respect to the spin-weighted $\Sigma_c^{(\ast)} \bar D^{(\ast)}$ threshold by $\varepsilon_{3/2}$ and $\varepsilon_{1/2}$.
The static Hamiltonian $\mathbf{H}_0$ restricted to the subspace spanned by those ground states is a diagonal  $7\times 7$ matrix whose diagonal entries are either $\varepsilon_{3/2}$ or $\varepsilon_{1/2}$.

We can approximate the energies of the seven pentaquarks using first-order perturbation theory in the spin-splitting potential $\mathbf{V}_\mathrm{SS}$ in Eq.~\eqref{VSS}.
The $J$ block of the spin-splitting potential $\mathbf{V}_\mathrm{SS}$ is given by Eq.~\eqref{<V1SS>^J} with $j = \frac{3}{2}$ or $\frac{1}{2}$, $j_\mathrm{B} = 1$, and $j_\mathrm{M} = \frac{1}{2}$. 
If $\mathbf{V}_\mathrm{SS}$ is treated as a first-order perturbation, the energies of the seven pentaquarks can be calculated by diagonalizing $\mathbf{H}_0 + \mathbf{V}_\mathrm{SS}$.
If $J^P = \frac{5}{2}^-$, the triangle relations for $(j,S_{Q \bar Q})J$ allow only $(j^\pi, S_{Q\bar Q}) =(\frac{3}{2}^+,1)$.
The energy of the $\frac{5}{2}^-$ pentaquark is the sum of $\varepsilon_{3/2}$ and the entry of $\mathbf{V}_\mathrm{SS}$ with $j = \frac{3}{2}$ and $S_{Q \bar Q} = S_{Q \bar Q}^\prime=1$, which is linear in $\Delta_\mathrm{B}$ and $\Delta_\mathrm{M}$.
If $\lvert \varepsilon_{1/2}-\varepsilon_{3/2}\rvert$ is small compared to the spin splittings $\Delta_\mathrm{M}$ and $\Delta_\mathrm{B}$, the energies of the pentaquarks with $J^P = \frac{3}{2}^-$ or $\frac{1}{2}^-$ must be calculated using degenerate perturbation theory.
If $J^P = \frac{3}{2}^-$, the triangle relations allow $(j^\pi, S_{Q\bar Q}) =(\frac{3}{2}^+,0)$, $(\frac{3}{2}^+,1)$, and $(\frac{1}{2}^+,1)$. 
The $J = \frac{3}{2}$ block of the static Hamiltonian is the diagonal matrix $\mathbf{H}_0  = \diag(\varepsilon_{3/2}, \varepsilon_{3/2}, \varepsilon_{1/2})$.
The $J = \frac{3}{2}$ block of  $\mathbf{V}_\mathrm{SS}$ is a $3\times3$ matrix given by Eq.~\eqref{<V1SS>^J}.
The energies of the three $\frac{3}{2}^-$ pentaquarks are the eigenvalues of the $J = \frac{3}{2}$ block of $\mathbf{H}_0 + \mathbf{V}_\mathrm{SS}$.
If $J^P=\frac{1}{2}^-$, the triangle relations allow $(\frac{3}{2}^+,1)$, $(\frac{1}{2}^+,0)$, and $(\frac{1}{2}^+,1)$.
The $J = \frac{1}{2}$ block of the static Hamiltonian is the diagonal matrix $\mathbf{H}_0  = \diag(\varepsilon_{3/2}, \varepsilon_{1/2}, \varepsilon_{1/2})$.
The $J = \frac{1}{2}$ block of $\mathbf{V}_\mathrm{SS}$ is a $3\times3$ matrix given by Eq.~\eqref{<V1SS>^J}.
The energies of the three $\frac{1}{2}^-$ pentaquarks are the eigenvalues of the $J = \frac{1}{2}$ block of $\mathbf{H}_0 + \mathbf{V}_\mathrm{SS}$.
The energies of the three $\frac{3}{2}^-$ pentaquarks and the three $\frac{1}{2}^-$ pentaquarks are complicated functions of $\varepsilon_{1/2}$, $\varepsilon_{3/2}$,  $\Delta_\mathrm{B}$, and $\Delta_\mathrm{M}$.

Molecular dominance with ideal mixing is the special case in which the two ground-state energies are equal:  $\varepsilon_{1/2} = \varepsilon_{3/2}$.
The expressions for the energies of the seven pentaquarks simplify dramatically, because the static Hamiltonian $\mathbf{H}_0$ is diagonal.
The spin-splitting potential $\mathbf{V}_\mathrm{SS}$ is diagonal in the dimeson basis $(J_\mathrm{B},J_\mathrm{M})$ with $J_\mathrm{B}=\frac{1}{2},\frac{3}{2}$ and $J_\mathrm{M}=0,1$.
Its diagonal entries are given by Eq.~\eqref{<V1SS>^J}.
The energies of the seven pentaquarks are 
\begin{subequations}
\begin{align}
E\bigl(\tfrac{1}{2}^-\bigr) &= \varepsilon_{3/2} - \tfrac{3}{4} \Delta_\mathrm{M} -\tfrac{2}{3} \Delta_\mathrm{B},
\\
E\bigl(\tfrac{3}{2}^-\bigr) &= \varepsilon_{3/2}  - \tfrac{3}{4} \Delta_\mathrm{M} + \tfrac{1}{3} \Delta_\mathrm{B},
\\
E\bigl(\tfrac{3}{2}^{-\prime}\bigr) &= E\bigl(\tfrac{1}{2}^{-\prime}\bigr)  = \varepsilon_{3/2}   +\tfrac{1}{4} \Delta_\mathrm{M} -\tfrac{2}{3} \Delta_\mathrm{B},
\\
E\bigl(\tfrac{5}{2}^-\bigr) &= E\bigl(\tfrac{3}{2}^{-\prime\prime}\bigr) = E\bigl(\tfrac{1}{2}^{-\prime\prime}\bigr)
= \varepsilon_{3/2} +\tfrac{1}{4} \Delta_\mathrm{M} + \tfrac{1}{3} \Delta_\mathrm{B}.
\end{align}
\label{dEmolecule}%
\end{subequations}
Their spin-weighted average is zero.
The numerical values of the spin splittings for $\Delta_\mathrm{M}=142.1$~MeV and $\Delta_\mathrm{B}=66.5$~MeV are given in Table~\ref{tab:molecule}.

The amplitudes for the pentaquarks with molecular dominance simplify dramatically in the limit of  ideal mixing $\varepsilon_{1/2} = \varepsilon_{3/2}$.
The four spin splittings in Eqs.~\eqref{dEmolecule} are equal to the spin splittings of the $\Sigma_c\bar{D}$, $\Sigma_c^\ast\bar{D}$, $\Sigma_c\bar{D}^\ast$, and $\Sigma_c^\ast\bar{D}^\ast$ thresholds, respectively.
This suggests that the $\frac{1}{2}^-$ pentaquark is a $\Sigma_c\bar{D}$ molecule,  the $\frac{3}{2}^-$ pentaquark is a $\Sigma_c^\ast\bar{D}$ molecule, the $\frac{3}{2}^{-\prime}$ and $\frac{1}{2}^{-\prime}$ pentaquarks are $\Sigma_c\bar{D}^\ast$ molecules, and the $\frac{5}{2}^-$, $\frac{3}{2}^{-\prime\prime}$, and $\frac{1}{2}^{-\prime\prime}$ pentaquarks are $\Sigma_c^\ast\bar{D}^\ast$ molecules.
In the dihadron basis, the constituents are labeled by $( J_\mathrm{B},J_\mathrm{M})$, where $J_\mathrm{B} = \frac{1}{2},\frac{3}{2}$ and $J_\mathrm{M} = 0,1$.
For the pentaquark $\mathrm{P}$ with spin $J$ and constituents B\,M, the only nonzero amplitude in the dihadron basis can be chosen to be $A^\mathrm{P}_{( J_\mathrm{B},J_\mathrm{M} )}  = +1$.

The amplitudes in the $\bm{8}$-hadron basis $( j^\pi, S_{Q \bar Q})$ can be obtained from those in the  dihadron basis by using the recoupling formula in Eq.~\eqref{di-penta}.
For the pentaquark $\mathrm{P}$ with spin $J$ whose constituents have spins $J_\mathrm{B}$ and $J_\mathrm{M}$, the amplitudes in the  $\bm{8}$-hadron basis are
 \begin{equation}
A^\mathrm{P}_{( j^\pi, S_{Q \bar Q})} = (-1) \sqrt{ \tilde{j}\, \tilde{S}_{Q \bar Q}\, \tilde{J}_\mathrm{B}\, \tilde{J}_\mathrm{M}}
\begin{Bmatrix}
j_\mathrm{B}  & \frac{1}{2}      &  J_\mathrm{B} \\
j_\mathrm{M} & \frac{1}{2}      & J_\mathrm{M} \\
j                      & S_{Q \bar Q} &   J \\
\end{Bmatrix} .
\label{APBM-Molecule}
\end{equation} 
This expression was used to calculate the amplitudes for the $\bm{8}$-hadron basis $(j^\pi, S_{Q \bar Q})$ in Table~\ref{tab:molecule}.

\subsection{Strange Hidden-charm Pentaquarks}
\label{app:StrangeSpinSplittings}

We identify the observed $c \bar{c}s$ pentaquarks as members of an HQSS multiplet corresponding to the ground-state for a $c \bar c$ pair in the  $\bigl( \frac{1}{2}^{+\prime} \bigr) \frac{1}{2}$ potential with $(I,S) = \bigl( 0, -1 \bigr)$.
We denote the energy of the ground-state with respect to the spin-weighted $\Xi_c \bar D^{(\ast)}$ threshold by $\varepsilon_{1/2}$.
The HQSS multiplet consists of a spin-singlet state with $J^P=\frac{1}{2}^-$ and spin-triplet states $\frac{1}{2}^-$ and $\frac{3}{2}^-$.
The static Hamiltonian $\mathbf{H}_0$ restricted to the subspace spanned by those three states is a diagonal  $3\times 3$ matrix whose diagonal entries are $\varepsilon_{1/2}$.

We calculate the energies of the three pentaquarks using first-order perturbation theory in the spin-splitting potential $\mathbf{V}_\mathrm{SS}$ given by 
Eq.~\eqref{VSS} with $\Delta_\mathrm{B} = 0$.
The $J$ block of $\mathbf{V}_\mathrm{SS}$ is given by Eq.~\eqref{<V1SS>^J} with $j = \frac{1}{2}$, $j_\mathrm{B} = 0$, and $j_\mathrm{M} = \frac{1}{2}$.
The spin splitting of the pentaquark is the expectation value of $\mathbf{V}_\mathrm{SS}$, which is linear in $\Delta_\mathrm{M}$. 
If $J^P=\frac{3}{2}^-$, the triangle relations allow only $(j^\pi, S_{Q\bar Q}) =(\frac{1}{2}^+,1)$ .
If $J^P =\frac{1}{2}^-$, the triangle relations allow $(\frac{1}{2}^+,0)$ and $(\frac{1}{2}^+,1)$.
The energies of the three pentaquarks are
\begin{subequations}
\begin{align}
E\bigl(\tfrac{1}{2}^-\bigr) &= \varepsilon_{1/2} - \tfrac{3}{4} \Delta_\mathrm{M},
\label{dEst-12}
\\
E\bigl(\tfrac{1}{2}^{-\prime}\bigr) &= E\bigl(\tfrac{3}{2}^-\bigr) = \varepsilon_{1/2} +\tfrac{1}{4} \Delta_\mathrm{M}.
\label{dEs-12p32}
\end{align}
\label{dEstrange}%
\end{subequations}
Their spin-weighted average is zero.
The numerical values of the spin splittings for $\Delta_\mathrm{M}=142.1$~MeV are given in Table~\ref{tab:strange}.

The spin splittings in Eq.~\eqref{dEstrange} can be obtained most easily by realizing that the spin-splitting potential $\mathbf{V}_\mathrm{SS}$ is diagonal in the dihadron basis $(J_\mathrm{B},J_\mathrm{M})$.
If we set $j_\mathrm{B}=0$ and $j_\mathrm{M}=\frac{1}{2}$, the eigenvalues in Eq.~\eqref{VSS-diag} reduce to $[J_\mathrm{M} (J_\mathrm{M} + 1) - 3]\Delta_\mathrm{M}/2$.
The two spin splittings in Eq.~\eqref{dEstrange} are the same as those of the $\Xi_c \bar D$ and $\Xi_c \bar D^\ast$ thresholds.
This suggests that the $\frac{1}{2}^-$ pentaquark is a $\Xi_c \bar D$ molecule and the $\frac{1}{2}^{-\prime}$ and $\frac{3}{2}^-$ pentaquarks are $\Xi_c \bar D^\ast$ molecules.
In the dihadron basis, the constituents are labeled by $( J_\mathrm{B},J_\mathrm{M})$, where $J_\mathrm{B} = \frac{1}{2}$ and $J_\mathrm{M} = 0,1$.
For the pentaquark $\mathrm{P}$  with spin $J$ whose constituents have spins $J_B$ and $J_M$, the only nonzero amplitude  in the dihadron basis can be chosen to be $A^\mathrm{P}_{( J_\mathrm{B},J_\mathrm{M} )}  = +1$.
The amplitudes for  $\mathrm{P}$ in the $\bm{8}$-hadron basis $( j^\pi, S_{Q \bar Q})$ are then given by Eq.~\eqref{APBM-Molecule}.
The analytic results for the amplitudes in both the dihadron basis and the $\bm{8}$-hadron basis  are given in Table~\ref{tab:strange}.

\section{Pentaquark Decays}
\label{app:PentaquarkDecays}

In this Appendix, we derive approximate model-independent ratios of decay rates of hidden-charm pentaquarks into heavy-hadron pairs whose thresholds are lower than those of the constituents. 
They involve angular-momentum algebra with Wigner $6j$ and $9j$ coefficients.

\subsection{Decays into Lower-Energy Charm-Hadron Pairs}
\label{app:LowerEnergyCharmHadrons}

The  model-independent ratios of decay rates in Ref.~\cite{Braa24a} apply to pentaquarks that are either pure spin singlet ($S_{Q \bar Q}=0$) or pure spin triplet ($S_{Q \bar Q}=1$).
Many of the pentaquarks in Tables~\ref{tab:nonstrange}, \ref{tab:molecule}, and \ref{tab:strange} are linear combinations of spin singlet and spin triplet.
There may still be model-independent ratios but they may be more complicated.

We consider a pentaquark $\mathrm{P}$ associated with an $S$-wave heavy-quark pair.
The angular-momentum state of the pentaquark $\mathrm{P}$ with spin quantum numbers $(J,M_J$) can be expanded in the $\bm{8}$-hadron basis: 
\begin{equation}
\vert \mathrm{P} (M_J)\rangle 
= \sum_{j,S_{Q \bar Q}} A^\mathrm{P}_{(j^\pi,S_{Q \bar Q})}  \vket{(j^\pi,S_{Q \bar Q})}{J}{M_J} ,
\label{P-(j,SQQ)}
\end{equation} 
where $A^\mathrm{P}_{j^\pi,S_{Q \bar Q}} $ is the amplitude for the pentaquark $\mathrm{P}$ to have light-QCD quantum numbers $j^\pi$ and heavy-quark-pair spin $S_{Q \bar Q}$.
The sum is constrained by the triangle relations for $(j,S_{Q \bar Q})J$.

The pentaquark can decay into a heavy baryon $\mathrm{B}$ and a heavy meson $\mathrm{M}$ whose threshold is lower than that of the pentaquark constituents.
The baryon and meson belong to the HQSS multiplets of a  $\bm{3}$-baryon and a $\bm{\bar{3}}$-meson with light-QCD quantum numbers ${j_\mathrm{B}^\prime}^{\pi_\mathrm{B}^\prime}$ and ${j_\mathrm{M}^\prime}^{\pi_\mathrm{M}^\prime}$. 
They have spin quantum numbers $(J_\mathrm{B},m_\mathrm{B})$ and $(J_\mathrm{M},m_\mathrm{M})$.
The decay rate of the pentaquark $\mathrm{P}$ at rest into the heavy baryon $\mathrm{B}$ and the heavy meson $\mathrm{M}$ is
\begin{equation}
\Gamma[ \mathrm{P}  \to \mathrm{B\,M}]
= \frac{p_\mathrm{P,BM}}{32 \pi^2 M_\mathrm{P}^2} \sum_{m_\mathrm{B},m_\mathrm{M}} \int \mathrm{d} \Omega\,
\Big\vert \bigl\langle \mathrm{BM}(\Omega,m_\mathrm{B}, m_\mathrm{M}) \big\vert  T \big\vert \mathrm{P}(M_J)\bigr\rangle \Big\vert^2 ,
\label{Gamma[PtoBM]}
\end{equation}
where $M_\mathrm{P}$ is the mass of the pentaquark, $p_\mathrm{P,BM}$ is the momentum of either $\mathrm{B}$ or $\mathrm{M}$ in the rest frame of $\mathrm{P}$, and $T$ is the relativistically normalized $T$-matrix for the decay.
The sum is over the projections of the spins of $\mathrm{B}$ and $\mathrm{M}$.
The decay rate does not depend on the projection $M_J$ of the pentaquark spin.
The momentum $p_\mathrm{P,BM}$ is determined by $M_\mathrm{P}$ and the masses $M_\mathrm{B}$ and $M_\mathrm{M}$ of $\mathrm{B}$ and M:
\begin{equation}
p_\mathrm{P,BM} = \sqrt{\lambda(M_\mathrm{P}^2, M_\mathrm{B}^2, M_\mathrm{M}^2)} / ( 2 M_\mathrm{P}) ,
\label{pBM}
\end{equation}
where $\lambda$ is defined after Eq.~\eqref{vPBM}. 
The integral over the solid angle in Eq.~\eqref{Gamma[PtoBM]} can be expressed as a sum over orbital-angular-momentum quantum numbers $( L_\mathrm{BM} ,m_\mathrm{BM})$:
\begin{equation}
\Gamma[ \mathrm{P}  \to \mathrm{B\,M}] 
= \frac{p_\mathrm{P,BM}}{8\pi M_\mathrm{P}^2} \sum_{m_\mathrm{B},m_\mathrm{M}} \sum_{L_\mathrm{BM},m_\mathrm{BM}}
 \Big\vert \bigl\langle \mathrm{BM} \bigl( L_\mathrm{BM} ,m_\mathrm{BM},m_\mathrm{B},m_\mathrm{M} \bigr)  \big\vert  T \big\vert \mathrm{P} (M_J) \bigr\rangle \Big\vert^2 .
\label{Gamma[PtoBM]-L}
\end{equation}
The sum over $m_\mathrm{B}$ and $m_\mathrm{M}$ in Eq.~\eqref{Gamma[PtoBM]-L} can be expressed as a sum over the total spin $S$ of $\mathrm{B}$ and $\mathrm{M}$ and its projection $M_S$.
The sum over $m_\mathrm{BM}$ and $M_S$ can further be expressed as a sum over the total angular momentum $J^\prime$ of $\mathrm{B}$ and $\mathrm{M}$ and its projection $M_J^\prime$.
Thus Eq.~\eqref{Gamma[PtoBM]-L} becomes
\begin{equation}
\Gamma[ \mathrm{P}  \to \mathrm{B\,M}] 
= \frac{p_\mathrm{P,BM}}{8\pi M_\mathrm{P}^2}  \sum_{S,L_\mathrm{BM}} \sum_{J^\prime, M_J^\prime} 
 \Big\vert \bigl\langle \mathrm{BM} \bigl( L_\mathrm{BM} ,S, J^\prime,M_J^\prime \bigr)  \big\vert  T \big\vert \mathrm{P} (M_J) \bigr\rangle \Big\vert^2 .
\label{Gamma[PtoBM]-J}
\end{equation}
After inserting the expression for the pentaquark state in Eq.~\eqref{P-(j,SQQ)}, the decay rate in Eq.~\eqref{Gamma[PtoBM]-J} can be simplified by using conservation of $J$ and $M_J$.
The resulting expression for the decay rate of a pentaquark $\mathrm{P}$ with spin $J$ reduces to
\begin{equation}
\Gamma[ \mathrm{P}  \to \mathrm{B\,M}] 
= \frac{p_\mathrm{P,BM}}{8\pi M_\mathrm{P}^2} \sum_{S,L_\mathrm{BM}}
\Biggl\lvert  \sum_{j,S_{Q \bar Q}} A^\mathrm{P}_{(j^\pi,S_{Q \bar Q})}
\vbra{\bigl( (J_\mathrm{B}, J_\mathrm{M}) S, L_\mathrm{BM} \bigr)}{J}{M_J}
T \vket{(j^\pi,S_{Q \bar Q})}{J}{M_J} \Biggr\rvert^2.
 \label{Gamma[PtoBM]-S}
\end{equation} 

\subsection{Recoupling of Angular Momenta}
\label{app:Recoupling}

In the heavy-quark limit, the $T$ matrix has heavy-quark-spin symmetry (HQSS) and it conserves the B\nobreakdash-O angular momentum $L$.
In order to exploit HQSS and the conservation of $L$, it is necessary to reorder the angular momenta of the baryon-meson pair in Eq.~\eqref{Gamma[PtoBM]-S} so that those that couple to the total angular momentum $J$  are the total light-QCD angular momentum $j^\prime$ and the heavy-quark-pair spin $S_{Q \bar Q}^\prime$.
The change of the angular-momentum basis can be carried out conveniently in two steps.
The first step is the recoupling of $j_\mathrm{B}^\prime$,  $j_\mathrm{M}^\prime$, and the spins $s_1=\frac{1}{2}$ and $s_2=\frac{1}{2}$ of the $Q$ and $\bar Q$ using an identity involving a Wigner $9j$ symbol:
\begin{multline}
\vket{\bigl((j_\mathrm{B}^\prime,s_1)J_\mathrm{B},(j_\mathrm{M}^\prime,s_2)J_\mathrm{M}\bigr)}{S}{M_S}
=
(-1)
\sqrt{ \tilde{J}_\mathrm{B} \tilde{J}_\mathrm{M} }
\sum_{j^\prime,S_{Q \bar Q}^\prime} \sqrt{\tilde{j}^\prime \tilde{S}_{Q \bar Q}^\prime}
\\
 \times 
\begin{Bmatrix}
j_\mathrm{B}^\prime               & s_1                      & J_\mathrm{B} \\
j_\mathrm{M}^\prime                & s_2                    & J_\mathrm{M} \\
j^\prime & S_{Q \bar Q} ^\prime& S \end{Bmatrix}
\vket{\bigl((j_\mathrm{B}^\prime ,j_\mathrm{M}^\prime )j^\prime, (s_1,s_2)S_{Q \bar Q}^\prime \bigr)}{S}{M_S}.
\label{di-tetra1}
\end{multline}
The sign $(-1)$ takes into account the change in the order of fermionic operators associated with the $\bm{\bar{3}}$-meson and the heavy quark.
The second step in the change of the angular-momentum basis is the recoupling of $j^\prime$, $S_{Q \bar Q}^\prime$, and  $L_\mathrm{BM}$ using an identity involving a Wigner $6j$ symbol:
\begin{multline}
\vket{\bigl( (j^\prime ,S_{Q \bar Q}^\prime)S, L_\mathrm{BM} \bigr)}{J}{M_J}
=
\sqrt{ \tilde{S}} \sum_L (-1)^{S_{Q \bar Q}^\prime + S  + L_\mathrm{BM} + L} \sqrt{  \tilde{L}} 
\\
\times 
\begin{Bmatrix}
j^\prime & S_{Q \bar Q}^\prime & S \\
J  & L_\mathrm{BM} & L  \end{Bmatrix}
\vket{\bigl( (j^\prime,L_\mathrm{BM}) L, S_{Q \bar Q} ^\prime \bigr)}{J}{M_J}.
\label{ketjSL-ketjLS}
\end{multline}
After the changes in the angular-momentum bases in Eqs.~\eqref{di-tetra1} and \eqref{ketjSL-ketjLS}, the baryon-meson angular-momentum state in Eq.~\eqref{Gamma[PtoBM]-S} becomes
\begin{multline}
\vket{\bigl( (J_\mathrm{B}, J_\mathrm{M}) S, L_\mathrm{BM} \bigr)}{J}{M_J}
= (-1)^{ L_\mathrm{BM} +1} \sqrt{\tilde{J}_\mathrm{B}  \tilde{J}_\mathrm{M} } 
\sum_{j^\prime,S_{Q \bar Q}^\prime} (-1)^{S_{Q \bar Q}^\prime} \sqrt{\tilde{j}^\prime \tilde{S}_{Q \bar Q}^\prime }
 \sum_L (-1)^{S+L } \sqrt{\tilde{S} \tilde{L} }
\\
\times 
\begin{Bmatrix}
j ^\prime&  S_{Q \bar Q}^\prime & S  \\
J &  L_\mathrm{BM} & L \end{Bmatrix}
\begin{Bmatrix}
j_\mathrm{B}^\prime           & \frac{1}{2}                     & J_\mathrm{B}  \\
j_\mathrm{M} ^\prime          & \frac{1}{2}                     & J_\mathrm{M}  \\
j^\prime &  S_{Q \bar Q} ^\prime& S \end{Bmatrix}
\vket{\bigl( ( ( j_\mathrm{B}^\prime ,j_\mathrm{M}^\prime )j^\prime, L_\mathrm{BM} ) L, S_{Q \bar Q}^\prime \bigr)}{J}{M_J}.
\label{di-penta12}
\end{multline}
After making the substitution in Eq.~\eqref{di-penta12} for the baryon-meson state in the $T$-matrix element in Eq.~\eqref{Gamma[PtoBM]-S}, we can use conservation of the B\nobreakdash-O angular momentum to deduce $L=j$ and  conservation of the heavy-quark-pair spin to deduce $S_{Q \bar Q}^\prime = S_{Q \bar Q}$.
The $T$-matrix element between the angular-momentum states of the pentaquark and the baryon-meson pair can be expressed in terms of a reduced $T$-matrix element:
\begin{multline}
\vbra{\bigl( ( ( {j_\mathrm{B}^\prime}^{\! \pi_\mathrm{B}^\prime} ,{j_\mathrm{M}^\prime}^{\! \pi_\mathrm{M}^\prime} )j^\prime, L_\mathrm{BM} ) L, S_{Q \bar Q}^\prime \bigr)}{J}{M_J} T \vket{(j^\pi,S_{Q \bar Q})}{J}{M_J}
 \\
     =  \delta_{L,j} \, \delta_{S_{Q \bar Q}^\prime,S_{Q \bar Q} }\,  
 \bigl\langle  \bigl( ( {j_\mathrm{B}^\prime}^{\! \pi_\mathrm{B}^\prime} ,{j_\mathrm{M}^\prime}^{\! \pi_\mathrm{M}^\prime} )j^\prime, L_\mathrm{BM} \bigr) j \big\Vert T \big\Vert j^\pi\bigr\rangle .
\end{multline} 
The resulting expression for the decay rate of the pentaquark in Eq.~\eqref{Gamma[PtoBM]-S} is 
\begin{multline}
\Gamma[ \mathrm{P}  \to \mathrm{B\,M}] 
= \frac{p_\mathrm{P,BM}}{8\pi M_\mathrm{P}^2} \, \tilde{J}_\mathrm{B}  \tilde{J}_\mathrm{M}  \sum_{S,L_\mathrm{BM}} \tilde{S}\, 
\Biggl\lvert \sum_{j,S_{Q \bar Q}} A^\mathrm{P}_{(j^\pi,S_{Q \bar Q})} (-1)^{j + S + S_{Q \bar Q}} \sqrt{\tilde{j} \tilde{S}_{Q \bar Q} } \\
\times 
\sum_{j^\prime} \sqrt{\tilde{j}^\prime} 
\begin{Bmatrix}
j ^\prime&  S_{Q \bar Q} & S  \\
J &  L_\mathrm{BM} & j \end{Bmatrix}
\begin{Bmatrix}
j_\mathrm{B}^\prime           & \frac{1}{2}                     & J_\mathrm{B}  \\
j_\mathrm{M}^\prime           & \frac{1}{2}                     & J_\mathrm{M}  \\
j^\prime &  S_{Q \bar Q} & S \end{Bmatrix}
 \bigl\langle  \bigl( ( {j_\mathrm{B}^\prime}^{\! \pi_\mathrm{B}^\prime} ,{j_\mathrm{M}^\prime}^{\! \pi_\mathrm{M}^\prime} ) j^\prime, L_\mathrm{BM} \bigr) j \big\Vert T \big\Vert j^\pi \bigr\rangle \biggr\rvert^2 ,
\label{Gamma[PtoBM]-reduced-T}
\end{multline}
where $A^\mathrm{P}_{(j^\pi,S_{Q \bar Q})}$ is the amplitude for the pentaquark $\mathrm{P}$ in the $\bm{8}$-hadron basis $(j^\pi,S_{Q \bar Q})$.

\subsection{Model-Independent Ratios}
\label{app:ModelIndependentPredictions}

The reduced $T$-matrix element $\langle (j^\prime, L_\mathrm{BM} \bigr) j  \Vert T \Vert  j^\pi \rangle$ in Eq.~\eqref{Gamma[PtoBM]-reduced-T} scales like $v_\mathrm{P,BM}^{L_\mathrm{BM}}$ at leading order in $v_\mathrm{P,BM}$, where $v_\mathrm{P,BM}$ is the relative velocity between the heavy baryon and the heavy meson defined in Eq.~\eqref{vPBMrel}.
If $v_\mathrm{P,BM}$ is sufficiently small, the $T$-matrix element will be dominated by the term in the sum over $L_\mathrm{BM}$ with the smallest possible value $L_\mathrm{min}$.
That minimum value $L_\mathrm{min} $ is determined by the triangle relations for $(j_\mathrm{B}^\prime,j_\mathrm{M}^\prime) j^\prime$ and $(j^\prime, L_\mathrm{BM} ) j$ and by the parity constraint $\pi = \pi_\mathrm{B}^\prime \pi_\mathrm{M}^\prime (-1)^{L_\mathrm{BM}}$.
There will be approximate model-independent ratios of decay rates if the triangle relations and the parity constraint allow only a single pair of values of $j^\prime$ and $j$.
The reduced $T$-matrix element $\langle  ( j^\prime, L_\mathrm{min} \bigr) j \Vert T \Vert j^\pi \rangle$ is then a multiplicative factor in the $T$-matrix element  that scales as $v_\mathrm{P,BM}^{L_\mathrm{min}}$.
In the absence of spin splittings of the pentaquarks, heavy baryons, and heavy mesons,
the square of the reduced $T$-matrix element would cancel in ratios of decay rates.
The phase space factors $p_\mathrm{P,BM}/\bigl( 8\pi M_\mathrm{P}^2 \bigr)$ in Eq.~\eqref{Gamma[PtoBM]-reduced-T} , which provide another factor of $v_\mathrm{P,BM}$, would also cancel.
The velocity $v_\mathrm{P,BM}$ is most sensitive to the spin splittings through the multiplicative  factor of $M_\mathrm{P} - M_\mathrm{B}  - M_\mathrm{M}$ in $\lambda(M_\mathrm{P}^2,M_\mathrm{A}^2,M_\mathrm{B}^2)$ in Eq.~\eqref{vPBMrel}.
We assume the leading effects of the spin splittings are determined by the factor $v_\mathrm{P,BM}^{2 L_\mathrm{min} + 1}$, where $v_\mathrm{P,BM}$ is the relative velocity defined in Eq.~\eqref{vPBMrel}.

We proceed to identify the conditions for approximate model-independent ratios when only the coupled potentials $(j^\pi) j \big/ \ldots \big/ \frac{1}{2}$ have bound states.
At large $r$, the potentials approach the threshold for a  $\bm{3}$-baryon and a $\bm{\bar{3}}$-meson with light-QCD quantum numbers $j_\mathrm{B}^{\pi_\mathrm{B}}$ and $j_\mathrm{M}^{\pi_\mathrm{M}}$.
For simplicity, we assume the parities of the $\bm{3}$-hadrons at the threshold for the decay products are the same as those at the threshold for the bound states: $\pi_\mathrm{B}^\prime = \pi_\mathrm{B}$ and $\pi_\mathrm{M}^\prime = \pi_\mathrm{M}$.
The parity constraint then reduces to the condition that $L_\mathrm{min}$ is even.
The triangle relations for $(j_\mathrm{B}^\prime,j_\mathrm{M}^\prime) j^\prime$ and $(j^\prime, L_\mathrm{min} ) j$ imply that here is a unique value of $j^\prime$ only if $L_\mathrm{min}=0$, $j_\mathrm{B}^\prime=0$, $\lvert j_\mathrm{B}^\prime - j_\mathrm{M}^\prime\rvert = j + L_\mathrm{min}$, or $\lvert j - L_\mathrm{min}\rvert = j_\mathrm{B}^\prime + j_\mathrm{M}^\prime$, in which case the unique value is $j$, $j _\mathrm{M}^\prime$, $\lvert j_\mathrm{B}^\prime - j_\mathrm{M}^\prime \rvert$, or $j_\mathrm{B}^\prime + j_\mathrm{M}^\prime$, respectively.

There are approximate model-independent ratios of the decay rates in Eq.~\eqref{Gamma[PtoBM]-reduced-T} for a given $j^\pi$ only if there is a unique value of $j^\prime$ for which the reduced $T$-matrix element with minimal orbital angular momentum $L_\mathrm{min}$ is nonzero.
The sums over $L_\mathrm{BM}$, $j$, and $j^\prime$ can then be restricted to those values.
The reduced $T$-matrix element, which scales like $v_\mathrm{P,BM}^{L_\mathrm{min}}$, can be factored out of the remaining sums over $S$ and $S_{Q \bar Q}$.
The factors in the decay rate for $\mathrm{P} \to \mathrm{B}\mathrm{M}$ in Eq.~\eqref{Gamma[PtoBM]-reduced-T} that depend on  angular momenta and the relative velocity therefore reduce to
\begin{multline}
\Gamma[ \mathrm{P}  \to \mathrm{B M}] 
\propto v_\mathrm{P,BM}^{2L_\mathrm{min} +1}\, \tilde{J}_\mathrm{B} \,  \tilde{J}_\mathrm{M}\,\tilde{j}\,  \tilde{j}^\prime\,  \sum_S \tilde{S}\, 
\Biggl\lvert
\sum_{S_{Q \bar Q}} A^\mathrm{P}_{(j^\pi,S_{Q \bar Q})}  (-1)^{S_{Q \bar Q}} \sqrt{\tilde{S}_{Q \bar Q} }
\\
\times 
\begin{Bmatrix}
j^\prime&  S_{Q \bar Q} & S  \\
J &  L_\mathrm{min} & j \end{Bmatrix}
\begin{Bmatrix}
j_\mathrm{B}^\prime           & \frac{1}{2}                     & J_\mathrm{B}  \\
j_\mathrm{M}^\prime           & \frac{1}{2}                     & J_\mathrm{M}  \\
j^\prime &  S_{Q \bar Q}& S \end{Bmatrix} \Biggr\rvert^2.
\label{Gamma[PtoBM]-A}
\end{multline}
This is the product of a velocity factor $v_\mathrm{P,BM}^{2L_\mathrm{min} +1}$ and a number that is determined by the amplitudes $A^\mathrm{P}_{(j^\pi,S_{Q \bar Q})}$ in the $\bm{8}$-hadron basis and by angular-momentum algebra.
The model-independent ratio of the decay rates for $\mathrm{P} \to \mathrm{B\,M}$ and $\mathrm{P^\prime} \to \mathrm{B^\prime\,M^\prime}$ is the ratio of two such factors.
These expressions with $j= \frac{3}{2}$, $j^\prime = \frac{1}{2}$, $j_\mathrm{B}^\prime = 0$, $j_\mathrm{M}^\prime = \frac{1}{2}$, and $L_\mathrm{min}=2$ were used with amplitudes $A^\mathrm{P}_{(3/2^+,S_{Q \bar Q})}$ from Table~\ref{tab:nonstrange} to calculate the ratios  in Eqs.~\eqref{nonstrangeGammaRatios(*)} of the decay rates of $c \bar c$ pentaquarks into $\Lambda_c\, \bar{D}$ and $\Lambda_c\, \bar{D}^\ast$.
The same expressions were also used with the amplitude $A^\mathrm{P}_{3/2^+,1} = -1$ from Table~\ref{tab:molecule} to calculate the branching ratio in Eq.~\eqref{BR[5/2->D,D*]}  of the $\frac{5}{2}^-$ pentaquark into $\Lambda_c\, \bar{D}$ and $\Lambda_c\, \bar{D}^\ast$ in the case of molecular dominance with ideal mixing.

If there are pentaquarks for which $A^\mathrm{P}_{(j^\pi,S_{Q \bar Q})}$ is nonzero for only $S_{Q \bar Q} = 0$ or only  $S_{Q \bar Q} = 1$, then that amplitude can be set equal to 1.
For those pentaquarks, the factors in the decay rate for $\mathrm{P}  \to \mathrm{B\,M}$ in Eq.~\eqref{Gamma[PtoBM]-A} that depend on angular momenta and the relative velocity reduce to
\begin{equation}
\Gamma[ \mathrm{P}  \to \mathrm{B M}] 
\propto v_\mathrm{P,BM}^{2L_\mathrm{min} +1} \, \tilde{J}_\mathrm{B} \,  \tilde{J}_\mathrm{M}\,  \tilde{j}\, \tilde{j}^\prime\, \tilde{S}_{Q \bar Q} \, 
\sum_S \tilde{S}\, \begin{Bmatrix}
j ^\prime&  S_{Q \bar Q} & S  \\
J &  L_\mathrm{min} & j \end{Bmatrix}^2
\begin{Bmatrix}
j_\mathrm{B}^\prime           & \frac{1}{2}                     & J_\mathrm{B}  \\
j_\mathrm{M}^\prime           & \frac{1}{2}                     & J_\mathrm{M}  \\
j^\prime &  S_{Q \bar Q}& S \end{Bmatrix}^2 .
\label{Gamma[PtoBM]-6j9j}
\end{equation}
The factor multiplying $v_\mathrm{P,BM}^{2L_\mathrm{min} +1}$ is completely determined by angular-momentum algebra.
The simpler model-independent ratios for those pentaquarks were given previously in Ref.~\cite{Braa24a}.

The expression for the angular-momentum dependent term in Eq.~\eqref{Gamma[PtoBM]-A} simplifies dramatically if the minimal orbital angular momentum is $L_\mathrm{min} = 0$, which implies $j^\prime=j$.
The $6j$ symbol can be simplified by using the identity
\begin{equation}
\begin{Bmatrix}
j_1 &  j_2 & j_3  \\
j_4 &  0    & j_5 \end{Bmatrix}
= \delta_{j_1,j_5}\, \delta_{j_3,j_4} \frac{1}{\sqrt{ \tilde{j}_1 \tilde{j}_3 }}\,  (-1)^{j_1 +  j_2 + j_3} \, \{j_1 \;  j_2 \; j_3\},
\label{6j-0}
\end{equation}
where $\{j_1 \;  j_2 \; j_3\}$ is the triangle symbol, which is equal to 1 if the triangle relation for $(j_1 ,  j_2) j_3$ is satisfied and 0 otherwise.
The expression for the decay rate in Eq.~\eqref{Gamma[PtoBM]-A} for a pentaquark $P$ with spin $J$ reduces to
\begin{equation}
\Gamma[ \mathrm{P}  \to \mathrm{B M}] 
\propto v_\mathrm{P,BM}\, \bigl\lvert A^\mathrm{P}_{(J_B,J_M)} \bigr\rvert^2,
\label{Gamma[PtoBM]-A^2}
\end{equation}
where the amplitude $A^\mathrm{P}_{(J_B,J_M)}$ is
\begin{equation}
A^\mathrm{P}_{(J_B,J_M)}  = (-1) \sqrt{ \tilde{J}_\mathrm{B} \tilde{J}_\mathrm{M} }
\sum_{S_{Q \bar Q}} \sqrt{ \tilde{j} \tilde{S}_{Q \bar Q} } \, \begin{Bmatrix}
j_\mathrm{B}^\prime           & \frac{1}{2}                     & J_\mathrm{B}  \\
j_\mathrm{M}^\prime           & \frac{1}{2}                     & J_\mathrm{M}  \\
j &  S_{Q \bar Q}& J \end{Bmatrix} A^\mathrm{P}_{(j^\pi,S_{Q \bar Q})} .
\label{AJBJM-AjSQQ}
\end{equation}
The model-independent ratio of the decay rates for $\mathrm{P} \to \mathrm{B\,M}$ and $\mathrm{P^\prime} \to \mathrm{B^\prime\,M^\prime}$ is the product of the velocity ratio $v_\mathrm{P,BM}/v_{\mathrm{P}^\prime,\mathrm{B}^\prime \mathrm{M}^\prime}$ and the probability ratio $\bigl\lvert A^\mathrm{P}_{(J_B,J_M)}\bigr\rvert^2/  \bigl\lvert A^{\mathrm{P}^\prime}_{(J_{B^\prime},J_{M^\prime})}\bigr\rvert^2$. 
These expressions with $j= \frac{1}{2}$, $j_\mathrm{B}^\prime = 0$, and $j_\mathrm{M}^\prime = \frac{1}{2}$ were used with amplitudes $A^\mathrm{P}_{(1/2^+,S_{Q \bar Q})}$ from Table~\ref{tab:strange} to calculate  the ratio in Eq.~\eqref{Gamma4459/4338} of the rates for the $c \bar c s$ pentaquark decays $P_{c \bar{c} s} (4338) \to \Lambda_c\, \bar{D}_s$ and $P_{c \bar{c} s} (4459) \to \Lambda_c\, \bar{D}_s^\ast$.

The expression for the amplitude $A^\mathrm{P}_{(J_B,J_M)}$ in Eq.~\eqref{AJBJM-AjSQQ} differs from that in Eq.~\eqref{Ap-Adh} for changing the pentaquark amplitude from the $\bm{8}$-hadron basis to the dihadron basis only in that $j_\mathrm{B}$ and $j_\mathrm{M}$ are replaced by $j_\mathrm{B}^\prime$ and $j_\mathrm{M}^\prime$ and there is no sum over $j$.
This can be used to obtain the  model-independent ratios for $c \bar c s$ pentaquarks more simply.
Since the only possible value of $j$ for B\nobreakdash-O potentials that approach the $\Xi_c \bar{D}^{(*)}$ threshold at large $r$ is $j=\frac{1}{2}$ and since  $j_\mathrm{B}^\prime = j_\mathrm{B} = 0$ and $j_\mathrm{M}^\prime = j_\mathrm{M} =\frac{1}{2}$, the amplitude $A^\mathrm{P}_{(J_B,J_M)}$ in Eq.~\eqref{Ap-Adh} coincides with the dihadron amplitude for the pentaquark $\mathrm{P}$ with spin $J$ and constituents with spins $J_\mathrm{B}$ and $J_\mathrm{M}$.
For each of the three pentaquarks in Table~\ref{tab:strange}, there is only one nonzero dihadron amplitude and its value is $A^\mathrm{P}_{(J_B,J_M)} = +1$.
The right side of Eq.~\eqref{Gamma[PtoBM]-A^2} therefore reduces to $v_\mathrm{P,BM}$.
The model-independent ratio of the decay rates for $\mathrm{P} \to \mathrm{B\,M}$ and $\mathrm{P^\prime} \to \mathrm{B^\prime\,M^\prime}$ reduces to the velocity ratio $v_\mathrm{P,BM}/v_{\mathrm{P}^\prime,\mathrm{B}^\prime \mathrm{M}^\prime}$.

We now consider the conditions for approximate model-independent ratios  under the assumption of molecular dominance.
All the coupled potentials $(j^\pi) j \big/ \ldots \big/ \frac{1}{2}$ with $j$ ranging from $\bigl\lvert j_\mathrm{B} - j_\mathrm{M} \bigr \rvert$ to $j_\mathrm{B} + j_\mathrm{M}$ are assumed to have bound states.
The triangle relations and the parity constraint now require a unique value of $j$ as well as a unique value of $j^\prime$.
We will identify the conditions for approximate model-independent ratios only in the case  where the minimal orbital angular momentum is $L_\mathrm{min} = 0$.
The triangle relations for $(j_\mathrm{B}^\prime,  j_\mathrm{M}^\prime) j$ and $(j_\mathrm{B},j_\mathrm{M}) j$ imply that there is a unique value of $j = j^\prime$ only if $j_\mathrm{B}^\prime = 0$, $j_\mathrm{B}=0$, $\lvert j_\mathrm{B}^\prime - j_\mathrm{M}^\prime \rvert= j_\mathrm{B} + j_\mathrm{M}$, or $\lvert j_\mathrm{B} - j_\mathrm{M} \rvert = j_\mathrm{B} ^\prime+ j_\mathrm{M}^\prime$, in which case that unique value is $j _\mathrm{M}^\prime$, $j_\mathrm{M}$, $j_\mathrm{B} + j_\mathrm{M}$, or $j_\mathrm{B}^\prime + j_\mathrm{M}^\prime$, respectively.

In the case of molecular dominance, there are approximate model-independent ratios of the decay rates in Eq.~\eqref{Gamma[PtoBM]-reduced-T} only if there is a unique pair of values of $j^\prime$ and $j$ for which the reduced $T$-matrix element with minimal orbital angular momentum $L_\mathrm{min}$ is nonzero.
The sums over $L_\mathrm{BM}$,  $j$, and $j^\prime$ can then be restricted to those values.
The reduced $T$-matrix element, which scales like $v_\mathrm{P,BM}^{L_\mathrm{min}}$, can be factored out of the remaining sums over $S$ and $S_{Q \bar Q}$.
The factors in the decay rate for $\mathrm{P} \to \mathrm{B}\,\mathrm{M}$ in Eq.~\eqref{Gamma[PtoBM]-reduced-T} that depend on angular momenta and the relative velocity therefore reduce to Eq.~\eqref{Gamma[PtoBM]-A}.
This expression with $j= \frac{1}{2}$, $j_\mathrm{B}^\prime = 0$, and $j_\mathrm{M}^\prime = \frac{1}{2}$ was used with amplitudes $A^\mathrm{P}_{(1/2^+,S_{Q \bar Q})}$ from Table~\ref{tab:molecule} to calculate the ratios  in Eqs.~\eqref{moleculeBRatios(*)} of the decay rates of $c \bar c$ pentaquarks into $\Lambda_c\, \bar{D}$ and $\Lambda_c\, \bar{D}^\ast$ in the case of molecular dominance with ideal mixing.

\bibliography{bibliography}
\bibliographystyle{JHEP}
	
\end{document}